\documentclass[10pt]{article}

\usepackage{mathrsfs}

\renewcommand{\Re}{{\rm Re \, }}
\newcommand{\im}{{\rm Im \, }}
\newcommand{\re}{{\rm Re \, }}

\usepackage{geometry}
\usepackage{verbatim}
\usepackage{prettyref}
\usepackage{amsmath}
\usepackage{amssymb}
\usepackage{graphicx}
\usepackage{setspace}
\usepackage{esint}
\usepackage{color}
\definecolor{darkgreen}{rgb}{0,0.5,0}
\definecolor{darkblue}{rgb}{0,0,0.6}
\definecolor{purple}{rgb}{0.4,.2,0.7}

\begin{document}
  
\titlepage

\vspace*{1cm}

\begin{center}
{\LARGE \bf Holographic renormalization in no-boundary \\ quantum cosmology

 }

\vspace{2mm}

\end{center}

\begin{center}

Lihui Liu 

\vspace{2mm}

{\it Instituut voor Theoretische Fysica, KU Leuven, Celestijnenlaan 200D}

{\it B-3001 Leuven, Belgium}

\vspace{2mm}

{\tt lihui@itf.fys.kuleuven.be}

\vspace{1.8cm}

{\large \bf Abstract}

\end{center}

\vspace{1mm}

\noindent Recent results of Hartle-Hawking wave functions on asymptotic dS boundaries show non-normalizability, while the bulk origin is not clear. This paper attempts to addresse this problem by studying (Kerr-)dS$_3$ cosmology in Einstein gravity deformed by a minimally coupled scalar field.  Various saddle-point contributions $\exp[i(\text{saddle-point action})]$ to the Hartle-Hawking wave functions are computed with mini-superspace formalism. The saddle-point actions are first obtained in the spacetime bulks by direct computation, then on the asymptotic dS boundary by holographic renormalization. It is found that the imaginary part of the saddle-point actions, as functions of scalar field deformation, are generally bounded in the bulk, but can diverge to $-\infty$ on the boundary. This can probably be a source of the scalar field-related non-normalizability of the boundary Hartle-Hawking wave functions. For Kerr-dS$_3$ cosmology, some saddle-point actions have imaginary parts diverging to $-\infty$ both in the bulk and on the boundary, when the boundary $T^2$ is stretched to infinitely long. This can be an origin of non-normalizability of the boundary Hartle-Hawking wave functions related to temperature divergence. Finally the holographic renormalization is extended to scalar-deformed dS$_{d+1}$ ($d\geq 3$) cosmologies for the imaginary part of the saddle-point actions. The result is tested on a 5d example, and saddle points leading to divergent contribution to the boundary Hartle-Hawking wave function are shown to exist.



\vspace{1cm}

\newpage 
\tableofcontents

\newpage

\section{Introduction}

Over the past dozen years with the observation of early cosmology attaining higher and higher sophistication, 
there has been a continuous trend of theoretical effort based on gauge/gravity duality to explore the quantum gravity mechanism behind the observational results and their implications. Among all the theoretical frameworks proposed for this purpose, the dS/CFT correspondence \cite{Strominger:2001pn,Strominger:2001gp,Maldacena:2002vr} is perhaps the most straightforward one, since its bulk side is suitable for directly describing an inflationary universe.
Compared to the famous AdS/CFT correspondence, one other widely used holographic framework in cosmology, the groundwork of dS/CFT is far less settled. While for the former, general consent has been comfortably granted to the formulation of \cite{Witten:1998qj,Witten:1998zw,Klebanov:1999tb,Witten:2001ua}, a similar consensual point for the latter has not yet been reached by the holography community \cite{Strominger:2001gp,Maldacena:2002vr,Witten:2001kn,Klemm:2001ea,Balasubramanian:2002zh,Dyson:2002nt,Hertog:2011ky,Hartle:2013vta}.  

One branch among all the lines of thinkings which has been fruitful, advocates a simple transplantation of the AdS/CFT principle into the dS/CFT realm. It proposes the identification of the Hartle-Hawking wave function \cite{Hartle:1983ai,Hawking:1983hj} of the bulk side, with the partition function of the dual field theory on the boundary \cite{Maldacena:2002vr}, where the latter can be non-unitary \cite{Strominger:2001pn,Banerjee:2013mca}. A concrete microscopic realization of this prescription was set forth recently in \cite{Anninos:2011ui,Ng:2012xp}, which relates the Vasiliev higher spin gravity theories \cite{Vasiliev:1990en} in the dS$_4$ bulk to a critical grassmannian scalar field theory of $Sp(N)$ symmetry on the boundary, which is the dS/CFT cousin of the Vasiliev/$O(N)$ correspondence based on AdS holography \cite{Klebanov:2002ja,Giombi:2012ms}. 

Soon after, this concrete prescription was used to obtain the Hartle-Hawking wave function of the Vasiliev universe via computing the functional determinants of the boundary $Sp(N)$ theory \cite{Anninos:2012ft,Anninos:2013rza}. With some proper regularization procedure, the wave functions are obtained on dS boundaries of various topologies, and one intriguing feature uncovered is the omnipresent non-normalizability of the Hartle-Hawking wave functions. In particular, they can diverge either as a function of scalar deformation (scalar divergence) or as a function of the geometry of the dS boundary (temperature divergence).

Another recent result that can be placed in the same line is \cite{Castro:2012gc}, where the one-loop Hartle-Hawking wave function of the empty Kerr-dS$_3$ universe is obtained on its $T^2$-boundary, and is argued to be full-loop exact. 
The wave function depends on the modular parameter $\tau=\tau_1+i\tau_2$ of the boundary torus. It turns out to be exponentially divergent with $\tau_2$ tending to $\infty$ (temperature divergence), where the divergence arises at the loop level.

One tempting interpretation of these divergences is the instability of dS spacetimes, already suggested by studies from different approaches (see discussion and references in \cite{Castro:2012gc}).
However this speculation is much based on the quantum mechanics knowledge about a quantum system prepared in a false vacuum, which states that its decay process is described by an exponentially divergent wave function (see for example \cite{Hatano:2007uz}). Indeed without explicit computation it is hard to see to what extent this quantum mechanical common sense applies to the wave function of the universe. Quantitative study is thus needed for unveiling the actual bulk mechanism responsible for the divergence of the boundary wave function.

The recent development in another field of research can pave the way for moving towards this goal. It is the series of work on the quantum probability measure of classical universes \cite{Hartle:2007gi,Hartle:2008ng,Hartle:2012tv,Hartle:2010vi}, part of the decades-long quest for a probabilist explanation of why inflation happens \cite{Hawking:1987bi}. Powerful tools have been developed for computing Hartle-Hawking wave functions at tree-level in the spacetime bulk. Therefore it appears a promising direction to apply these tools to produce boundary results from the bulk, in order to gain insight into the results from direct dS boundary computations.

Indeed some efforts in this regard have already been made on the analytical level. In \cite{Anninos:2012ft}, having obtained the Hartle-Hawking wave function on the $S^1\times S^2$ boundary of the Vasiliev universe, divergent with the shrinking of $S^1$ (temperature divergence), a bulk computation at tree level in Einstein gravity is performed, with the dS boundary of the same topology $S^1\times S^2$. The result shows that some bulk saddle-point contribution diverges in a very similar way that the Vasiliev universe wave function does. This computation is also performed in \cite{Banerjee:2013mca} in a different context. In \cite{CHinProg} a more elaborated analysis of this Einstein gravity model shows that this divergence is related to some unphysical classical bulk configuration. These works in the bulk shed light on the origin of the boundary wave function divergences.

\vspace{2mm}

The work to be presented in this paper is in the continuation of the work mentioned above, with a substantial use of the numerical methods in \cite{Hartle:2010vi} (the numerical scheme can be traced back to \cite{Hertog:2004dr,Hertog:2004rz,Hertog:2005hu}). Yet it tackles the problem somewhat from another angle, in that the attention will be focused on 3d models. The advantage of this is that one can enrich the physical content of the model to quite some extent, while still keep exact computations possible. In the following, I present a summary of the key logic of the work.

\subsubsection*{\underline{\it What to compute (saddle-point actions) and why (their imaginary parts matter)}}

Concretely, the models to be considered are 3d closed universes. Supposed to be at weak coupling, the bulk physics is governed by Einstein gravity with a positive cosmological constant. A minimally coupled scalar field is introduced, and moreover, the universes should have an asymptotic dS region. Two topologies for the asymptotic boundary will be considered: $T^2$ and $S^2$. For each case, the essential part of the work can be encapsulated in the following two successive steps: 

\begin{itemize}

\item First, the Hartle-Hawking wave function is computed in the spacetime bulk directly
with the defining path integral \cite{Hartle:1983ai}. However, due to the difficulty in defining and evaluating such path integrals in general,\footnote{See discussions in \cite{Hertog:2011ky,PIIndGrav,Halliwell:1989dy}. The examples of exact computation on models based on empty (A)dS$_3$ spaces \cite{Castro:2012gc,Maloney:2007ud,Giombi:2008vd} should not be considered as counter examples. First it is a one-loop computation whose result is claimed to be full-loop exact using a symmetry claim based on \cite{Brown:1986nw}. Second although the computation deals directly with degrees of freedom in the bulk, it is done over the full (A)dS space and the result is still a wave (partition) function on the asymptotic (A)dS boundary.} in this paper the work is focused on the saddle-point contributions, assuming that loop contributions are highly suppressed. 
The problem now reduces to computing the saddle-point actions, to be denoted by $\cal S$. The associated tree-level contribution to the wave function is thus $e^{i{\cal S}}$. Generically, ${\cal S}$ is complex in the context of no-boundary quantum cosmology, so that the saddle point contribution acquires a non-trivial factor of amplitude in addition to the factor of a pure phase: $e^{i{\cal S}}=e^{i\re {\cal S}}\times e^{-\im {\cal S}}$ \cite{Hawking:1983hj,Hartle:2008ng,DeWitt:1967yk}.

\item Second, the bulk result is sent onto the asymptotic dS boundary. This step is supposed to yield the version of wave function to be directly identified with the partition function of the dual boundary field theory, in case the boundary dual exists. If we focus on the tree level, the problem is a simple application of the standard holographic renormalization scheme to the dS bulk action $\cal S$, which eliminates the infrared (IR) divergence and yields a finite renormalized action $\tilde {\cal S}$.  The associated tree-level contribution to the boundary Hartle-Hawking wave function is therefore $e^{i\tilde {\cal S}}=e^{i\re \tilde {\cal S}}\times e^{-\im \tilde {\cal S}}$. 
For the 3d models, $\tilde {\cal S}$ can be worked out explicitly and the results can be numerically computed. 
In this step an important thing to examine for the purpose of the paper, is the behavior of ${\rm Im} {\cal S}$ and $\im\! \tilde {\cal S}$. Their divergence to $-\infty$ is considered as possible source of non-normalizability of the Hartle-Hawking wave function.

\end{itemize}

\subsubsection*{\underline{\it About the saddle points: how to find them and what do they look like}}

Since it is clarified in the first step above that the Hartle-Hawking wave functions will be computed at tree-level, there is now one more step to be prepended to the whole agenda: the search for contributing saddle points. The latter are compact classical spacetimes with on-shell fields living in them and matching the pre-ordered values on the spacetime boundary. The search for saddle points will be done numerically using the numerical scheme in \cite{Hartle:2008ng}.
Simply stated, one first assigns the wanted values of the fields on the boundary, and then integrate the classical bulk equations of motion until the boundary, adjusting some defining parameters of the saddle point, to let the fields match the boundary condition. This done, we will have the raw material to feed into the main process to let it produce the results of $\cal S$ and $\tilde {\cal S}$, among some other physical quantities.  Some important features of this part of the work are:
\begin{itemize}
	\item The search will be carried out in mini-superspace formalism in order for it to be possible. That is, we require the spacetime have a time coordinate which realizes homogeneous isotropic slicing. This will make us loose saddle points but will not cause error to the result of $\cal S$ and $\tilde {\cal S}$.
	\item Then it turns out that the mini-superspace coordinate time is necessarily complex so that the entire history of a saddle point is a complex curve.
	\item Given a boundary condition, infinitely many saddles exist with time contours containing different amount of Euclidean history and/or ending up on different layers in the scalar field's Riemann surface. 
\end{itemize}


The rest of the paper will be organized as follows. Sec.\ref{FFR} will present the frame work and technicalities for studying the scalar-deformed Kerr-dS$_3$, including the action principle for the saddle points, the computation of saddle-point actions in the bulk, and the holographic renormalization, all formulated in mini-superspace formalism. 

In Sec.\ref{PERTT2} the search of saddle points and the computation of their actions will be done analytically, with scalar deformation set perturbative. It is shown how different mini-superspace time contours lead to different saddle points fitting the same boundary conditions. 

Sec.\ref{NUMT2} is the prolongation of the work of Sec.\ref{PERTT2} into the domain of finite scalar deformation. The saddle-point actions $\cal S$ (bulk) and $\tilde {\cal S}$ (boundary) are traced both against scalar deformation and against boundary geometry deformation, and divergence to $-\infty$ of $\im\!\cal S$ and $\im\!\tilde {\cal S}$ are observed. Meanwhile the Riemann surfaces of the scalar field are studied, which shows empirically some intriguing connection between the scalar divergence of $\im\!\tilde {\cal S}$ and the pattern that singularities move around in the Riemann surfaces.

In Sec.\ref{COSMOS2}, all the study of the scalar-deformed Kerr-dS$_3$ in the previous sections are carried out on scalar-deformed dS$_3$, including the perturbative analytical study, the exact holographic renormalization and the non-perturbative numerical computations. Possible orgin of scalar divergence in $\im\!\tilde {\cal S}$ is also found.

Sec.\ref{FTIS} gives a very brief look into two immediate extensions of the work, which can be interesting possible future directions. It is first shown that the work can be extended to models with potentials other than quadratic. The second part explores the possibility of extension to higher dimensions, where a concrete model, scalar-deformed dS$_5$ is studied and similar features as the 3d models are uncovered.

The summary and perspectives are presented in Sec.\ref{DCS}.

\section{Framework for boundary topology $T^2$}\label{FFR}

This section serves to establish the notations, vocabularies and the general formalisms for one of the models of interest: the Kerr-dS$_3$ universes deformed by a minimally coupled scalar field, which have an asymptotic boundary of topology $T^2$. Universes with asymptotic spacelike boundaries of topology $S^d$ ($d=2,3,\dots$) will also be considered later in the paper. However adapting the formalisms developed in this section to the $S^d$-cases will be straightforward enough.

Sec.\ref{ANSATZ} presents the spacetime geometry near the $T^2$-asymptotic boundary, and then introduces the Hartle-Hawking wave function for the very model in question. Sec.\ref{EOMOA} is focused on the tree-level parts of the Hartle-Hawking wave function, and establishes numerically operable  formalisms for their computation. In Sec.\ref{HR_OSA}, the saddle-point actions are computed with the boundary of the saddle points sent to the asymptotic dS boundary. Holographic renormalization is then performed to remove the infrared (IR) divergences to obtain the saddle-point actions on the asymptotic dS boundary. 

\subsection{Spacetime setup and Hartle-Hawking wave function} \label{ANSATZ}

\subsubsection*{Action and spacetime in general}

With bulk physics set at weak coupling, its action should be that of Einstein gravity with a positive cosmological constant and a minimally coupled scalar field:
\begin{align}
	& S={1\over 16\pi G}\int_{\cal M} \! \! d^3x\, \sqrt{-g}\, \big[R-2\ell^{-2} + (\partial \Phi)^2 -2 V(\Phi)\big] +S_{\rm b}, \label{tot1} 
\end{align}
where $\cal M$ denotes the spacetime manifold, $g_{\mu\nu} $ ($\mu,\nu=0,1,2$) is the 3-metric with $R$ its Ricci scalar, $\ell$ is the dS radius, and $V(\Phi)$ is the scalar potential. We will also need to include some appropriate boundary term $S_{\rm b}$ defined on $\partial \cal M$ in accordance with the boundary conditions.

 
The universes to be considered are supposed to asymptote to a (locally) dS space with asymptotic boundary of topology $T^2$. It will be assumed that some Fefferman-Graham type expansion (see for example \cite{Anninos:2012qw}) is available in the asymptotic dS region with the presence of the scalar field:
\begin{align}
	\ell^{-2}g_{\mu\nu}dx^\mu dx^\nu \sim -d\lambda^2+q_{ij}(\lambda,x^k) dx^i dx^j, \ \ {\rm where}\ \  q_{ij}(\lambda,x^k)\sim e^{2\lambda}q^{(0)}_{ij}(x^k)+(\text{sub-leading orders}), \label{eq02}
\end{align}
where $\lambda$ is some coordinate time and the asymptotic dS boundary, to be denoted by $I^+$, is located at $\lambda=\infty$; $i$, $j$ and $k$ are spatial indices taking the values $1$ or $2$; $q^{(0)}$ specifies the geometry of a 2-torus. The subleading terms are not shown explicitly in Eq.(\ref{eq02}). Due to the scalar field back reaction their fall-off is slowed down with respect to those in the standard Fefferman-Graham series. Such behavior has already been described in the AdS/CFT context \cite{Hertog:2004dr,Henneaux:2002wm,Martinez:2004nb,Henneaux:2004zi}. The sub-leading terms are presented in appendix \ref{ABAsydS} and will later be very useful for holographic renormalizaion. Throughout the paper the dS radius $\ell$ will be used as the length unit, and factorized as the global coefficient of the metric, so that $x^\mu$, $\lambda$, $q_{ij}$ is dimensionless.

\begin{figure}
\begin{center}
	\includegraphics[height=6cm]{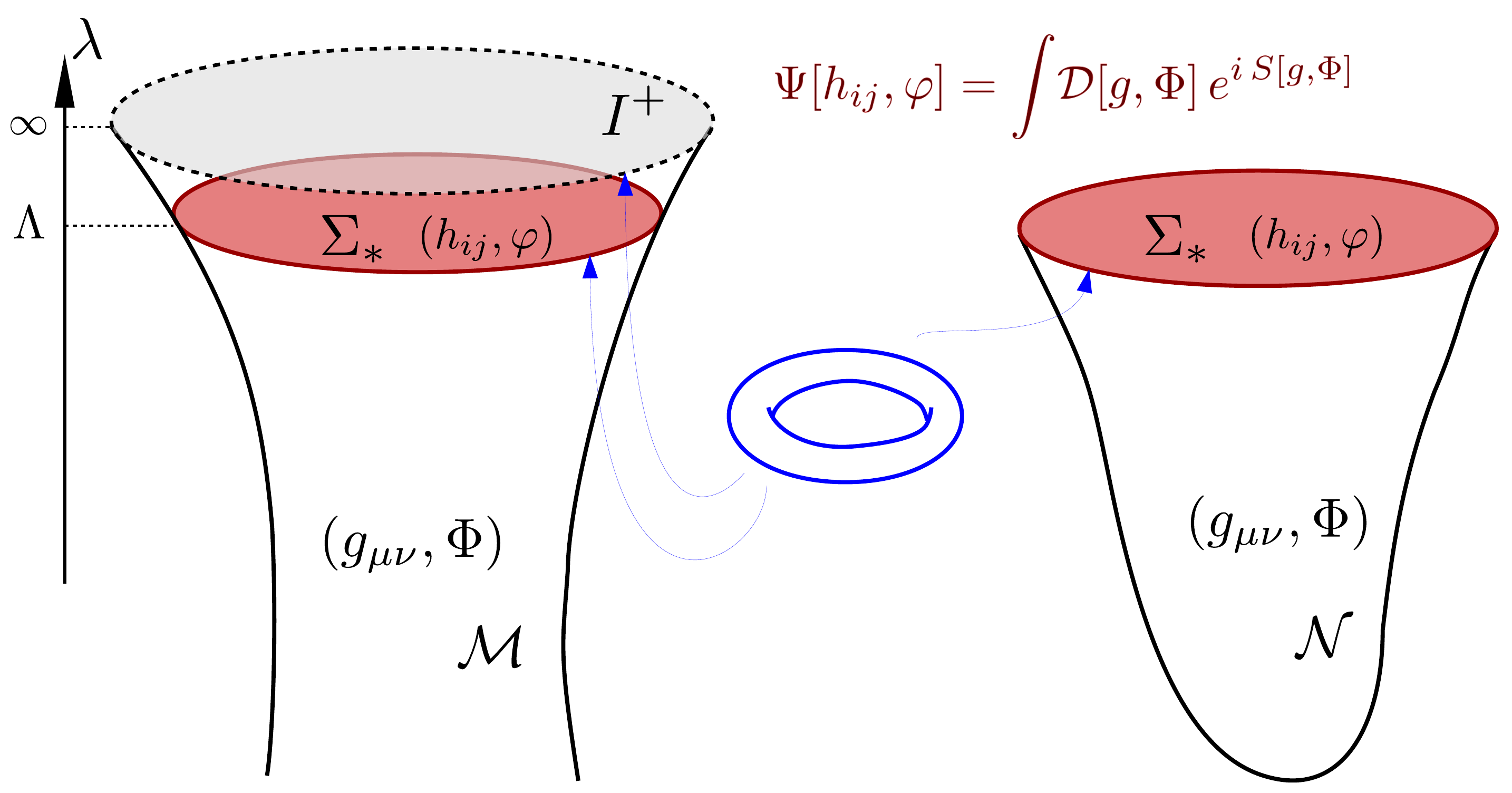}
	\caption{\footnotesize {\bf Left:} The asymptotic dS region of the of 3d universe considered. The spacetime manifold is denoted by  $\cal M$, which supports a 3-metric $g_{\mu\nu}$ and a scalar field $\Phi$. The asymptotic boundary $I^+$ has topology $T^2$, and the asymptotic region is sliced into a stack of $T^2$ against some time parameter $\lambda$. The Hartle-Hawking wave function will be computed with respect to the slice $\lambda=\Lambda$, denoted by $\Sigma_*$, whose induced 2-metric and scalar field are denoted by $h_{ij}$ and $\varphi$, so that the wave function is $\Psi[h_{ij},\varphi]$.  {\bf Right:} The spacetime $\cal N$ to be summed up in the defining path integral of $\Psi[h_{ij},\varphi]$. They are compact and have $\Sigma_*$ as boundary: $\partial {\cal N}=\Sigma_*$, and have field contents $(g_{\mu\nu},\Phi)$ regular everywhere in $\cal N$, fitting the values $(h_{ij},\varphi)$ on the boundary. $\cal N$ do not necessarily have a slicing into a stack of $T^2$.} \label{slicing}
\end{center}
\end{figure}

\subsubsection*{The Hartle-Hawking wave function in the bulk field representation} \label{SC22}

We can now move on to the quantum aspect to introduce the Hartle-Hawking wave function. As stated in the introduction, we will first need to compute directly the Hartle-Hawking wave in the spacetime bulk at tree level. Therefore the function will be defined on some spacelike hypersurface in the bulk of the universe, and is a function(al) of the field configuration and the geometry of this hypersurface. To be relevant to the holographic computation later, this hypersurface is chosen in the asymptotic dS region of the universe and is just one of the equal-$\lambda$ surface as given in Eq.(\ref{eq02}). Let this hypersurface correspond to $\lambda=\Lambda<\infty$ and be denoted by $\Sigma_*$, and let the induced 2-metric of this slice be denoted by $h_{ij}$ and the scalar field value confined on it by $\varphi$. An illustration of this setup is shown in the left part of Fig.\ref{slicing}. It is still necessary to have some more specifications of the hypersurface $\Sigma_*$ to avoid too abstract talking about the Hartle-Hawking wave function. 

It will be convenient to use the language of complex geometry to parameterize $h_{ij}$. Therefore $\Sigma_*$ will be characterized in terms of its area, given by the K\"ahler modulus $\cal A$, and its shape, specified by the complex structure modulus $\tau=\tau_1+i\tau_2$. By a reparameterization within the slice $\Sigma_*$, its 2-metric can be rendered to the form\footnote{In doing so the resulting $h_{ij}$ as in Eq.(\ref{hijAT13}) will generically no longer be that appearing in the Feffmann-Gramm form because the reparameterization can produce a nonzero shift vector. However this does not prevent us from formally defining the Hartle-Hawking wave function(al) in terms of $h_{ij}$.}
\begin{align}
	 h_{ij}dx^idx^j=&\, \ell^2{ {\cal A} \, \tau_2^{-1}} \Big(  d\xi_2^2+2\, \tau_1\, d\xi_1d\xi_2+ |\tau|^2 d\xi_1^2 \Big), \label{hijAT13}
\end{align}
where $\xi^1$ and $\xi^2$ are both periodic with periodicity $2\pi$. Due to the factorization of $\ell^2$, $\cal A$ is dimensionless. 
The advantage of using the complex structure modulus $\tau$ is to easily formulate the T-duality of the 2-torus: a symmetry of the geometry of the torus when its complex structure modulus undergoes the following transform
\begin{align}
	\tau\longrightarrow \tau'={e\tau+f \over c\tau+d}, \ \ {\rm where}\ \ c,d,e,f\in \mathbb Z,\ {\rm and}\ e d-fc=1, \label{TDUAL}
\end{align}
These transforms form the group $SL(2,\mathbb Z)$. The path integral to be introduced immediately below, Eq.(\ref{PIforHH}), should pick up the contribution from all inequivalent 2-tori lying on the $SL(2,\mathbb Z)$ orbit.

With the above preliminaries, we can introduce the Hartle-Hawking wave function \cite{Hartle:1983ai} on a concrete basis. It will be denoted by $\Psi$, and computed against $\Sigma_*$, and is thus a function(al) of $h_{ij}$ and $\varphi$. Its formal definition is
\begin{align}
	\Psi[h_{ij},\varphi]=\Psi[{\cal A},\tau,\varphi]=\int {\cal D}\left[  g,\Phi \right] \exp\Big(i\, S[g,\Phi]\Big). \label{PIforHH}
\end{align}
Here $S[g,\Phi]$ is given by Eq.(\ref{tot1}) with a proper boundary term to be specified very shortly. The integral ${\cal D}[g,\Phi]$ runs over all compact 3-spacetimes bounded by $\Sigma_*$, with the fields $g_{\mu\nu}$ and $\Phi$ fitting the boundary values $(h_{ij},\varphi)$. Such spacetimes are different from the actual physical universe $\cal M$ in Eq.(\ref{tot1}), and they will be denoted by $\cal N$ to mark the difference. Therefore we have $\partial {\cal N}=\Sigma_*$, and if we further let $F$ be the imbedding map of $\Sigma_*$ into $\cal N$, then $F^*g=h$, and $F^*\Phi=\varphi$. These conditions for the path integral are just the {\it no-boundary proposal} \cite{Hartle:1983ai}. An illustration of such no-boundary spacetimes is given in the right half of Fig.\ref{slicing}.

Now the action in Eq.(\ref{PIforHH}) can be fully specified. Given that we are fixing $h_{ij}$ and $\varphi$ on the boundary $\Sigma_*$, the boundary conditions are Dirichlet. Therefore the corresponding boundary term in the total action should be the Gibbons-Hawking term \cite{Gibbons:1976ue}. Thus the total action is
\begin{align}
	S={1\over 16\pi G}\int_{\cal N} \! \! d^3x\, \sqrt{-g}\, \big[R -2\ell^{-2} + (\partial \Phi)^2 -2 V(\Phi)\big] +{1\over 8\pi G}\int_{\Sigma_*}\!\! d^2x\, \sqrt h\, K, \label{tot2} 
\end{align}
where $K=h^{ij}K_{ij}$ with $K_{ij}$ the extrinsic curvature of $\Sigma_*$.

\subsection{The action principle for saddle-points}  \label{EOMOA}

Let us move on to the computational level. The goal is to compute the path integral (\ref{PIforHH}) at tree level, and that to achieve this we need to first find out the saddle points. 
This subsection will introduce the formalism for the search of saddle points. Recall that the saddle points are on-shell compact spacetimes satisfying the no-boundary proposal. Therefore the formalism to be presented is essentially about setting up a calculable action principle. 

An important restriction has to be introduced to make the actual computation feasible: the mini-superspace truncation of degrees of freedom. This means we only deal with the saddle points whose spacetime has a globally defined coordinate time which slices the spacetime into homogeneous isotropic hypersurfaces, all having the same topology $T^2$ as the boundary (see Fig.\ref{NBSDmss}).

In the contents that follow, it will be explained how to formulate the variation principle of the saddle points in the mini-superspace, including the implementation of the mini-superspace formalism, the corresponding boundary conditions, the action and the equations of motion. Finally a preliminary simplification of the saddle-point action will be given for the use of next subsection after a holographic setup is introduced.





\subsubsection*{Mini-superspace reduction}

To implement the mini-superspace formalism, we can start with the 3-metric of the no-boundary spacetime $\cal N$, requiring that the saddle points have the metric which takes the form of $2+1$ decomposition into a stack of homogeneous isotropic 2-tori:
\begin{align}\label{gxAT0}
	g_{\mu\nu}dx^{\mu}dx^{\nu}= &\, -\ell^2 d\chi^2+q_{ij}(\chi)dx^i dx^j=-\ell^2d\chi^2+\ell^2A(\chi) \, T_2(\chi)^{-1} \big|  d\xi_2+T(\chi) d\xi_1 \big|^2.
\end{align}
Here the spatial coordinates $\xi_1$ and $\xi_2$ are just those in Eq.(\ref{hijAT13}), which have periodicity $2\pi$ separately. A global time coordinate $\chi$ is introduced such that the lapse function is $1$, and later on whenever the term ``coordinate time'' or simply ``time'' is used for some saddle point, it always refers to this mini-superspace coordinate time. For each $\chi$-slice the 2-metric is $q_{ij}(\chi)dx^i dx^j=\ell^2A(\chi) \, T_2(\chi)^{-1} \big|  d\xi_2+T(\chi) d\xi_1 \big|^2$, a flat metric of a torus of area (K\"ahler modulus) $A(\chi)$ and complex structure $T(\chi)=T_1(\chi)+i\, T_2(\chi)$. Yet a further restriction turns out to be necessary for the quantitative computation to be possible: we need $T_1$ to be constant in $\chi$, or can be transformed to constant in time through Eq.(\ref{TDUAL}). The 3-metric thus becomes
\begin{align}\label{gxAT}
	\ell^{-2}g_{\mu\nu}dx^{\mu}dx^{\nu}= &\, -d\chi^2+A(\chi) \, T_2(\chi)^{-1} \Big[  d\xi_2^2+2\, T_1\, d\xi_1d\xi_2+ \big(T_1^2+T_2(\chi)^2 \big) d\xi_1^2 \Big]\nonumber \\ =&\, -d\chi^2 + A(\chi) \, T_2(\chi) d\xi_1^2 +A(\chi) \, T_2(\chi)^{-1} \, d\big( \xi_2+T_1 \xi_1)^2 .
\end{align}
A simpler form can be achieved if we let 
\begin{align}
	&a(\chi)^2={A(\chi)\over T_2(\chi)},\ \ b(\chi)^2=A(\chi)T_2(\chi); \label{rel18}\\
	& \zeta_1=\xi_1,\ \ \zeta_2=\xi_2+T_1\xi_1,
\end{align}
with which Eq.(\ref{gxAT}) further reduces to
\begin{align}\label{gxaebe}
	\ell^{-2}g_{\mu\nu}dx^{\mu}dx^{\nu}=  - d\chi^2+a(\chi)^2  d\zeta_2^2+b(\chi)^2 d\zeta_1^2 \, .
\end{align}
The periodicities of the angular coordinates are \begin{align} \label{PRDeta} (\zeta_1,\zeta_2)\sim (\zeta_1,\zeta_2)+2\pi (1,T_1)\sim  (\zeta_1,\zeta_2)+2\pi (0,1). \end{align}

\subsubsection*{Coordinate time range: between the boundary and the south pole}


Then let us take a closer look at the coordinate time, supposing it elapses from a starting point $\chi_o$ to an end point $\chi_*$. With the lapse function being set to $1$, it follows that we cannot universally assign the same endpoints $\chi_o$ and $\chi_*$ to all cases when solving for classical solutions, but should have them determined case by case using the boundary conditions. Later we will see that for each classical solution, $\chi_*-\chi_o$ is different and is generically complex.\footnote{Had we left the lapse function $N$ general, then we should fix the time range $\chi_*-\chi_o$, and let the lapse $N$ be determined by boundary conditions, and $N$ is generally complex.} One end of the time interval should correspond to the spacetime boundary $\partial {\cal N}=\Sigma_*$ and we can let it be $\chi_*$. Thus at $\chi_*$, the 2-metri $q_{ij}(\chi)$ in Eq.(\ref{gxAT0}) should evaluate to $h_{ij}$, and the scalar field $\Phi$ should evaluate to $\varphi$. That is
\begin{align}
	&A_*=a_*b_*={\cal A},\ \ (T_2)_*={b_*\over a_*}=\tau_2, \ \ \Phi_*=\varphi, \label{rel18**}
\end{align}
where the subscript ``$*$'' means evaluated at $\chi_*$: $a_*=a(\chi_*)$ etc. There is still the spectator parameter $(T_1)_*\equiv T_1(\chi)=\tau_1$ which can be arbitrarily chosen without having concrete effect on any other results. On the other hand, there should be no boundary present at the starting point $\chi_o$, and we can implement this condition by requiring 
\begin{align}
	& a(\chi_o+\epsilon)\sim i\epsilon,\ \  b(\chi_o)=b_o\ \ {\rm for} \ \ b_o,\epsilon\in \mathbb C\ {\rm and}\ |\epsilon| \sim 0; \label{NBCab}\\ 
	& \Phi(\chi_o)=\Phi_o \in \mathbb C,\label{NBCPhi}
\end{align}
where $b_o$ and $\Phi_o$ are finite complex numbers.  Demanding Eq.(\ref{NBCPhi}) is because the scalar field should be regular all over $\cal N$. Moreover, when Eq.(\ref{NBCab}) is satisfied, the leading behavior of the 3-metric Eq.(\ref{gxaebe}) near $\chi_o$ is
\begin{align}
	\ell^{-2} g_{\mu\nu}dx^{\mu}dx^{\nu}\Big|_{\chi= \chi_o+\epsilon}\sim -\big(d\epsilon^2 + \epsilon^2 d\zeta_2^2\big) +b_o^2 \, d\zeta_1^2.
\end{align}
This states that the $a$-circle smoothly shrinks to zero size at $\chi_o$, therefore ensuring the absence of spacetime boundary. In full generality, the smooth capping of any linear combination of the $a$-circle and the $b$-circle at $\chi_o$ can remove the spacetime boundary. However since these cases can be obtained by doing an $SL(2,\mathbb Z)$ transform Eq.(\ref{TDUAL}) on the case of shrinking $a$-circle, we do not lose generality by only considering Eq.(\ref{NBCab}).  By the way Eq.(\ref{NBCab}) shows that the fields and the time coordinate are necessarily complexified as a result of no-boundary proposal. 

The starting moment of minisuperspace cosmic time $\chi_o$ is named ``\underline{\it south pole}'' in the literature, c.f. for example \cite{Hertog:2011ky,Hartle:2008ng}, and this paper will also use this term. 
In Fig.\ref{NBSDmss} a sketch of such no-boundary spacetime configuration is presented, whose difference from the right half of Fig.\ref{slicing} is that now $\cal N$ allows a homogeneous isotropic slicing into 2-tori.




\subsubsection*{Action principle for mini-superspace saddle points}

Now we are at the point of putting the metric Eq.(\ref{gxaebe}) and the scalar field $\Phi$ on-shell in order to let $({\cal N},g_{\mu\nu},\Phi)$ become the saddle point. For this purpose let us derive the action principle. Inserting the metric ansatz (\ref{gxaebe}) into the total action (\ref{tot2}), assuming the scalar field be homogeneous on the equal-$\chi$ surfaces, after partial integral to eliminate second order derivatives, we have
\begin{align}
	S=&\,{1\over \kappa}\int_{\chi_{{}_o}}^{\chi_{{}_*}}\!\!d\chi \left[-\dot a \dot b - ab+{ab\over 2}\left(\dot \Phi^2- 2\, \ell^2 \, V(\Phi) \right)\right]-{1\over \kappa}\left( \dot a b+ a\dot b \right)_{\!\chi_{{}_o}}, \label{Itotabp}
\end{align}
where we introduced the coupling constant $\kappa$ defined as
\begin{align} {\pi \ell \over 2 G}={1\over \kappa}.  \label{kpaCC} \end{align} 
From now on, we set $\kappa=1$ for simplicity. In Eq.(\ref{Itotabp}) the second term on the righthand side is a ``boundary'' term (or rather ``south pole term'') at $\chi_o$ produced by partial integral. There is also a same boundary term at $\chi_*$, but it is canceled by the Gibbons-Hawking term. The equations of motion obtained from varying the action (\ref{Itotabp}) are
\begin{align}
	&{\dot a\over a} {\dot b \over b}-1-{1 \over 2} \dot \Phi^2- \ell^2\, V(\Phi)=0\, , \label{Eabp}\\
	&{\ddot a\over a}-1+{1 \over 2} \dot \Phi^2-\ell^2\, V(\Phi) =0\, , \label{Eaph}\\
	&{\ddot b\over b}-1+{1\over 2} \dot \Phi^2-\ell^2 \,V(\Phi)=0\, , \label{Ebph}\\ 
	&\ddot \Phi+ {\dot a \over a} \dot\Phi+ {\dot b \over b} \dot\Phi+\ell^2 \, V'(\Phi)=0\, , \label{Ephi}
\end{align}
where the dots stand for the derivative with respect to $\chi$: $\dot a={da\over d\chi}$ etc. 
In appendix \ref{ACEQGEN} the action and the equations of motion are derived for a more general case where the equal-time slices of the mini-superspace model have the topology $S^{d_1}\times S^{d_2}$. This will be useful in Sec.\ref{COSMOS2}, and also here Eqs (\ref{Itotabp})---(\ref{Ephi}) can be obtained by setting $d_1=d_2=1$ in Eqs (\ref{abAction})---(\ref{EqVphi}).


In looking for the saddle points, in principle we first assign the wanted boundary values of fields at $\Sigma_*$: $(h_{ij},\varphi)=({\cal A},\tau,\varphi)$. We then integrate the equations (\ref{Eabp})---(\ref{Ephi}) for $\{a(\chi),b(\chi),\Phi(\chi)\}$ starting from the south pole $\chi=\chi_o$ to the boundary $\chi=\chi_*$. At the south pole $\chi_o$ we need to input the starting-up data $(b_o,\Phi_o)$, as introduced in Eqs (\ref{NBCab}) and (\ref{NBCPhi}), such that the solutions  fit the boundary condition at the final moment $\chi_*$ as Eq.(\ref{rel18**}). There are three complex conditions at $\Sigma_*$ given by Eq.(\ref{rel18**}), exactly what is needed to determine the three complex parameter: $(b_o,\Phi_o)$ and $\chi_*-\chi_o$.

\subsubsection*{Saddle-point actions preliminary}

The mini-superspace formulation used so far makes the evaluation of saddle-point actions very simple. After partial integral on $a$ and $b$ in the action (\ref{Itotabp}) to recover their second order derivatives: $\dot a\dot b={1\over 2} (a\dot b+\dot a b)\,\dot{}-{1\over 2} (\ddot a b+a\ddot b)$, it becomes 
\begin{align}
	S=&\,{1\over 2}\int_{\chi_{{}_o}}^{\chi_{{}_*}}\!\!d\chi \left[   \ddot a b+a \ddot b -2 ab+ab\left(\dot \Phi^2- 2\, \ell^2 \, V(\Phi) \right)\right]
	-{1\over 2}\left[\left( \dot a b+ a\dot b \right)_{\!\chi_{{}_*}}+\left( \dot a b+ a\dot b \right)_{\!\chi_{{}_o}}\right]. \label{Itotabpbis}
\end{align}
The bulk part is precisely the sum of the equations of motion (\ref{Eaph}) and (\ref{Ebph}), vanishing on-shell. Therefore the saddle-point action becomes simply a sum of two boundary terms:
\begin{align}
	S[a, b, \Phi]_\text{on-shell}= -{1\over 2} \left( \dot a_* b_*+ a_*\dot b_*+ \dot a_o b_o+ a_o \dot b_o \right) :={\cal S}({\cal A},\tau,\varphi) , \label{Itotabpbbis}
\end{align}
where $a_*=a(\chi_*)$, $a_o=a(\chi_o)$, etc is understood. Also since $a$, $b$ and $\Phi$ are the solutions to Eqs (\ref{Eabp})--(\ref{Ephi}) and satisfy the boundary condition Eqs (\ref{rel18**}),
the resulting on-shell action is a function of $\cal A$, $\tau$ and $\varphi$, and thus can be denote by ${\cal S}({\cal A},\tau,\varphi)$. Then since $a_o=0$ and $\dot a_o=i$ as in Eq.(\ref{NBCab}), Eq.(\ref{Itotabpbbis}) is further reduced to
\begin{align}
	{\cal S}({\cal A} ,\tau,\varphi) = -{1\over 2}\left(\dot a_* b_* + a_*\dot b_*+i \, b_o \right). \label{ItotBS}
\end{align}
Note that the results Eqs (\ref{Itotabpbbis}) and (\ref{ItotBS}) requires no knowledge of the detail of the scalar potential.

On the other hand, we can still apply the commonly used approach in the literature to compute the on-shell action, where one inserts the constraint (\ref{Eabp}) into the action and obtains
\begin{align}
	S[a, b, \Phi]_\text{on-shell}=&\,-\int_{\chi_{{}_o}}^{\chi_{{}_*}}\!\!d\chi \, 2\, a  b \left[1+\ell^2 \, V\big(\Phi\big) \right]- \left( \dot{ a}  b+  a\dot{ b} \right)_{\chi_{{}_o}}={\cal S}({\cal A},\tau,\varphi).\label{IabpOnSh}
\end{align}
However the numerical error when using this integral expression can go easily out of control in certain cases basically because it counts on the cancellation between two large terms, $ab$ and $abV(\Phi)$, to produce a finite result. In the numerical results presented later, only Eq.(\ref{Itotabpbbis}) will be actually used. 
%




\begin{figure}
\begin{minipage}[t]{0.49\textwidth}
\begin{center}
\hspace*{0cm}\includegraphics[width=0.8\textwidth]{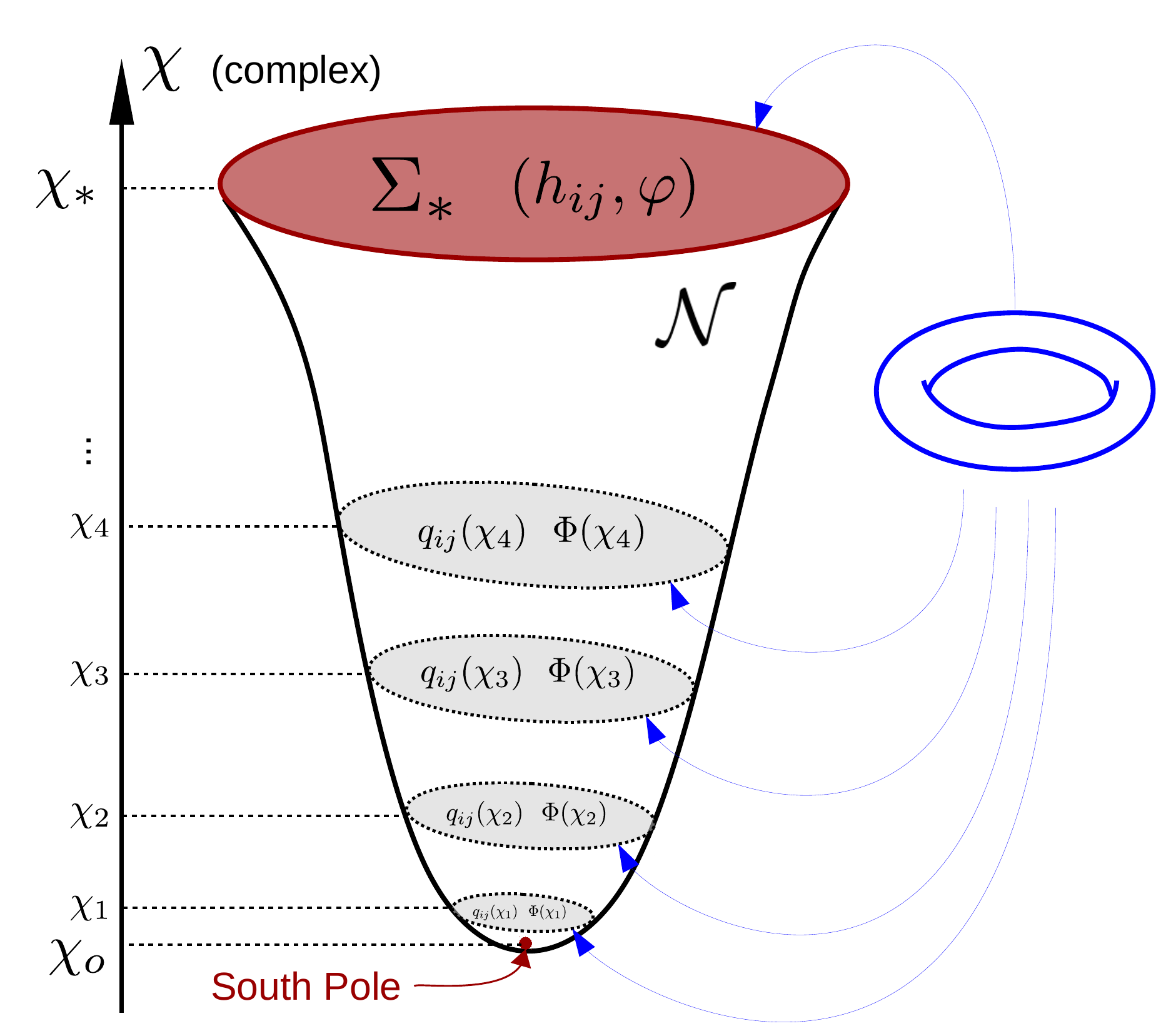}
\end{center}
\vspace*{-0cm}
\caption{\footnotesize No-boundary spacetime $\cal N$ allowing a mini-superspace formulation as Eq.(\ref{gxaebe}). The coordinate time $\chi$ elapses from $\chi_o$ to $\chi_*$, with each equal-$\chi$ surface a homogeneous isotropic 2-torus. Some sample slices $\chi=\chi_1,\dots,\chi_4$ are shown. The induced 2-metrics $q_{ij}(\chi_k)$ are as given in Eq.(\ref{gxAT0}). The $\chi=\chi_*$ slice $\Sigma_*$ must have real metric components $h_{ij}$ and real scalar field $\varphi$, while for other slices the fields $q_{ij}(\chi_k)$ and $\Phi(\chi_k)$ ($k=1,\dots,4$) can be complex. At $\chi_o$, one of the circles of the torus caps smoothly, so that there is no spacetime boundary there. The locus $\chi_o$ is referred to as ``south pole''. Note its difference from the right half of Fig.\ref{slicing}. Here $\cal N$ has a special time coordinate globally giving rise to a homogeneous isotropic $T^2$-slicing, while in Fig.\ref{slicing} the only requirement is that $\partial \cal N$ be Riemannian and of topology $T^2$.} \label{NBSDmss}
\end{minipage}
\hspace{2mm}
\begin{minipage}[t]{0.49\textwidth}
\begin{center}
\includegraphics[width=0.8\textwidth]{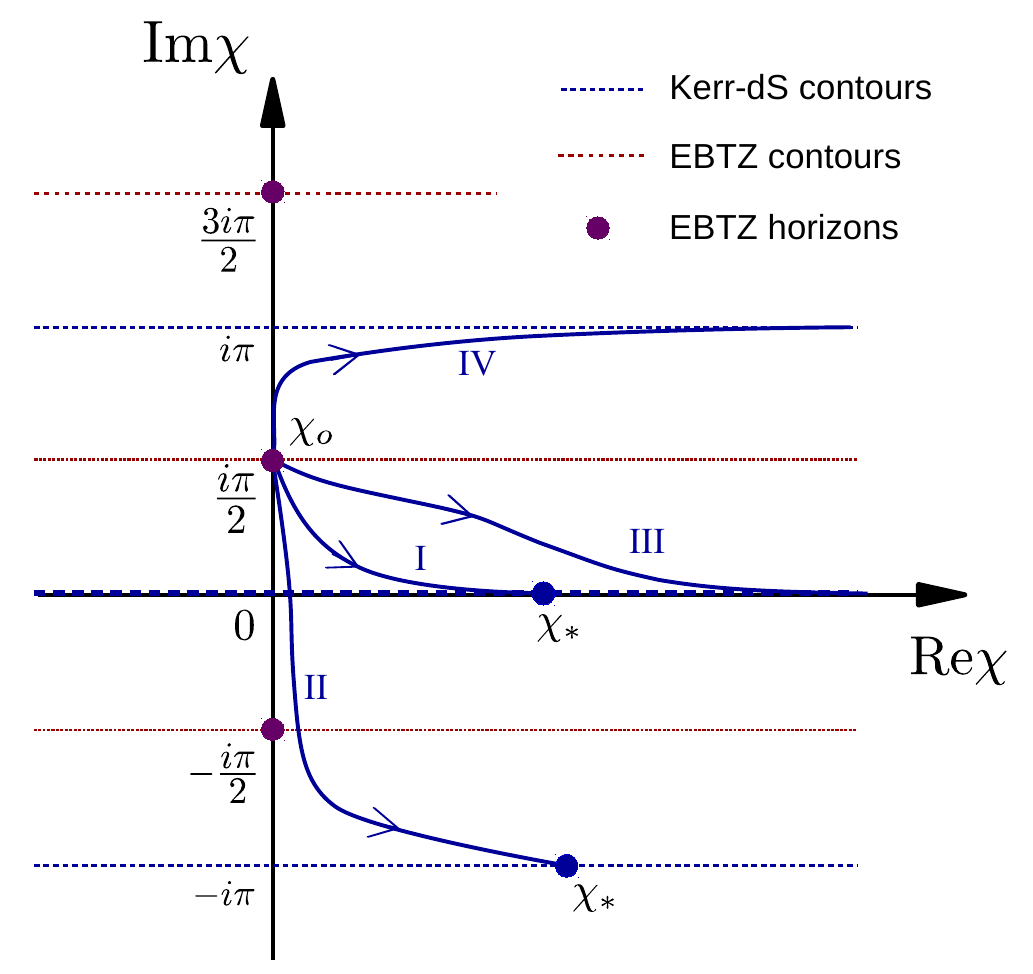}
\end{center}
	\caption{\footnotesize The metric of empty spacetime saddle points Eq.(\ref{3Met}) represented on the complex time plane. When Eq.(\ref{3Met}) is confined on the dashed blue or red lines it takes the form of Kerr-dS or EBTZ. The south pole $\chi_o={i\pi \over 2}$ is the horizon of an EBTZ contour. Saddle points are obtained by confining Eq.(\ref{3Met}) to a contour starting from $\chi_o$ and terminates at some point $\chi_*$ on a Kerr-dS contour, such as the contours I through IV. The saddle points I and II terminate in the bulk of the Kerr-dS spacetime with $\re \chi_*<\infty$, while the saddles III and IV terminate on the future boundary $I^+$ of the Kerr-dS space with $\re \chi_*=\infty$. This paper only covers the latter type of saddle points. Saddle points like I and III ($\im\chi_*=0$) are most commonly considered in the literature for Hartle-Hawking wave functions, here referred to as ``fundamental''.
}\label{KerrdSEBTZ}
\end{minipage}
\end{figure}

\subsection{Saddle-point actions and holographic renormalization} \label{HR_OSA}

Last subsection has offered a preview of saddle-point action evaluation, showing that it can be drastically simplified, reduced to merely some boundary terms with no integral involved. This subsection will pursue further the computation of the saddle-point actions to work out their formal results in a holographic setting, which are furthermore numerically computable. 

In order to introduce the holographic context, the defining surface $\Sigma_*$ of the Hartle-Hawking wave function will be sent to $I^+$. Then it will turn out that the infrared (IR) divergences will emerge in the bulk saddle-point actions $\cal S$ under the limit $\Sigma_* \rightarrow I^+$. Then holographic renormalization will be performed which removes the IR divergences with local counter terms, and leads to a finite saddle-point action on the boundary $\tilde {\cal S}$. 

In the following paragraphs, first the asymptotic behaviors of the fields of the saddle points will be analyzed, so as to display the IR divergences in $\cal S$, and second the counter terms will be introduced to have them removed. A numerically implementable result is then obtained and the numerical calculation will be presented in Sec.\ref{NUMT2}.

\subsubsection*{Infrared divergences of the on-shell action}  \label{T2INFD}

In Sec.\ref{EOMOA} it is mentioned that the time coordinate $\chi$ is generally complex (we will see concrete examples only in the next section). Therefore we need to assume that the  future boundary of the asymptotic de Sitter space $I^+$ corresponds to the coordinate time $\chi=i\delta_*+\infty$ with $\delta_*$ a finite real number. We will come to the point of how to determine $\delta_*$ in later sections. With $I^+$ thus set, the limit of $\Sigma_*\rightarrow I^+$ can be formulated as
\begin{align}
	\Sigma_*\rightarrow I^+ \ \ \Leftrightarrow \ \ \chi_*=i\delta_*+\Lambda \ (\Lambda\rightarrow \infty), \label{SigmatoI}
\end{align}
and this formulation together with the notations $\delta_*$ and $\Lambda$ will be used to all models to be covered. The real part $\Lambda$ can be identified with that used in Fig.\ref{slicing}, and the imaginary part $\delta_*$ will be called the ``\underline{\it Euclidean shift}'' of the boundary $\Sigma_*$. 

The expression Eq.(\ref{ItotBS}) makes the analysis of IR divergences extremely simple. We will only need the asymptotic expressions of the involved fields evaluated at $\chi_*=i\delta_*+\Lambda$ for large $\Lambda$. They are presented in the appendix \ref{ABAsydS}. We need to be notified that they are obtained for any minimally coupled scalar field with potential of the form
\begin{align}
	V(\Phi)={1\over 2} \ell^{-2}m^2 \Phi^2+O(\Phi^4), \ \ {\sqrt{3}\over 2} < m<1,\label{Vph2O}
\end{align}
where $m$ is the mass parameter of the scalar field. This paper only focuses on the mass range ${d^2-1\over 4} < m^2<{d^2\over 4}$, where $d$ is the spatial dimension which is $2$ for the current model. The upper and the lower bounds are not included, although these are very interesting cases. I would like to leave them to future work and only give some punctual comments when there is something interesting to mention. With the formulation Eq.(\ref{SigmatoI}) for the near dS boundary limit, the behaviors of the fields on $\Sigma_*$ when $\Sigma_*\rightarrow I^+$ are given by Eqs (\ref{Phi152})--(\ref{163afbtm2}) with $u=e^{-\Lambda}$. The expressions we will need are
\begin{align}
	& \Phi_*=\varphi \sim \hat \alpha \, e^{-\Delta_- \Lambda}+ \hat \beta \, e^{-\Delta_+  \Lambda}+\dots, \ \ {\rm with}\ \  \Delta_{\pm}=1\pm \sqrt{1-m^2}, \label{Phi*38} \\
	&{\cal A}=a_*b_*=C_aC_b\, e^{2\Lambda} \left( 1- {\hat \alpha^2 \over 2}e^{-2\Delta_- \Lambda}+{\hat \alpha^4 \over 16}e^{-4\Delta_- \Lambda} -m^2 \hat \alpha \hat \beta e^{-2\Lambda}\right)+\dots ,  \label{calACC}  \\
	&{1\over 2}\big(a_*\dot b_*+\dot a_* b_* \big)=C_aC_b\, e^{2\Lambda} \left[ 1+(\Delta_- -1){\hat \alpha^2 \over 2}e^{-2\Delta_- \Lambda} +(1-2\Delta_-){\hat \alpha^4 \over 16} e^{-4\Delta_- \Lambda} \right]+\dots, \label{a*b*CC}\\
	&\tau_2 ={b_*\over a_*} ={C_b \over C_a} +\dots, \label{tauCC}
\end{align}
where $0<C_a,C_b<\infty$, and $\hat \alpha$, $\hat \beta$ are constants which can be complex. Due to the range of scalar mass ${\sqrt 3\over 2}<m<1$, we have $0<\Delta_--\Delta_+ <1$, ensuring that Eq.(\ref{Phi*38}) has indeed included the leading terms. Also there is one short-coming regarding the expansion Eq.(\ref{Phi*38}), in that the coefficients $\hat \alpha$ and $\hat \beta$ are not $SL(2,\mathbb Z)$-invariant, because $\Lambda$ is not. We can improve this by rewriting this asymptotic expansion against $\cal A$:
\begin{align}
	&   \varphi \sim  \alpha \,{\cal A}^{-{\Delta_- \over 2} }+\beta \, {\cal A}^{-{\Delta_+ \over 2}} , \label{Phi*40}  \ \ {\rm where}\\ 
	&\alpha = (C_aC_b)^{\Delta_- \over 2} \hat \alpha,\ \ \beta = (C_aC_b)^{\Delta_+ \over 2} \hat \beta. \label{afbthat} 
\end{align}
Now since both $\varphi$ and $\cal A$ are modular invariant, so are $\alpha$ and $\beta$.


One important property that Eq.(\ref{a*b*CC}) implies is the reality of the IR divergences in the on-shell action Eq.(\ref{ItotBS}). Here the IR divergences in $\cal S$ are those arising when we take the limit $\Lambda\rightarrow \infty$ keeping the other parameters fixed. The fixed parameters include $\tau$, $\alpha$ and $\beta$, or equivalently, $b_o$ and $\Phi_o$. Therefore from Eq.(\ref{ItotBS}) we see that all IR divergent terms are included in Eq.(\ref{a*b*CC}) and are manifestly real. Therefore $\im\cal S({\cal A},\tau,\varphi)$ is finite under the limit $\Lambda\rightarrow \infty$, and it tends to a function of $\tau$ and $\alpha$. 
Therefore if we let
\begin{align}
	\re {\cal S}\big({\cal A}, \tau,\varphi\big)={\cal S}_{\rm R}\big({\cal A}, \tau,\varphi\big),\ \ \im {\cal S}\big({\cal A}, \tau,\varphi\big)={\cal I}\big({\cal A}, \tau,\varphi\big), \label{ReSImSSplit}
\end{align}
then we can express the reality of IR divergence with
\begin{align}
	\lim_{{\cal A}\rightarrow \infty} {\cal I}\Big({\cal A}, \tau,\alpha {\cal A}^{-{\Delta_- \over 2}}+ \beta {\cal A}^{-{\Delta_+ \over 2}} \Big) :={\cal I} \big(\tau,\alpha  \big). \label{ImOSFin}
\end{align}
The disappearing of $\cal A$-dependence is obvious, while the $\varphi$-dependence being replaced by $\alpha$-dependence is because of the expansion (\ref{Phi*40}) as well as the fact that when $\Phi$ is on shell, $\beta$ is a function of $\tau$ and $\alpha$.  In the following up sections when it comes to the imaginary part of the on-shell action in the bulk, it will always be referring to ${\cal I} \big(\tau,\alpha  \big)$. $\cal I$ for finite ${\cal A}$ will not be studied since it is not relevant to the purpose of the paper. An example of this study can be found in the appendix of \cite{Hartle:2008ng}.

Now we can write down the expression of the on-shell action which display the IR divergences:
\begin{align}
	{\cal S}\big({\cal A},\tau,\varphi \big)= -C_aC_b\, e^{2\Lambda} \left[ 1+(\Delta_- -1){\hat \alpha^2 \over 2}e^{-2\Delta_- \Lambda} +(1-2\Delta_-){\hat \alpha^4 \over 16} e^{-4\Delta_- \Lambda} \right]-{i\over 2} b_o +\dots, \label{ALTGTH}
\end{align}
or, taking the real and imaginary parts:
\begin{align}
	&\,{\cal S}_{\rm R}({\cal A},\tau,\varphi)= -C_aC_b\, e^{2\Lambda} \left[ 1+(\Delta_- -1){\hat \alpha^2 \over 2}e^{-2\Delta_- \Lambda} +(1-2\Delta_-){\hat \alpha^4 \over 16} e^{-4\Delta_- \Lambda} \right]+{1\over 2} \im b_o+\dots \label{splitRe} \\
	&\, {\cal I}(\tau,\alpha)=-{1\over 2} \re b_o, \label{splitIm}
\end{align}
where the dots in Eqs (\ref{ALTGTH}) and (\ref{splitRe}) are exponentially suppressed terms.

\subsubsection*{Cancellation of infrared divergence}

With Eq.(\ref{ALTGTH}) we are able to carry out the holographic renormalization procedure. The steps to follow is standard: adding the local counter term needed to cancel the leading IR divergence; updating the coefficients of the rest of the IR divergences; adding counter terms which cancel the now leading IR divergence, and so  on until all divergence is removed.

Regarding Eq.(\ref{ALTGTH}), we need first add to it $\cal A$ to cancel the volume divergence $C_aC_be^{2\Lambda}$, where $\cal A$ should be carefully expanded into power series of $e^\Lambda$ using Eq.(\ref{calACC}). After this step, we will find that the next-to-leading order is $-{1\over 2} \Delta_- \hat \alpha^2 e^{(2-2\Delta_- )\Lambda}$. Therefore we continue the procedure adding ${1\over 2} \Delta_- {\cal A}\varphi^2$, which will cancel $-{1\over 2} \Delta_- \hat \alpha^2 e^{(2-2\Delta_- )\Lambda}$ and meanwhile induce an infinite sum of terms each of order $e^{(2-2n\Delta_-)\Lambda}$ with $n=2,3,4,\dots$ Due to the setting of mass range in this paper, these terms are all exponential suppressed\footnote{When the scalar mass saturates the lower bound $m={\sqrt 3 \over 2}$, it seems that the $\hat \alpha^4 e^{(2-4\Delta_-)\Lambda}$-term is finite and we do not know whether to introduce further counter term to remove it or not. However straightforward computation shows that when $\hat \alpha^2 e^{(2-2\Delta_-)\Lambda}$-divergence are cancelled, the $\hat \alpha^4 e^{(2-4\Delta_-)\Lambda}$-term is automatically removed. Thus the counter terms presented below Eq.(\ref{CTTM}) actually works for $m= {\sqrt 3\over 2}$.} and hence the holographic renormalization is done. The counter term action is thus the well familiar result of holographic renormalizaion with the presence of a scalar field \cite{Skenderis:2002wp}:
\begin{align}
	{\cal S}_{\rm ct}={\cal A}+{\Delta_- \over 2} {\cal A}\, \varphi^2 .  \label{CTTM}
\end{align}
%
%
Adding this to Eq.(\ref{ALTGTH}), we have the renormalized saddle-point action, to be denoted by $\tilde {\cal S}$ here (the same notation will be used for other models to be studied later):
\begin{align}
	\tilde {\cal S}:=&\, {\cal S}+{\cal S}_{\rm ct}=-{i\over 2}b_o - C_a C_b\, e^{2\Lambda} \left[ 1+(\Delta_- -1){\hat \alpha^2\over 2} e^{-2 \Delta_- \Lambda} \right]\nonumber \\ &+ C_a C_b\, e^{2\Lambda} \left( \Delta_- {\hat \alpha^2\over 2} e^{-2 \Delta_- \Lambda}+\Delta_- \hat \alpha \hat\beta  \right) +C_aC_b\, e^{2\Lambda} \left( 1- {\hat \alpha^2 \over 2}e^{-2\Delta_- \Lambda} -m^2 \hat \alpha \hat \beta\, e^{-2\Lambda} \right) +\dots \nonumber \\ =& \,   - {i\over 2}b_o+ \left( \Delta_- -m^2 \right) C_a C_b\, \hat\alpha\, \hat\beta+\dots = - {i\over 2}b_o + \left( \Delta_- -m^2 \right) \alpha\, \beta+\dots \label{HolRen1}
\end{align}
Here the dots in each line represent the exponentially suppressed terms. In the first line after the second equality is the bulk action ${\cal S}$, and the second line are terms from $\cal A$ and ${\Delta_-\over 2}  {\cal A}\, \varphi^2$. In the last line, Eq.(\ref{afbthat}) was used. The final result for the renormalized on-shell action, when ${\cal A}\rightarrow \infty$ is fully implemented, is\footnote{This is the result in $\alpha$-representation, which is related to that of the $\beta$-representation by a Legendre transform. The $\beta$-representation result is actually $\tilde {\cal S}(\tau,\beta)= - {i\over 2}b_o(\tau,\beta)+ \left( \Delta_+ -m^2 \right) \alpha(\tau,\beta)\, \beta.$}
\begin{align}
	\tilde {\cal S}(\tau,\alpha)= - {i\over 2}b_o(\tau,\alpha)+ \left( \Delta_- -m^2 \right) \alpha\, \beta(\tau,\alpha). \label{HolRen2}
\end{align}
It will be useful to introduce the notations in parallel with Eq.(\ref{ReSImSSplit}):
\begin{align}
	\re \tilde {\cal S}\big( \tau,\alpha\big)=\tilde {\cal S}_{\rm R}\big(\tau,\alpha\big),\ \ \im \tilde {\cal S}\big(\tau,\alpha\big)= \tilde {\cal I}\big(\tau,\alpha\big), \label{RItSSplit}
\end{align}
and thus Eq.(\ref{HolRen2}) is split into
\begin{align}
	& \tilde {\cal S}_{\rm R}\big(\tau,\alpha\big)={1\over 2}\im b_o(\tau,\alpha)+ \left( \Delta_- -m^2 \right) \alpha\, \re\beta(\tau,\alpha), \label{SS53} \\
	& \tilde {\cal I}\big(\tau,\alpha\big) =-{1\over 2}\re b_o(\tau,\alpha)+ \left( \Delta_- -m^2 \right) \alpha\, \im \beta(\tau,\alpha). \label{II54}
\end{align}
This is an exact result of holographic renormalization for finite scalar deformation, with the limitation that the deformation must be homogeneous and isotropic. If the computation is correct, then we should have the relation of ``one-point function'' generation \cite{Klebanov:1999tb}
\begin{align}
 {\partial 	\tilde{\cal S} \over \partial \alpha} =(\Delta_--\Delta_+)\beta. \label{OPFunc}
\end{align}
This is not at all straightforward seen from Eq.(\ref{HolRen2}) but it will be verified, either perturbatively in Sec.\ref{PERTT2} or non-perturbatively by numerics in Sec.\ref{NUMT2}.

\subsubsection*{Holographic renormalization as generalized Legendre transform} \label{GFTHH}



So far the result of holographic renormalization $\tilde {\cal S}$ is automatically considered as the quantity to be directly identified with the boundary field theory generating function (in case boundary dual available), which seems natural enough. There is however an issue arising, if we remember that the initial motivation is to address the problem raised in \cite{Anninos:2012ft}. The latter have a seemingly different opinion, which states that the bulk-computed quantity to be identified with boundary QFT partition function, is the Hartle-Hawking wave function, but the one having been Fourier-transformed into the representation of dS boundary data (as if $\tau$ and $\alpha$ for our case). 


Therefore, suppose our bulk wave function $\Psi({\cal A},\tau,\varphi)$, when transformed to the boundary data representation, becomes $\Psi_{\!{}_{I^+}}(\tau,\alpha)$ with its tree-level approximation $\Psi_{\!{}_{I^+}}\!\sim \exp(i{\cal S}_{\!{}_{I^+}})$, then we need to have $\tilde {\cal S}={\cal S}_{\!{}_{I^+}}$, in order that the conclusions about divergence drawn from $\tilde{\cal S}$ are on the same comparing ground with those drawn in \cite{Anninos:2012ft}. Indeed although a rigorous proof is not yet found for the very model being studied, the properties of $\tilde {\cal S}$ strongly suggests that the equality $\tilde {\cal S}={\cal S}_{\!{}_{I^+}}$ should hold.


Basically we see that on one hand  ${\cal S}_{\!{}_{I^+}}$ is supposed to be the on-shell value of an action well-defined for the variation principle with $(\tau,\alpha)$ fixed on the boundary $I^+$, while on the other hand $\tilde {\cal S}$ fits the main criteria to be such a quantity. First, it is finite and second it satisfies the relation (\ref{OPFunc})  (to be verified in the next two sections), obligatory for the on-shell action in the $\alpha$-representation, knowing that $(\Delta_- -\Delta_+)\beta$ is the conjugate momentum of $\alpha$.\footnote{For the range of mass chosen in this paper, the $(\alpha, \beta)$ space inherits a natural symplectic structure from the phase space of the bulk scalar field. Let $\Pi_\Phi$ be the conjugate momentum of $\Phi$ and thus $\Pi_\varphi= \Pi_\Phi(\chi_*)$. Using the expression of the action (\ref{Itotabp}) and Eq.(\ref{Phi*40}), we have asymptotically $ \Pi_{\varphi} =a_* b_*\, \dot \Phi_* \sim - \Delta_-\alpha\, {\cal A}^{\Delta_+ /2}- \Delta_+ \beta\, {\cal A}^{\Delta_-/ 2}$, 
from which we can derive explicitly the transform between $(\varphi,\Pi_{\varphi})$ and $\big(\alpha,(\Delta_- -\Delta_+)\beta \big)$, and we will find that it is a canonical transform since the transformation matrix has unit determinant. We can further on to derive the type one generation function of this transform and will find that it is, up to negative powers of $\cal A$, just $-{1\over 2}\Delta_- {\cal A} \varphi^2$, one of the counter terms in the holographic renormalization (c.f. for example \cite{Ghandour:1986ec} for the relation between generating functions of canonical transforms and the kernels of Fourier transforms). All this is still true when $m$ saturates the upper bound $m=1$, and the scalar asymptotic behavior becomes $\varphi\sim \alpha {\cal A}^{-\Delta_-/2}\ln {\cal A}+\beta  {\cal A}^{-\Delta_+/2}$.
} 

A very relevant work in this regard has been presented in \cite{Papadimitriou:2010as} which reveals that the holographic renormalization applies to a wide category of theories possessing an asymptotic boundary, regardless of whether a boundary dual theory exists or not; from the bulk point of view, it is a procedure of figuring out the right action which have the variation principle well defined in terms of the asymptotic data characterizing the classical solutions. Therefore according to \cite{Papadimitriou:2010as} the holographic renormalization just performed to arrive at Eq.(\ref{HolRen2}) well corresponds to the procedure for figuring out the right action for the variation principle in terms of $\tau$ and $\alpha$. A theoretical framework has been set forward in \cite{Papadimitriou:2010as}, and probably it can be applied here to achieve a solid demonstration of $\tilde {\cal S}={\cal S}_{\!{}_{I^+}}$.

\section{Saddle points of empty spacetimes and scalar perturbations} \label{PERTT2}

In the previous section the framework has been worked out for quantitatively computing the physical properties of the no-boundary saddle points. However before having them implemented numerically, it is important to obtain an analytic preview by treating the scalar field perturbatively. The result will serve as
the starting points for the numerical search of saddle points subjected to finite scalar deformations. This section is devoted to this task and the numerical computation will be carried out in the next section.
A key point is that since the perturbation theory is for approximating a theory that takes scalar backreaction into full account, the spacetime background should not be set strictly rigid. Instead, its deformation should be consistently truncated to the order preset in the perturbative framework, and here it is truncated to the quadratic order. The following contents will be presented in this section.

Sec.\ref{EPTNBSD} works out the saddle points of empty spacetime.
Emphasis will be on the necessary complexification of the coordinate time, and also on the existence of infinitely many saddle points that fit some boundary condition assigned, differing in the amount of Euclidean time covered by their complex time contours.


In Sec.\ref{HPES} a homogeneous scalar perturbation is switched on, and is explicitly solved out as a function of the complexified coordinate time. The Riemann surface of the solution is investigated, which shows that it has infinite number of layers due to the presence of logarithmic singularities. It thus follows that saddle points with a perturbative scalar field can differ one from another not only by the amount of Euclidean time they experience, but further by the Riemann sheet that their time contours terminate on.

Sec.\ref{OSactPert} and Sec.\ref{MVT2} compute the saddle-point actions to the quadratic order of scalar perturbation, with holographic renormalization performed. Sec.\ref{OSactPert} focuses on saddle points differing in the Eulidean time length that their time contours cover, while Sec.\ref{MVT2} focuses on the saddle points differing in the way their complex time contours circle around branch points. These different saddle points generally lead to different actions.

Sec.\ref{Sec55} will very briefly cover the issue of inhomogeneous scalar perturbation, which aims to point out that for each inhomogeneous mode against the Fourrier basis of the torus, the qualitative properties of the Riemann surface and of the saddle-point actions are essentially the same as those of the homogeneous mode.

\subsection{Empty spaceeime saddle points and its dS or EAdS contours} \label{EPTNBSD}

In the absence of the scalar field, the wave function is reduced to a function of the geometry of $\Sigma_*$: $\Psi\! \left({\cal A},\tau \right)$, so is its tree-level approximation is $e^{i{\cal S}_0({\cal A},\tau)}$ with ${\cal S}_0({\cal A},\tau)$ the on-shell value of Eq.(\ref{tot2}) evaluated on the saddle point with $\Phi=0$ and boundary $\Sigma_*$ characterized by $({\cal A},\tau)$. 

\subsubsection*{Infinite number of saddle points and the ``fundamental'' saddle point}

In the mini-superspace setup, as summarized in the second paragraph below Eq.(\ref{Ephi}), the saddle points are obtained by integrating Eqs (\ref{Eabp})---(\ref{Ebph}) with $\Phi=0$ from the south pole $\chi_o$ respecting the conditions (\ref{NBCab}) and (\ref{NBCPhi}), until some final moment $\chi_*$ where the field values match the assigned boundary data $({\cal A},\tau)$ as Eq.(\ref{rel18**}). The south pole data $b_o$, as well as the final moment $\chi_*$ are determined through this matching. Since it is the difference $\chi_*-\chi_o$ that really matters, we can fix the south pole. Here it is fixed at $\chi_o={i\pi \over 2}$, and the solution to Eqs (\ref{Eabp})---(\ref{Ebph}) is trivial:
\begin{align}
	& a(\chi)= \cosh \chi ,\ \ b(\chi)=-ib_o \sinh\chi,  \label{Sabphi=0}
\end{align}
where the absence of spacetime boundary at $\chi_o$ is seen from $a(\chi)=\cosh\chi$, in that $a(\chi_o+\epsilon)\sim i\epsilon$ ($\epsilon\sim 0$) is satisfied. Now we can enforce the boundary condition at $\Sigma_*$ to determine $b_o$ and $\chi_*$, which are, inferred from Eq.(\ref{rel18**}):
\begin{align}
	{\cal A}=a_*b_*=-ib_o \cosh \chi_* \sinh \chi_*, \ \ \tau_2={b_*\over a_*}= -ib_o \tanh \chi_*. \label{AtauMatch}
\end{align}
In the context of $\Sigma_*\rightarrow I^+$, with $\chi_*$ parameterized by Eq.(\ref{SigmatoI}), we have ${\cal A}\sim e^{2\Lambda}$, and the only possibility for Eq.(\ref{AtauMatch}) to hold is
\begin{align}
	& \delta_*=k\pi,\ \ {\rm where}\ \ k\in\mathbb Z; \ \ {\rm and}\ \ b_o=i\tau_2 \coth\Lambda\sim i\tau_2, \label{AtauMatched}
\end{align}
where $\delta_*=\im \chi$ is the Euclidean shift as in Eq.(\ref{SigmatoI}). Therefore there are infinitely many saddle points of empty spacetime, labeled by integer $k$ in the Euclidean shift $\delta_*$, corresponding to different amount of Euclidean time that the saddle point experiences in completing its complex history in terms of $\chi$. However the metric of all these saddle points are the same, given by Eq.(\ref{gxaebe}) with $a(\chi)$ and $b(\chi)$ just obtained above:
\begin{align}
	\ell^{-2}ds^2 = -d\chi^2+\cosh^2 \!\chi \, d\zeta_2^2+\tau_2^2 \sinh^2 \! \chi\, d\zeta_1^2, \label{3Met}
\end{align}
where the coordinates $\zeta_{1,2}$ are periodically identified according to Eq.(\ref{PRDeta}). 

To summarize, a complete specification of a saddle point of empty spacetime consists of the metric Eq.(\ref{3Met}), and a $\chi$-contour which starts from the south pole and ends at $\chi_*=ik\pi+\infty$ with some specific integer $k$, and this contour can be arbitrarily deformed between the fixed two ends since with no singularity present on the $\chi$-plane, such different contours, starting from the same south pole data $b_o$, will result in the same boundary data at $\Sigma_*$ and the same on-shell action. A schematic illustration of these saddle points is given in Fig.\ref{KerrdSEBTZ}.

Among all the saddle points labeled by $k$, we can single out one type of them that are commonly considered in the literature as the contributing saddle point of the Hartle-Hawking wave function  (for example in the original paper \cite{Hertog:2011ky,Hartle:1983ai,Hartle:2008ng}). These are saddle points experiencing a minimum amount of positive Euclidean time, such as the saddle points I and III in Fig.\ref{KerrdSEBTZ} which have $\delta_*=0$. Such saddle points are to be referred to as ``fundamental saddle points'' in this paper. Later we will see, when scalar field is switched on, that they yield very different physical outcomes from the non-fundamental saddle points.

\subsubsection*{Kerr-dS universe and EBTZ black hole}

An important property to mention about the saddle points is that its spacetime can be either asymptotically dS or asymptotically AdS according to the choice of complex time contour. 

We first notice that when $\chi$ runs on the horizontal lines $\chi=ik\pi+\lambda$ where $k\in\mathbb Z$ is fixed and $\lambda\in \mathbb R$ varies, the metric Eq.(\ref{3Met}) becomes
\begin{align}
	\ell^{-2}ds^2 = -d\lambda^2+\cosh^2 \!\lambda \, d\zeta_2^2+\tau_2^2 \sinh^2 \! \lambda\, d\zeta_1^2, \label{3Met2}
\end{align}
and this is just the Kerr-dS metric. To see more clearly we introduce the time variable $t$ defined as
\begin{align}
	\cosh^2\! \lambda ={t^2+\tau_2^2 \over |\tau|^2} \ \ {\rm or} \ \ \sinh^2\! \lambda ={t^2-\tau_1^2 \over |\tau|^2}
\end{align}
as well as the spatial coordinates $(\theta,\phi)$: 
\begin{align}
	\zeta_1={ |\tau|^2+\tau_1^2 \over \tau_2}\, \theta + \tau_1 \, \phi,\ \ \zeta_2={\tau_1\over \tau_2} \theta+\phi, \label{thtaphi}
\end{align}
whose periodicities are, referring to Eq.(\ref{PRDeta}):
\begin{align}
	(\Delta \phi,\Delta\theta)=2\pi(1,0) \ \ {\rm or}\ \ 2\pi \left( -{\tau_1\over |\tau|^2}, {\tau_2 \over |\tau|^2}\right) \label{PRDtAdS}
\end{align}
In terms of these variables, the 3-metric (\ref{3Met}) reads
\begin{align}
	\ell^{-2}ds^2 = -{t^2 dt^2 \over (t^2-\tau_1^2)(t^2+\tau_2^2)}+{(t^2-\tau_1^2)(t^2+\tau_2^2) \over t^2 } d\theta^2+t^2\left(d\phi-{\tau_1\tau_2 \over t^2}d\theta\right)^2, \label{3MetBis}
\end{align}
which is nothing but the Kerr-dS$_3$ metric in its standard form.

Another type of contours lead to asymptotic AdS behavior, which are the horizontal lines given by $\chi=i\left(k+{1\over 2} \right) \pi+\lambda$ where $k\in\mathbb Z$ is fixed and $\lambda\in \mathbb R$ varies. Along these contours the metric becomes Euclidean but with an overall negative sign, which can be absorbed into the dS length by defining $\ell=i\ell'$: 
\begin{align}
	\ell'^{-2}ds^2 = d\lambda^2+\sinh^2 \!\lambda \, d\zeta_2^2+\tau_2^2 \cosh^2 \! \lambda \, d\zeta_1^2, \label{3MetEBTZ}
\end{align}
This is the EBTZ black hole metric, whose event horizon is at $\lambda=0$ where the $\zeta_2$-circle shrinks smoothly to zero size. In fact the south pole $\chi_o={ i\pi \over 2}$ lies just on the $k=0$ contour and is the event horizon of that very EBTZ black hole. To cast Eq.(\ref{3MetEBTZ}) into the standard form, we introduce the radial coordinate $r$ such that
\begin{align}
	\sinh^2\! \lambda ={-r^2+\tau_2^2 \over |\tau|^2} \ \ {\rm or} \ \ \cosh^2\! \lambda =-{r^2+\tau_1^2 \over |\tau|^2},
\end{align}
as well as the same angular coordinates as in Eq.(\ref{thtaphi}). This leads to 
\begin{align}
	\ell'^{-2}ds^2 = {r^2 dt^2 \over (r^2+\tau_1^2)(r^2-\tau_2^2)}+{(r^2+\tau_1^2)(r^2-\tau_2^2) \over r^2 } d\theta^2+r^2\left(d\phi+{\tau_1\tau_2 \over r^2}d\theta\right)^2, \label{3MetBisEBTZ}
\end{align}
which describes a Euclidean BTZ black hole whose horizon is at $r=\tau_2$. With Eq.(\ref{PRDtAdS}) it is clear that $\tau$ is rather the modular parameter of its dual thermal AdS space \cite{Kraus:2006wn}, while it is $\tilde \tau=-\tau^{-1}$ that is the modular parameter of EBTZ black hole. Therefore we have the black hole temperature $\tilde \tau_2^{-1}$ and angular momentum $\tilde \tau_1$. 

The possibility of analytic continuation between Kerr-dS$_3$ metric and EBTZ metric is just an example of a more general feature that saddle points having an asymptotic dS region can be continued into one with an asymptotic EAdS region. This feature is used in \cite{Maldacena:2002vr} to obtain correlation functions on the dS boundary. Also it is used in \cite{Castro:2012gc} to obtain the locally dS$_3$ wave function from the locally AdS$_3$ partition function. In a more general context it has been used in \cite{Hertog:2011ky,Hartle:2013vta,Hartle:2012qb} to work out a holographic probability measure of classical dS cosmology, where the dual CFT lives on the EAdS boundary. In this paper, this feature can relate the cosmological problem to that of the thermodynamics of BTZ black holes with scalar hair.
%
%


\subsubsection*{Saddle-point action}

Plugging the solutions for $a(\chi)$ and $b(\chi)$ into Eq.(\ref{Itotabp}) putting $\Phi=0$, we obtain the saddle-point action
\begin{align}
	{\cal S}_0({\cal A},\tau)=&\int_{\chi_{{}_o}}^{\chi_{{}_*}}\!\!d\chi \left(-\dot a \dot b - ab\right)- \left( \dot a b+ a\dot b \right)_{\chi_{{}_o}} \nonumber \\ =&\,  \int_{i\pi\over 2}^{ik\pi+\Lambda} \!\!\!d\chi \Big(\! -\tau_2 \sinh \chi \cosh\chi - \tau_2 \cosh \chi \sinh\chi\Big)- \tau_2 \sinh^2{i\pi \over 2}  \nonumber \\ =&\, -\tau_2\sinh^2\Lambda=- \sqrt{\tau_2 \sinh \Lambda \cosh \Lambda\left(\tau_2 \sinh \Lambda \cosh \Lambda-\tau_2 \tanh \Lambda \right)}  \nonumber \\ =&\, - \sqrt{{\cal A} \left({\cal A}-\tau_2 \right)} = -{\cal A} +{\tau_2 \over 2} +O({\cal A}^{-1}).\label{ANpgPur}
\end{align}
The result consists of a volume divergent term and a finite term and other terms suppressed by negative powers of ${\cal A}$. Since it is real, the wave function at tree level $\Psi\sim e^{i{\cal S}_0({\cal A},\tau)}$ is a pure phase. Also, note that all the different saddle points, specified by $\chi_*=ik\pi+\Lambda$ with $k\in \mathbb Z$, give rise to the same on-shell action. 

Adding the counter terms Eq.(\ref{CTTM}) to Eq.(\ref{ANpgPur}), in fact only the first term $\cal A$ due to the absence of scalar field, we obtain the holographically renormalized saddle-point action:
\begin{align}
	\tilde {\cal S}_0(\tau)={\cal S}_0({\cal A},\tau)+{\cal A} = - \sqrt{{\cal A} \left({\cal A}-\tau_2 \right)}+{\cal A} ={\tau_2 \over 2} +O({\cal A}^{-1}).\label{ANpgPurBDR}
\end{align}
The corresponding saddle-point contribution to the boundary Hartle-Hawking wave function is $e^{i\tilde {\cal S}}=e^{i\tau_2\over 2}$, and is just a pure phase. This is unlike the 4d saddle points of empty spacetime of asymptotic boundary $S^1\times S^2$ considered in \cite{Anninos:2012ft,Banerjee:2013mca}, whose actions have non-trivial imaginary parts. Note that once we recover the coupling constant $\kappa$ as in Eq.(\ref{kpaCC}), we see that Eq.(\ref{ANpgPurBDR}) is just saddle-point action of the exact wave function of empty Kerr-dS obtained \cite{Castro:2012gc}, which can also be obtained in different ways along the EBTZ contour, for example in \cite{Kraus:2006wn,Carlip:1994gc,Teitelboim:1994az}, in the context of black hole thermodynamics. Indeed it is an interesting aspect to explore whether the result (\ref{HolRen2}) can be used address problems of black hole with scalar hair.




\subsection{Homogeneous scalar perturbation: infinitely layered Riemann surface} \label{HPES}

Now let us turn on a small scalar perturbation to the no-boundary saddle points. As a result the scalar field should also be an entry in the boundary conditions to impose on $\Sigma_*$. Since $\Sigma_* \rightarrow I^+$, it is convenient to assign the boundary conditions in terms of $({\cal A}, \tau, \alpha)$. With ${\cal A}\rightarrow \infty$, implying $\varphi=\alpha {\cal A}^{-\Delta_-/2}+\beta {\cal A}^{-\Delta_+/2}\approx \alpha {\cal A}^{-\Delta_-/2}$, we can let $\alpha$ be real to have $\varphi$ real. Note that without holographic renormalization, using $({\cal A}, \tau, \alpha)$ is just a reparameterization of the boundary conditions, while the wave function is still in the $({\cal A},\tau,\varphi)$-representation. Then we can solve the scalar equation Eq.(\ref{Ephi}) using its leading order in the expansion in terms of $\Phi$:
\begin{align} 
	&\ddot \Phi+2\coth(2\chi) \,\dot\Phi+ m^2 \Phi=0\, . \label{ScDyn}
\end{align}
Since the south pole is $\chi_o=i\pi/2$, for convenience we define $z= {1+\cosh(2\chi) \over 2}=\cosh^2\! \chi$, so that the south pole is at $z=0$. The equation (\ref{ScDyn}) now becomes a hypergeometric equation
\begin{align}
	z(1-z)\Phi''_{zz}+(1-2z)\Phi'_{z}+{ m^2\over 4} \Phi=0. \label{ScDyn2}
\end{align}
The solution of regular behavior at the south pole $z=0$ is 
\begin{align}
	\Phi(\chi)=C\, {}_2F_1\!\! \left({\Delta_- \over 2},{\Delta_+\over 2},1,z\right)= C\, {}_2F_1\!\! \left(u,v,w,\cosh^2\! \chi \right).\label{pcF}
\end{align}
where $C=\Phi(\chi_o)=\Phi_o$ is just the south pole data of the scalar field, and is to be determined by the boundary data $\alpha$. In the second step above we have set
\begin{align}
	u  =	{\Delta_- \over 2} ={1-\sqrt{1-m^2} \over 2},\ \ v={\Delta_+ \over 2} ={1+\sqrt{1-m^2} \over 2},\ \  w=1. \label{pcF2}
\end{align}
An important property of the analytical structure of the expression (\ref{pcF}), is that it has branch points at $\chi=ik \pi$ ($k\in \mathbb Z$), where the $b$-circle shrinks to zero size. When $\chi \sim ik \pi$ for any integer $k$, the hypergeometric function in Eq.(\ref{pcF}) has the asymptotic behavior  
\begin{align}
	 {}_2F_1\!\! \left(u,v,w,\cosh^2\! \chi \right)= -{\psi(u)+\psi(v)+2\gamma+i\pi +\ln (\cosh^2\! \chi-1)\over \Gamma(u) \Gamma(v)}+O(\cosh^2\! \chi-1), \label{eq73}
\end{align}
where we define the phase to be $-\pi<\arg(\cosh^2\! \chi-1)\leq\pi$.
With the choice of south pole at $\chi_{{}_o}={i\over 2} \pi$ the expression (\ref{pcF}) is valid only in the strip $0\leq {\rm Im}\chi<\pi$. When the scalar field is continued beyond this strip, the value of $\Phi$ depends on the complex $\chi$-contour that we choose connecting the south pole $\chi_{{}_o}$ and the evaluation point of $\Phi$.

\begin{figure}
\begin{center}
\begin{minipage}[t]{0.47\textwidth}
\begin{center}	
	\includegraphics[width=0.75\textwidth]{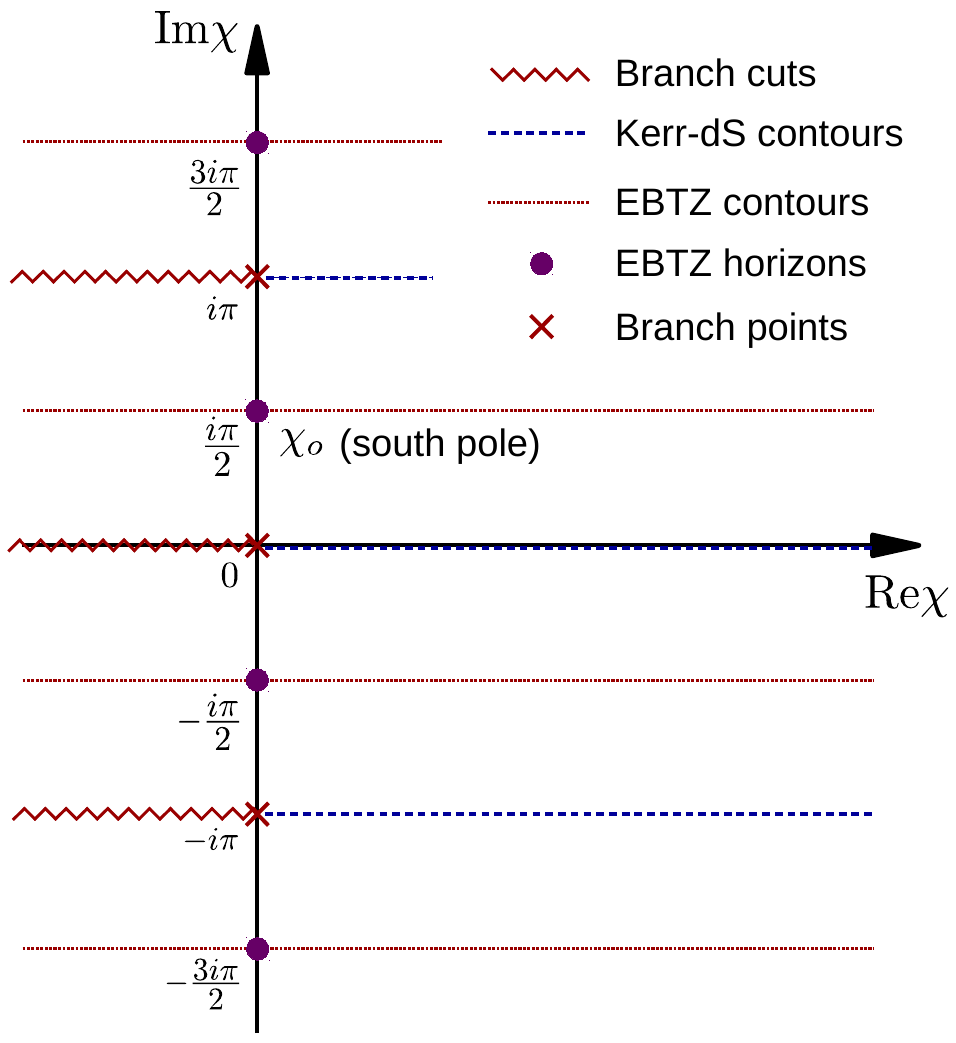}
	\caption{\footnotesize Riemann surface of the perturbative homogenous scalar field $\Phi(\chi)$ as a function of complexified $\chi$. The expression is given by (\ref{hatF}). The Kerr-dS contours and the EBTZ contours are still as in Fig.\ref{KerrdSEBTZ}. The south pole is still $\chi_o={i\pi\over 2}$. There are logarithmic branch points on each Kerr-dS contour at $ik\pi$ ($k\in \mathbb Z$) and the branch cuts are set to open towards $\Re \chi=-\infty$. The figure shows only the Riemann surface layer where the south pole is situated (fundamental layer). On other layers, accessible by crossing the brach cuts, $\chi=i\left(k+{1\over 2} \right)\pi$ ($k\in \mathbb Z$) can also become branch points.} \label{RiemannSurfaces}
\end{center}
\end{minipage}\hspace*{0.06\textwidth}
\begin{minipage}[t]{0.45\textwidth}
\begin{center}	
	 \includegraphics[width=0.66\textwidth]{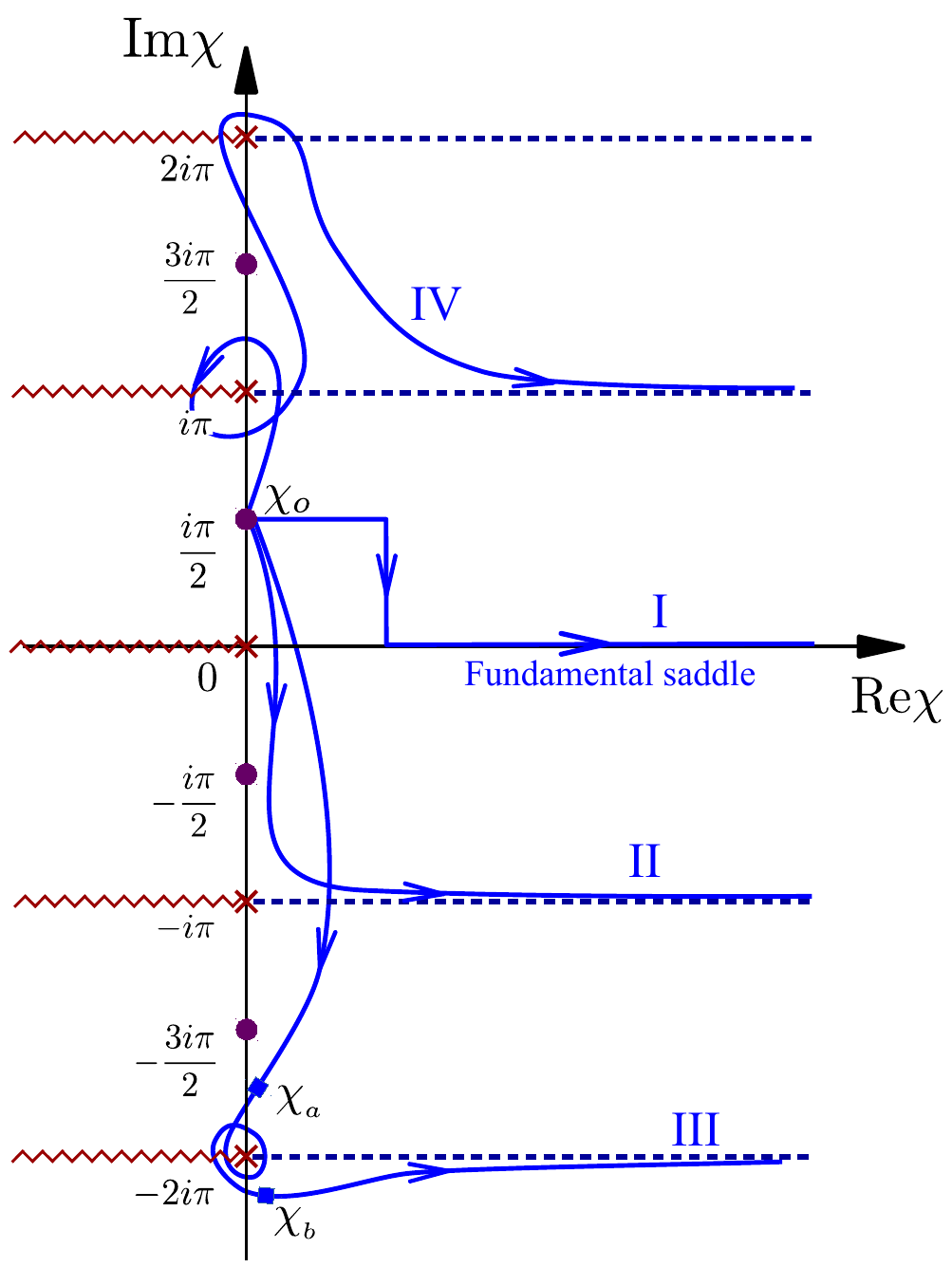}
	\caption{\footnotesize  Examples of saddle points with perturbative scalar field deformation, labeled by I through IV. 
The $\chi$-contours can either avoid all branch cuts and stay in the same layer of the Riemann surface as I and II; or circle around some branch point as III and IV and finish up in another layer.
The saddle point I has the feature of the saddle points most commonly considered in the Hartle-Hawking wave function literature: it has minimum amount of positive Euclidean time with the contour ending up in the same layer as the south pole. It will be qualified as the fundamental saddle point.} \label{AnalStructure}
\end{center}
\end{minipage}
\end{center}
\end{figure}

For the moment let the time be restricted to the ${\rm Re}\chi >0$ half plane for simplicity, while considering the other half plane of ${\rm Re}\chi <0$  will merely result in complex conjugate wave functions. Therefore all the branch cuts open towards the ${\rm Re}\chi <0$ half plane. Therefore we introduce the function
\begin{align}
	{\hat F}(\chi)=\left\{ \begin{array}{ll} {}_2F_1\!\! \left(u,v,w,\cosh^2\! \chi \right), & 0\leq {\rm Im}\chi<\pi, \\ \\ \begin{subarray}{l} \text{Continuation such that all branch} \\ \text{cuts open towards ${\rm Re}\chi <0$}, \end{subarray} & {\rm otherwise.}\end{array}\right. \label{hatF}
\end{align}
The scalar field on the whole complex $\chi$-plane is thus expressed as 
\begin{align}
	\Phi(\chi) = \Phi_o\, {\hat F}(\chi). \label{phiCF}
\end{align}
In Fig.\ref{RiemannSurfaces} a graph of the Riemann surfaces of the homogeneous perturbative scalar field is presented.

When matching $\Phi$ to its boundary condition at $I^+$, we will need the asymptotic behavior of $\hat F(\chi)$ for $\re \chi \rightarrow +\infty$. We can derive this asymptotic behavior from the following identity of hypergeometric function \cite{TbItgSrs}
\begin{align}
	{}_2F_1(u,v,w,x)=&\, {\Gamma(w)\Gamma(v-u) \over \Gamma(v)\Gamma(w-u)}\, {}_2F_1\!\left(u,w-v,u-v+1, {1\over 1-x}\! \right) \nonumber \\ &\hspace{2cm} + {\Gamma(w)\Gamma(u-v) \over \Gamma(u)\Gamma(w-v)} \,{}_2F_1\! \left(v,w-u,v-u+1, {1\over 1-x} \! \right),  \label{ASYF0}
\end{align}
together with the expansion ${}_2F_1(u,v,w,x)=1+{uv\over w}x+O(x^2)$, where the phase of the argument in Eq.(\ref{ASYF0}) is understood as $-\pi<\arg(1-x)\leq \pi$. Therefore we have for Eq.(\ref{hatF}),
\begin{align}
	{\hat F}(\chi)\sim &\, {\Gamma(w)\Gamma(v-u) \over \Gamma(v)\Gamma(w-u)}\,\left({1\over 4}e^{-i \pi}e^{2\chi}\right)^{-u} + {\Gamma(w)\Gamma(u-v) \over \Gamma(u)\Gamma(w-v)} \,\left({1\over 4}e^{-i \pi}e^{2\chi}\right)^{-v},  \ \ \re \chi \rightarrow +\infty, \label{ASYF1}
\end{align}
where we should be careful with the phase of $1-\cosh^2\chi$ to let the above expression match Eq.(\ref{ASYF0}) when $-\pi \leq \im \chi<\pi$. 
It can be easily figured out that when the scalar field is perturbative, the $\chi$-contours characterizing the saddle points are still as in the case of empty space, since their perturbative deformations merely result in beyond-leading-term contributions to the quantities to be computed. Therefore the $\chi$-contours start off from the south pole $\chi_o={i\pi\over 2}$ and ending up at the asymptotic boundary of a Kerr-dS contour $\chi_*=ik\pi+\Lambda$ with $k\in \mathbb Z$ and $\Lambda\rightarrow \infty$. Therefore we have\footnote{Note that when the scalar mass saturates the lower bound $m={\sqrt 3 \over 2}$, we have $2v-2u=1$. Then the sub-leading order of the first line of Eq.(\ref{ASYF0}) is of the same order as the leading order of the second line, and therefore the second term in Eqs (\ref{ASYF1}) and (\ref{ASYF2}) should receive an extra contribution. However this paper will not cover this case.} 
\begin{align}
	\varphi&\, =\Phi_*=\Phi_o{\hat F}(\chi_*) \nonumber \\&\sim \Phi_o {\Gamma(w)\Gamma(v-u) \over \Gamma(v)\Gamma(w-u)}\, \left[{1\over 4}e^{i (2k-1)\pi}e^{2\Lambda}\right]^{-u} +\Phi_o {\Gamma(w)\Gamma(u-v) \over \Gamma(u)\Gamma(w-v)} \,\left[ {1\over 4}e^{i (2k- 1)\pi}e^{2\Lambda}\right]^{-v}. \label{ASYF2}
\end{align}
Fitting this expression into the standard form Eq.(\ref{Phi*40}), 
provided that ${\cal A}=a_*b_*\sim{\tau_2 \over 4}e^{2\Lambda}$ according to Eqs (\ref{AtauMatch}) and (\ref{AtauMatched}), we obtain the coefficients:
\begin{align}
		&\alpha=\Phi_o\,\tau_2^{\, u} e^{iu (1-2k)\pi } {\Gamma(w)\Gamma(v-u) \over \Gamma(v)\Gamma(w-u)}, \ \ \beta=\Phi_o\,  \tau_2^{\, v}e^{iv(1-2k)\pi }  {\Gamma(w)\Gamma(u-v) \over \Gamma(u)\Gamma(w-v)}\, . \label{afbtPhi0}
\end{align}
The first relation serves to determine the south pole data $\Phi_o$ with the boundary value of $\alpha$ assigned at (near) $I^+$:
\begin{align}
		\Phi_o= \alpha\,\tau_2^{- u} e^{iu (2k-1)\pi } {\Gamma(v)\Gamma(w-u) \over \Gamma(w)\Gamma(v-u)}. \label{afdetPhi0}
\end{align}
Now the saddle point is fully specified, recalling that the other defining parameters: $b_o$ and $\delta_*$ have already been obtained in Eq.(\ref{AtauMatched}). Once again like in the previous subsection, infinite number of saddle points are obtained, here labeled by the integer $k$. 

In Fig.\ref{AnalStructure} some example saddle points are shown, represented by complex $\chi$-contours. Those relevant to the current and the next subsections are saddle points I ($k=0$) and II ($k=-1$), whose $\chi$-contours always stay in the same Riemann layer. Other saddle points have loops in their $\chi$-contours around the branch cuts, and this will be considered in Sec.\ref{MVT2}. Again we can single out the fundamental saddle point, which shares the features of the saddle points very commonly used in the literature. It is represented by the contour I and it lives through a shortest positive Euclidean history. However in the presence of the branch cuts in the $\chi$-plane, the fundamental saddle points should also have the time contours that do not circle around any branch point.

Plugging Eq.(\ref{afdetPhi0}) back into the second relation in Eq.(\ref{afbtPhi0}), we obtain $\beta$, which is complex in general. However as explained in the beginning of this subsection, since the contribution to $\varphi$ from $\beta$ is negligible when ${\cal A}\rightarrow \infty$, $\varphi$ can be practically considered as real. For later use, let us put
\begin{align}
	{\beta \over \alpha}=&  \left[ \tau_2\,  e^{i(1-2k)\pi}\right]^{v-u}{\Gamma(v)\Gamma(w-u)\Gamma(u-v)\over \Gamma(u) \Gamma(w-v)\Gamma(v-u)}= \tau_2^{\sqrt{1-m^2}} \rho(m,k), 
\end{align}
where we denoted the $\tau_2$-independent factor by $\rho$:
\begin{align}
	 \rho(m,k)= \exp\Big[i\, \pi(1-2k) \sqrt{1-m^2}\Big] \,{\Gamma\left(-\sqrt{1-m^2}\right)\Gamma\left({1+ \sqrt{1-m^2} \over 2}\right)^2\over \Gamma\left(\sqrt{1-m^2}\right)\Gamma\left({1- \sqrt{1-m^2} \over 2}\right)^2}. \label{rhomk}
\end{align}

\subsection{Saddle-point actions of homogeneous scalar peurtubations} \label{OSactPert}

\subsubsection*{In the naive bulk field representation}

For computing the on-shell action perturbatively we cannot use the expression (\ref{ItotBS}) or (\ref{IabpOnSh}), because they are obtained using the full equations (\ref{Eabp})---(\ref{Ebph}), while the perturbative solutions of saddle points do not satisfy them exactly. We need instead to follow the steps in the appendix of \cite{Hartle:2008ng}, dealing with gravity part ${\cal S}_g$ and the scalar part ${\cal S}_{\Phi}$ separately:
\begin{align}
	{\cal S} = {\cal S}_g &\, + {\cal S}_\Phi \, \ \ \ {\rm where} \label{calStot1} \\
	{\cal S}_g=&\int_{\chi_{{}_o}}^{\chi_{{}_*}}\!\!d\chi \left(-\dot a \dot b - ab\right)-\left( \dot a b+ a\dot b \right)_{\chi_{{}_o}}, \label{ANpg}
\\
		{\cal S}_\Phi=&\,{1\over 2}\int_{\chi_{{}_o}}^{\chi_{{}_*}} \!\!d\chi\, ab\left(\dot \Phi^2- \ell^2 m^2  \Phi^2 \right) 
		=	{1\over 2} \Big(ab \,\Phi\dot\Phi \Big)_{\! \chi_{{}_*}}. \label{ANsp}
\end{align}
In the last line the expression of ${\cal S}_\Phi$, the scalar field has been integrated by parts and the equation of motion (\ref{Ephi}) as well as the south pole condition $a(\chi_o)=0$ is used. 

The scalar part is easier to work out. We insert into Eq.(\ref{ANsp}) the asymptotic expansion Eq.(\ref{Phi*40}), and we have
\begin{align}
		{\cal S}_\Phi=&\, {1\over 2} {\cal A} \big[-\Delta_-\,\alpha^2\, {\cal A}^{-\Delta_-}- (\Delta_-+ \Delta_+)\, \alpha\beta\,  {\cal A}^{-\Delta_-- \Delta_+}- \Delta_+ \, \beta^2\, {\cal A}^{-\Delta_+} \big] \nonumber \\ 
	=&\,  - {\Delta_-\over 2} \alpha^2 {\cal A}^{\sqrt{1-m^2} } -{\Delta_-+ \Delta_+\over 2}\, \alpha \beta - {\Delta_+ \over 2}\, \beta^2{\cal A}^{ -\sqrt{1-m^2} } \nonumber \\
	=&\,- {\alpha^2\over 2} \bigg[\Delta_- {\cal A}^{\sqrt{1-m^2} } +(\Delta_-+ \Delta_+) \tau_2^{\sqrt{1-m^2}}\, \rho  \, \bigg]+\dots,
\label{Iphi1}
\end{align}
where $\rho$ is as in Eq.(\ref{rhomk}). In the last line, the first term in the bracket is IR divergent, the second term complex and finite, and the dots stand for terms suppressed by negative powers of $\cal A$. 

Next step we need to work out ${\cal S}_g$, the contribution from gravity, to the second order in $\alpha$. 
Let some perturbative correction be induced to the metric components due to scalar deformation:
\begin{align}
	a(\chi)\, \rightarrow \, a(\chi)+\delta_{{}_\Phi}a(\chi),\ \ \ b(\chi)\, \rightarrow \, b(\chi)+\delta_{{}_\Phi}b(\chi). \label{abPTBD}
\end{align}
where $a(\chi)$ and $b(\chi)$ are the unperturbed results Eq.(\ref{Sabphi=0}) with $b_o$ given by Eq.(\ref{AtauMatched}).
Since the south pole conditions (\ref{NBCab}) should always hold, we must have
\begin{align}
	\delta_{{}_\Phi}a_o =\delta_{{}_\Phi}\dot a_o=0, \label{Inidao}
\end{align}
where the subscript $o$ means evaluated at $\chi_o$. Varying the equations (\ref{Eabp})--(\ref{Ephi}) with $\delta_{{}_\Phi}$ we find that $\delta_{{}_\Phi}a$ and $\delta_{{}_\Phi}b$ must start from the order $\alpha^2$ in order to have the south pole conditions Eq.(\ref{Inidao}) respected, and we can also derive the condition 
\begin{align}
	\delta_{{}_\Phi}\dot b_o=0. \label{Inidbo}
\end{align}
Now plugging the perturbed quantities Eq.(\ref{abPTBD}) into the gravity sector action Eq.(\ref{ANpg}), using the explicit form of unperturbed $a(\chi)$ and $b(\chi)$, and also using the south pole conditions Eqs (\ref{NBCab}), (\ref{Inidao}) and (\ref{Inidbo}), we get
\begin{align}
	{\cal S}_g({\cal A},\tau,\alpha)={\cal S}_0({\cal A},\tau)-\dot a_* \, \delta_{{}_\Phi} b_*-\dot b_* \, \delta_{{}_\Phi} a_*, \label{SgS0BT}
\end{align}
where the subscript $*$ means evaluated at the final moment $\chi_*$. Also, $S_0({\cal A},\tau)$ is just the $\Phi=0$ result Eq.(\ref{ANpgPur}), where ${\cal A}$ is taken to be the product of unperturbed $a_*$ and $b_*$. Now we need to find out $\delta_{{}_\Phi} a_*$ and $\delta_{{}_\Phi} b_*$. For this purpose we turn to the result obtained in appendix \ref{ABAsydS}, where our perturbative results here correspond to putting $C_a={1\over 2}$, $C_b={\tau_2\over 2}$ and also $u=e^{-\Lambda}$ in Eqs (\ref{Phi152})--(\ref{163afbtm2}). From these expressions we can simply read off the variations of $a_*$ and $b_*$ up to the second order in $\alpha$:
\begin{align}
	& \delta_{{}_\Phi} a_*= -  {1 \over2 u}\left({\hat\alpha^2\over 4} u^{2\Delta_-}  -\hat a_2 u^2+{\hat \beta^2\over 4} u^{2\Delta_+}+\dots \right), \label{dtP99}\\ 
	& \delta_{{}_\Phi} b_*= -  {\tau_2 \over2 u}\left({\hat\alpha^2\over 4} u^{2\Delta_-}  - \hat b_2 u^2+{\hat \beta^2\over 4} u^{2\Delta_+}+\dots \right), \label{dtP100}
\end{align}
where $u=e^{-\Lambda}= \sqrt{{2\over \tau_2}{\cal A}\,}\,$, $\hat \alpha= \alpha\left({\tau_2\over 4}\right)^{-\Delta_-/2}$,  $\hat \beta= \beta\left({\tau_2\over 4}\right)^{-\Delta_+ /2}$ and
\begin{align}
	\hat a_2+ \hat b_2+m^2\hat\alpha \hat \beta=0. \label{a2b2afbtm2}
\end{align}
We also note that the variations of $a$ and $b$ in Eqs (\ref{dtP99}) and (\ref{dtP100}) start from quadratic order in $\alpha$, consistent with the discussion under Eq.(\ref{Inidbo}). Now we can insert Eqs (\ref{dtP99}) and (\ref{dtP100}) into Eq.(\ref{SgS0BT}) to obtain ${\cal S}_g$, which yields
\begin{align}
	S_g({\cal A},\tau,\alpha)=-\sqrt{ {\cal A} ({\cal A} - \tau_2 )} + {\alpha^2\over 2}  \bigg( {\cal A}^{\sqrt{1-m^2} } +2m^2  \tau_2^{\sqrt{1-m^2}} +   \tau_2^{2\sqrt{1-m^2}} \rho^2{\cal A}^{ -\sqrt{1-m^2} }  \bigg) \label{Sg99}
\end{align}
The total saddle-point action Eq.(\ref{calStot1}) is therefore the sum of Eqs (\ref{Iphi1}) and (\ref{Sg99}):
\begin{align}
	{\cal S}({\cal A},&\, \tau,\varphi)={\cal S}_g({\cal A},\tau,\varphi)+{\cal S}_\Phi ({\cal A},\tau,\varphi)  \nonumber \\ = &\,-\sqrt{ {\cal A} ({\cal A} - \tau_2 )} + {\alpha^2\over 2}  \bigg[(1-\Delta_-) {\cal A}^{\sqrt{1-m^2} } + \big(2m^2 -\Delta_-- \Delta_+ \big) \tau_2^{\sqrt{1-m^2}} \rho   \bigg]+\dots \nonumber \\  =& \bigg(\!\!-{\cal A}+ {1-\Delta_-\over 2}  {\cal A}^{\sqrt{1-m^2} }\alpha^2\bigg)+\!\bigg(\! {\tau_2 \over 2}+ {2m^2-\Delta_- - \Delta_+\over 2 }  \tau_2^{\sqrt{1-m^2}} \rho \,\alpha^2\! \bigg)+\dots. \label{IttBk1}
\end{align}
where in the last line the first parenthesis contains all the IR divergent terms, the second parenthesis contains the finite terms, and the dots are all higher order terms in $\alpha$ and those suppressed by negative powers of $\cal A$. The total action is obviously a complex number since $\rho$ is complex, but the IR divergences are manifestly purely real, just as discussed around Eq.(\ref{ImOSFin}). Therefore the imaginary part under the limit ${\cal A}\rightarrow \infty$ is\footnote{Had we regarded the spacetime background as strictly rigid, then the result would be ${\cal I}= -\tau_2^{\sqrt{1-m^2}} \,  {\rm Im} \rho\, \alpha^2 $, which however, does not approximate the model in case scalar deformation is finite.} 
\begin{align}
	{\cal I}( &\, \tau,\alpha)= {1\over 2}\, {\rm Im}\left[ \big(2m^2 -\Delta_- - \Delta_+ \big) \, \tau_2^{\sqrt{1-m^2}} \rho \,\alpha^2\right] = - \big(1-m^2 \big)\, \tau_2^{\sqrt{1-m^2}} \,  {\rm Im} \rho\, \alpha^2 ,
	\label{ImcalS}
\end{align}
where terms suppressed by negative powers of ${\cal A}$ are discarded.

\subsubsection*{Saddle-point actions holographically renormalized}

To do holographic renormalization on Eq.(\ref{IttBk1}), we simply apply the counter terms Eq.(\ref{CTTM}) perturbatively, where $\cal A$ should be understood as 
$(a_*+\delta_{{}_\Phi}a_*)(b_*+\delta_{{}_\Phi}b_*)=a_*b_*+a_*\delta_{{}_\Phi}b_*+b_*\delta_{{}_\Phi}a_*={\cal A}+\delta_{{}_\Phi}{\cal A}$, and where
\begin{align}
	\delta_{{}_\Phi}{\cal A} =a_*\delta_{{}_\Phi}b_*+b_*\delta_{{}_\Phi}a_*= -{\alpha^2\over 2}  \bigg( {\cal A}^{\sqrt{1-m^2} } +2m^2  \tau_2^{\sqrt{1-m^2}}   \bigg)+\dots \label{deltaA107}
\end{align}
Now using Eqs (\ref{IttBk1}) and (\ref{deltaA107}), the renormalization proceeds as
\begin{align}
	\tilde {\cal S} (\tau,\alpha)= &\, {\cal S} ({\cal A},\tau, \varphi)\! + \! \Big({\cal A}+\delta_{{}_\Phi}{\cal A} \Big)\! + \!{\Delta_- \over 2}{\cal A}\Bigg(  {\cal A}^{  -\Delta_- } \alpha^2+2 {\cal A}^{-{\Delta_++\Delta_- \over 2}} \alpha \beta  \Bigg)+\dots \nonumber \\ 
	=&\,   \bigg(\!\!-{\cal A}+ {1-\Delta_-\over 2}  {\cal A}^{\sqrt{1-m^2} }\alpha^2 \! \bigg) +\!\bigg( {\tau_2 \over 2}+ {2m^2-\Delta_- - \Delta_+\over 2 }  \tau_2^{\sqrt{1-m^2}} \rho \,\alpha^2\! \bigg)  \nonumber\\ &    + {\cal A} + {\alpha^2 \over 2}\Bigg[ \big(\Delta_- -1\big) {\cal A}^{  \sqrt{1-m^2}} +2\big( \Delta_--m^2\big)\tau_2^{\sqrt{1-m^2}} \rho  \Bigg]+\dots  \nonumber \\ =&\, {\tau_2 \over 2}+ {1\over 2}\big( \Delta_- -\Delta_+\big)\, \tau_2^{\sqrt{1-m^2}} \rho\, \alpha^2+\dots, \label{tItot90}
\end{align}
where the dots are terms suppressed by negative powers of $\cal A$ or by terms of higher orders than $\alpha^2$. The resulting $\tilde {\cal S}$ is well defined and is supposed to approximate Eq.(\ref{HolRen2}). This will be numerically verified in the next section. Now an immediate test is taking the derivative with respect to $\alpha$:
\begin{align}
	{\partial \tilde {\cal S} \over \partial \alpha}= (\Delta_- - \Delta_+)\, \tau_2^{\sqrt{1-m^2}}\rho\, \alpha=(\Delta_- - \Delta_+)\beta. \label{afbtConjPerturb}
\end{align}
We obtain precisely the ``one-point function generation'' Eq.(\ref{OPFunc}) on the perturbative level. 
Splitting the real and the imaginary parts, we have
\begin{align}
	\tilde {\cal S}_{\rm R}(\tau,\alpha) = {1\over 2}\big( \Delta_- -\Delta_+\big)\, \tau_2^{\sqrt{1-m^2}} \, \re\!\rho\, \alpha^2  \,;\ \ \ \  \tilde {\cal I}(\tau,\alpha) = {1\over 2}\big( \Delta_- -\Delta_+\big)\, \tau_2^{\sqrt{1-m^2}} \, \im\!\rho\, \alpha^2  . \label{ImtcS}\end{align}
Comparing the second expression and Eq.(\ref{ImcalS}), we find the relation 
\begin{align}
	 {\cal I}(\tau,\alpha)=\sqrt{1-m^2}\  \tilde {\cal I}(\tau,\alpha)\, \ \ \ (\text{scalar perturbation}). \label{IItdPT}
\end{align}
Assuming that the higher orders in the loop expansion are highly suppressed with respect to the tree level, this relation seems to imply that the holographic renormalization preserves the relative importance of different saddles. In particular, a dominating saddle for the Hartle-Hawking wave function in the bulk field representation stay dominating for the wave function in the boundary data representation. However we will see that Eq.(\ref{IItdPT}) is utterly abandoned when scalar deformation goes non-perturbative.

\subsubsection*{Observations concerning the sign of $\im \rho$}

Summarizing the results of the saddle-point contributions to the Hartle-Hawking wave functions:
\begin{align}
	&e^{i{\cal S}}=e^{i[{\cal A}+O(\alpha^2) ]} \times \exp\! \left[ \big(1-m^2\big)\tau_2^{\sqrt{1-m^2}} \, \im\! \rho\, \alpha^2+\dots \right], \ \ \ \ \text{(bulk)}; \label{TreeBLKPtb} \\
	&e^{i \tilde {\cal S}}=e^{i[{ \tau_2 \over 2}+O(\alpha^2)]}\times \exp\! \left[ \sqrt{1-m^2}\, \tau_2^{\sqrt{1-m^2}} \,\im\! \rho\,\alpha^2 +\dots\right],\ \ \text{(boundary)}. \label{TreeBDRPtb}
\end{align}

It seems that the sign of $\im \rho$ plays an important role in deciding whether these saddle point contributions are exponentially divergent or suppressed with the scalar deformation characterized by $\alpha$. However this is obviously not the case since these results are valid only for $\alpha \sim 0$. Whatever the sign of $\im \rho$, we will always need to trace the saddle-point contributions to finite $\alpha$ to decide whether they are divergent.

However the sign of $\im\! \rho$ is relevant to the ``high temperature'', or large $\tau_2$ behavior of these results. Note that Eqs (\ref{TreeBLKPtb}) and (\ref{TreeBDRPtb}) are perturbative only in terms of $\alpha$ never in $\tau$. Therefore they are viable for whatever positive value of $\tau_2$. Especially in case the saddle point has ${\rm Im}\rho>0$,  its tree-level contributions to the wave function diverges exponentially as $\tau_2\rightarrow \infty$, both in the bulk or on the boundary. Thus very likely ${\rm Im}\rho>0$ can be associated to non-normalizability of the Hartle-Hawking wave function in $\tau_2$, in case theis divergence is not cancelled by the contribution from other saddles. It will be important to see if this divergence still persists when $\alpha$ increases to the non-perturbative domain, and this will be investigated in the next section. 

In the left half of Fig.(\ref{ImRFLp}) $\im\! \rho$ is plotted for $k=0$ until $k=-5$, and it shows that for $k=0$, $\im\! \rho<0$ for all $m$ between ${\sqrt 3\over 2}$ and $1$, while for other $k$, $\im\! \rho$ can be positive for certain ranges of $m$.







\subsection{Loops circling branch points: multivaluedness explored} \label{MVT2}

This subsection looks into the cases where the saddle points have complex time contour containing loops around some branch points of the scalar field figured out in Sec.\ref{HPES}. Quantitative results will only be presented for the cases where only one branch point is involved. When several branch points are involved the discussion will be qualitative, since a detailed analysis will be tedious and there is not yet perspective seen that quantitive result can bring any conceptually important contribution to the purpose of this paper.

\subsubsection*{Loops around one branch point}

As is stated earlier in the paragraph under Eq.(\ref{eq73}), the branch points are $\chi^{[k]}=i k \pi$ with $k\in \mathbb Z$. Here let us consider the case where the $\chi$-contour circles around the $n$-th branch point $p$ times, where $p>0$ stands for clockwise and $p<0$ counterclockwise. As an example, in Fig.\ref{AnalStructure} the $\chi$-contour of the saddle point III circles the singularity $\chi^{[-2]}$ clockwisely twice ($n=-2$, $p=2$); the $\chi$-contour of the saddle IV circles $\chi^{[1]}$ clockwisely once ($n=1$, $p=1$) and then circles $\chi^{[2]}$ counterclockwisely once ($n=2$, $p=-1$).

\begin{figure}
\begin{center}
	\includegraphics[height=6.5cm]{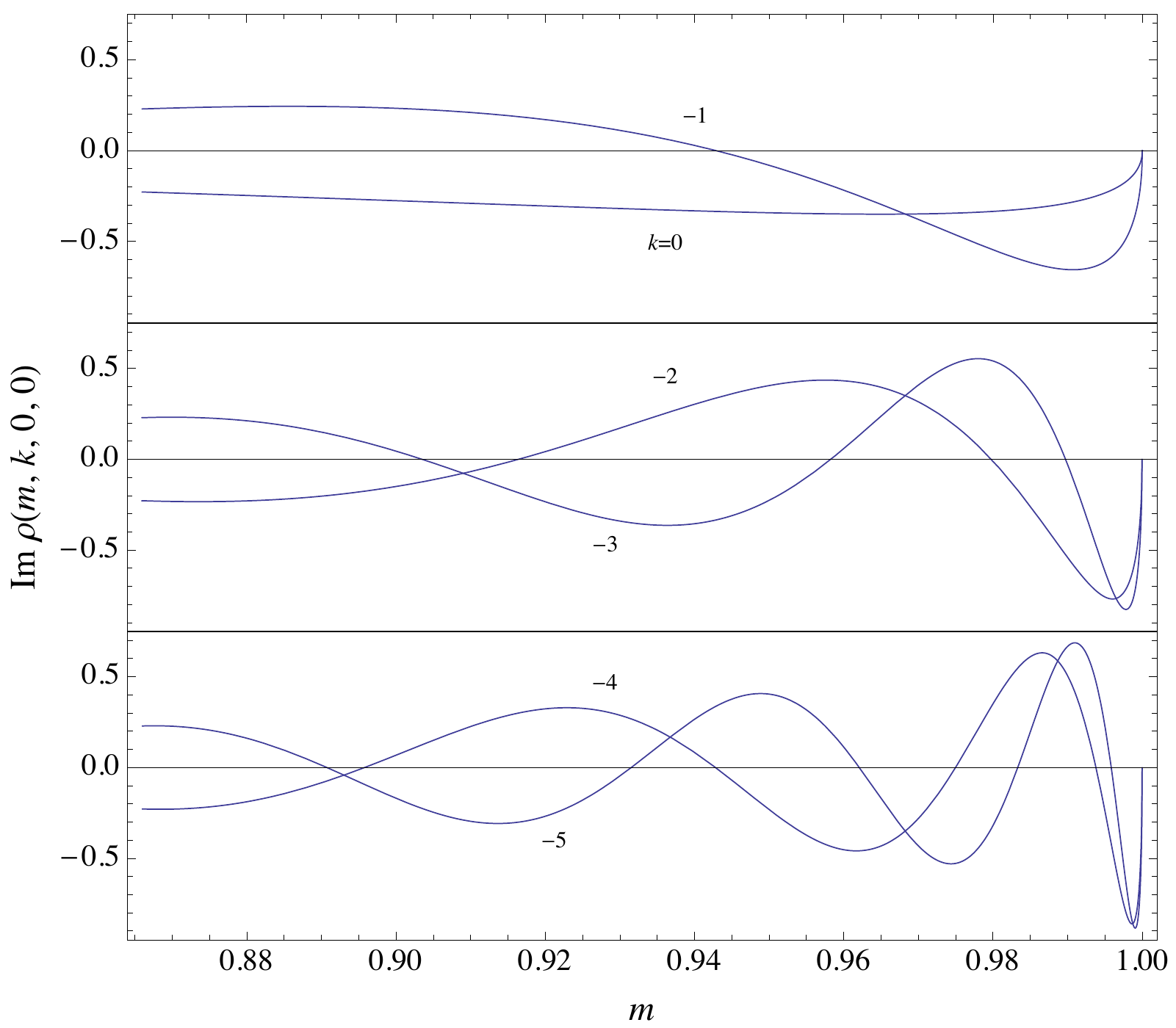}\hspace{14mm}\includegraphics[height=6.45cm]{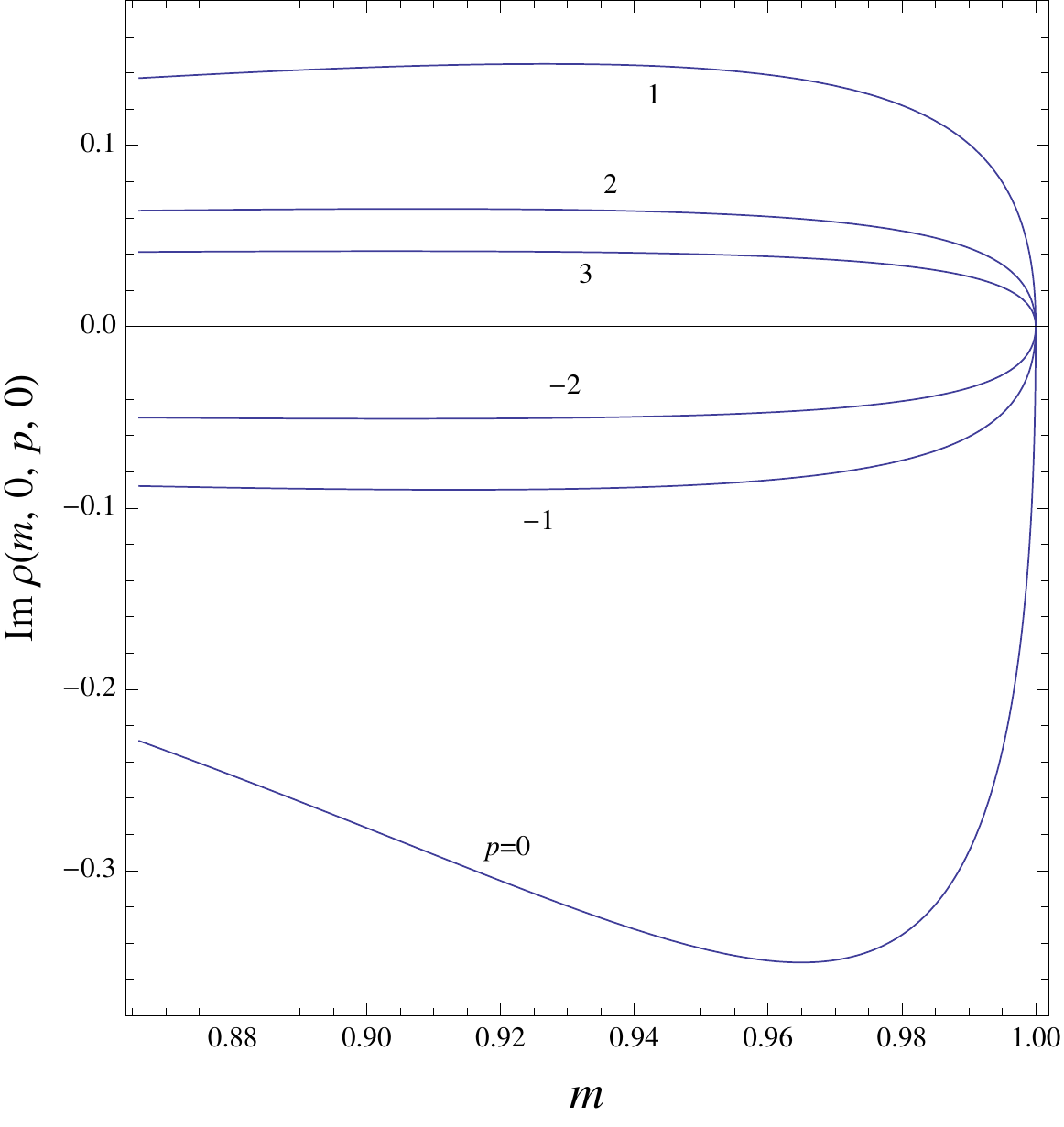}
\end{center}
\vspace*{-5mm}
 \caption{\footnotesize Plots of the coefficient $\im \rho$ of the quadratic term in the on-shell actions (${\cal I}\propto \tilde {\cal I} \propto \im\! \rho\,  \alpha^2$ for $\alpha\sim 0$) as shown in Eqs (\ref{ImcalS}) and (\ref{ImtcS}). 
 The expression of $\rho$ is given by the general formula Eq.(\ref{RHO122}). {\bf Left:} the stack of three plots shows $\im \rho$ for saddle points whose contours do not cross any branch cut, such as I ($k=0$) and II ($k=-1$) in Fig.\ref{AnalStructure}. In these cases $p=0$ and $n$ is irrelevant, so that $\rho$ is simply given by Eq.(\ref{rhomk}). For the fundamental saddle point $k=0$, $\im \rho<0$ for all mass values between ${\sqrt 3\over 2}$ and $1$ thus perurbatively the saddle's individual contribution to the Hartle-Hawking wave function is suppressed when $\tau_2\rightarrow \infty$. When $k=-1$, as shown in the upper frame, $\im \rho$ can become positive for certain ranges of mass; the same thing is true when $k$ becomes more negative. {\bf Right:} The value of $\im \rho$ for saddle points of contour containing minimum positive Euclidean history ($k=0$) and circling the brach point $\chi^{[0]}=0$ ($n=0$) from counterclockwisely twice ($p=-2$) to clockwisely three times ($p=3$). The contours have the look of the green contour in the left half of Fig.\ref{ctNPafPaf}, but the detail of the loop should be readapted according to the value of $p$.}\label{ImRFLp}
\end{figure}

Now we let the coordinate time approach $\chi^{[n]}$ until point $\chi_a$ and then make $p$ loops around it, stopping at point $\chi_b$ (c.f. the saddle III in Fig.\ref{AnalStructure}). For simplicity of reasoning, define $-\pi<\arg(\cosh^2\!\chi_a-1)\leq\pi$ and $-\pi<\arg(\cosh^2\! \chi_b-1)\leq\pi$. Then according to Eq.(\ref{eq73}) at $\chi_a$ the scalar field behaves as 
\begin{align}
	\Phi(\chi_a)=\Phi_o  \hat F(\chi_a )\sim - \, \Phi_o \times {\psi(u)+\psi(v)+2\gamma+i\pi +\ln (\cosh^2\! \chi_a-1)\over \Gamma(u) \Gamma(v)} . \label{eq74}
\end{align}
After completing the $p$ loops around $\chi^{[n]}$, the asymptotic expansion becomes
\begin{align}
	\Phi(\chi_b)\sim - \, \Phi_o \times {\psi(u)+\psi(v)+2\gamma+i\pi +\ln (\cosh^2\! \chi_b-1)\over \Gamma(u) \Gamma(v)}-\Phi_o \times{4p \, \pi i \over  \Gamma(u) \Gamma(v)} .
	\label{eq75}
\end{align}
That is, the expansion acquires a constant $-\Phi_o \times{4p \, \pi i \over  \Gamma(u) \Gamma(v)}$. It is $4\pi i$ in the numerator because $\cosh^2\! \chi-1\sim \left( \chi-\chi^{[n]}\right)^2$ when $\chi\sim \chi^{[n]}$, so that when the phase of $\chi-\chi^{[n]}$ increases by $2\pi$, that of $\cosh^2\! \chi-1$ increases by $4\pi$. Thus at this point the expression for $\Phi(\chi_b)$ is not $\Phi_o \hat F(\chi_b)$, but should be $\Phi_o \hat F(\chi_b)$ plus an extra term, and the leading term of this extra term when expanded, is $-\Phi_o \times{4p \, \pi i \over  \Gamma(u) \Gamma(v)}$. Since this extra term should be a solution of Eq.(\ref{ScDyn2}), the only choice is (cf  \cite{TbItgSrs} 9.153-7) 
\begin{align}-{4p \, \pi i \over  \Gamma(u) \Gamma(v)} \Phi_o \, \hat F\bigg(\chi_b -\chi^{[n]}+{i\pi \over 2}\bigg), \label{EXTTM} \end{align} 
and hence
\begin{align}
	\Phi(\chi_b)=\Phi_o \, \hat F(\chi_b)-{4p \, \pi i \over  \Gamma(u) \Gamma(v)} \Phi_o \, \hat F\bigg(\chi_b -\chi^{[n]}+{i\pi \over 2}\bigg). \label{eq76}
\end{align}
To summarize, we have the following expression for the scalar field 
\begin{align}
	\Phi(\chi)=\left\{ \begin{array}{ll} \displaystyle \Phi_o \hat F(\chi)\, ,& \text{$\chi$ earlier than $\chi_a$;}\\ \\ \displaystyle \Phi_o\hat F(\chi)-{4p \, \pi i \Phi_o \over  \Gamma(u) \Gamma(v)}  \hat F\bigg(\! \chi_b -\chi^{[n]}+{i\pi \over 2}\bigg)\, , & \text{$\chi$ later than $\chi_b$}. \end{array} \right. \label{phi+-}
\end{align}
The asymptotic behavior near $I^+$ is
\begin{align}
	\Phi(ik\pi +\Lambda)\sim\, & \hat \alpha\, e^{-2u\Lambda}+\hat \beta\, e^{-2v\Lambda}=\alpha \, {\cal A}^{-\Delta_-/2}+\beta \, {\cal A}^{-\Delta_+/2}, \ \ \ {\rm where}\\ 
	\alpha= \, &\Phi_o  \, \tau_2^{2u} e^{iu(1-2k)\pi } {\Gamma(w)\Gamma(v-u) \over \Gamma(v)\Gamma(w-u)}  \left[1-{4p\,\pi i \over \Gamma(u) \Gamma(v)}  e^{i(2n-1)u\,  \pi }\right], \label{AFA}
	 \\ \beta=\, &\Phi_o \, \tau_2^{2v} e^{iv(1-2k)\pi }{\Gamma(w)\Gamma(u-v) \over \Gamma(u)\Gamma(w-v)}  \left[1-{4p\,\pi i \over \Gamma(u) \Gamma(v)} e^{i (2n-1)v\, \pi }\right]. \label{BTA}
\end{align}
Thus if we specify $\alpha$ as the boundary condition, the south pole value of the scalar field $\Phi_o$ is determined by the assigned value of $\alpha$ through Eq.(\ref{AFA}):
\begin{align}
	\Phi_o=\alpha\, \tau_2^{-2u} e^{iu(2k-1)\pi } {\Gamma(v)\Gamma(w-u) \over \Gamma(w)\Gamma(v-u)}  \left[{\Gamma(u) \Gamma(v) \over \Gamma(u) \Gamma(v)-4p\,\pi i\,   e^{i(2n-1)u\,  \pi }}\right]. \label{Phioaf124}
\end{align}
The proportionality relation between the coefficients still formally reads $\beta=\alpha \,\tau_2^{\sqrt{1-m^2}} \rho$ but with
\begin{align}
	\rho=\rho(m,k,p,n)=e^{i(u-v)(2k-1)\pi}\, {\Gamma(v)\Gamma(u-v) \Gamma(w-u) \over \Gamma(u) \Gamma(v-u) \Gamma(w-v)}  {\Gamma(u) \Gamma(v) -4 p\,\pi i\, e^{(2n-1)v\, \pi i} \over \Gamma(u) \Gamma(v) -4 p\,\pi i\, e^{(2n-1)u\, \pi i} }\, . \label{RHO122}
\end{align}
The on-shell actions of these saddle points are still formally given by Eqs (\ref{IttBk1}) and (\ref{tItot90}), but with $\rho$ given by the above expression. The numerical calculation in the next section will cover the cases where the contour circles $\chi^{[0]}$ ($n=0$) from $p=-3$ to $p=2$ times. In the frame in the right part of Fig.\ref{ImRFLp} $\im \rho$ is plotted for these cases. When the $\chi$-contour circles $\chi^{[0]}=0$ clockwisely or counterclockwisely, $\im \rho$ is positive or negative for the whole range of scalar mass of ${\sqrt 3\over 2}<m<1$. Here the remarks concerning the sign of $\im \rho$ by the end of the previous subsection are still valid. 

\subsubsection*{Loops around several branch points}

When the $\chi$-contour circles around more than one branch point, the analysis proceeds in the same way as that is used to obtain Eq.(\ref{phi+-}), where in figuring out the extra term that $\Phi$ acquire we need to expand the whole expression of $\Phi(\chi)$ including the initial $\Phi_o\hat F(\chi)$ and the extra terms acquired after circling around the previous branch points. One important subtlety is that the new terms produced by circling around branch point can induce new branch points in the $\chi$-plane. Circling around these newly induced branch points will also bring $\Phi(\chi)$ to other Riemann surface layers. For example the new term Eq.(\ref{EXTTM}) induces logarithmic branch points at $\chi=i\left(k+ {1\over 2}\right)\!\pi$ ($k\in \mathbb Z$), which are not present on the Riemann surface layer in Fig.\ref{AnalStructure}. Stated more accurately, the reality is that the Riemann surface layer shown in Fig.\ref{AnalStructure}, is the one where the south pole lies. On this sheet, branch cuts are present at $\chi^{[k]}=ik\pi$ ($k\in \mathbb Z$), while on other layers which can be reached by passing through some branch cut shown in Fig.\ref{AnalStructure}, there can be branch points also at $\chi=i\left(k+ {1\over 2}\right)\!\pi$ ($k\in \mathbb Z$). Whatever the different ways the $\chi$-contour winds around the brach points, the bulk and boundary saddle-point actions always take the form of Eqs (\ref{IttBk1}) and  (\ref{tItot90}), but with $\rho$ to be derived case by case.

\subsection{Inhomogeneous scalar perturbation} \label{Sec55}

When we treat the scalar field perturbatively, solving the scalar equations against the spacetime background given by Eq.(\ref{3Met}), it is by all means feasible to consider inhomogeneous perturbations. To do this we need to solve the whole Klein-Gordan equation, and below is a sketch of the calculation, while in fact the calculation is already well known in the literature (cf for example \cite{Bousso:2001mw}). For the simplicity of solving the equation, let $r=\sinh \chi$ and $\tilde \zeta_1=\tau_2\, \zeta_1$, so that the metric Eq.(\ref{3Met}) takes the form
\begin{align}
	\ell^{-2}g_{\mu\nu}dx^{\mu}dx^{\nu}= - {dr^2 \over 1+r^2} + (1+r^2) d \zeta_2^2 +r^2 d\tilde \zeta_1^2.  \label{3mtcr}
\end{align}
The periodicities of the spatial coordinates are $\zeta_2+i\tilde\zeta_1\sim \zeta_2+i\tilde\zeta_1 +2\pi \sim \zeta_2+i\tilde\zeta_1+2\pi \tau$. The Klein-Gorden equation for our minimally coupled scalar field $\Phi$ is $(\square -m^2)\Phi=0$, whose explicit form against the metric (\ref{3mtcr}) is
\begin{align}
	\left[-{1\over r} \partial_r(1+r^2)r\partial_r+{\partial_2^2 \over 1+r^2}+{\tilde \partial_1^2\over r^2}- m^2\right]\Phi=0, \label{KGphi1}
\end{align}
where $\tilde \partial_1={\partial \over \partial \tilde \zeta_1}$ and $\partial_2={\partial \over \partial \zeta_2}$. Expanding the scalar field against Fourier modes on the 2-torus:
\begin{align}
	\Phi(r,\tilde \zeta_1,\zeta_2)=\sum_{p,q} \Phi_{pq}(r)\exp\! \left[ i\, q \,\zeta_2+i\,\left({p\over \tau_2}- {\tau_1 q\over \tau_2} \right) \tilde \zeta_1\right],
\end{align}
which we insert into the equation (\ref{KGphi1}) to obtain the equations for $\Phi_{pq}(r)$:
\begin{align}
	(1+r^2)\Phi_{pq}''+\left({1\over r} +3r \right) \Phi_{pq}'+ \left[{q^2 \over 1+r^2}+{(p-q\, \tau_1)^2 \over \tau_2^2 \,r^2}+m^2\right] \Phi_{pq}=0. \label{EQF}
\end{align}
Here the primes indicate the derivative with respect to $r$. The equation can be further converted into a more tangible form if we let \begin{align} \Phi_{pq}(r)=(1+r^2)^{q\over 2} r^{i\xi}y_{pq}(r), \ \ {\rm where} \ \ \xi={p-q\, \tau_1 \over \tau_2}\,, \end{align} and introduce the variable $z=r^2+1=\cosh^2\chi$ since we are interested in the solution regular at $r^2=-1$. The equation (\ref{EQF}) thus gives rise to a hypergeometric differential equation of $y_{pq}$:
\begin{align}
	z(1-z)(y_{pq})_{zz}+\left[1+q-(q+i\xi+2) z\right] (y_{pq})_{z}- {1\over 4}\left[m^2+(q+i\xi)(q+i\xi+2)\right] y_{pq}=0, \label{EQY}
\end{align}
of which the solution regular at the south pole $r^2=-1$ is 
\begin{align}
	&\Phi_{pq}(r)=(1+r^2)^{{|q|\over2}}\, r^{ i \xi}\, {}_2F_1 \left(u,v, w,1+r^2\right), \label{eq127}\\
	&\ \  u={1\over 2}\left(1+\sqrt{1-\ell^2m^2}+ |q|+ i\xi \right),\nonumber\\
	&\ \  v={1\over 2}\left(1-\sqrt{1-\ell^2m^2}+|q|+ i\xi \right),\nonumber\\
	&\ \  w=1+ |q| \nonumber,
\end{align}
where the zero mode $\Phi_{00}$ is just the homogeneous result in Sec.\ref{HPES}, Eq.(\ref{phiCF}). The whole solution for the scalar field, in terms of the variable $\chi$, is thus
\begin{align}
	\Phi\! \left(\chi,\tilde \zeta_1,\zeta_2 \right)= \sum_{p,q}A_{pq}\cosh^{|q|}\! \chi \ \sinh^{ i \xi}\! \chi\ {}_2F_1\! \big(u,v, w,\cosh^2 \!\chi\big) \exp\!\left[ i\, q \,\zeta_2+i\! \left({p\over \tau_2}- {\tau_1q\over \tau_2} \right)\! \tilde \zeta_1\right],
\end{align}
where the coefficients $A_{pq}$ are to be determined by the boundary condition for $\Phi$ on $\Sigma_*$.

Without going through every detail I would like to state qualitatively the main properties and results which are in line with the purpose of the work of the paper:
\begin{itemize}
	\item All modes inhomogeneous in the $\zeta_2$ direction vanish at the south pole: $\Phi_{pq}(\chi_o)=0$ for $q \neq 0$. Therefore the south pole value of the scalar field is $\Phi(\chi_o,\tilde \zeta_1,\zeta_2)=\sum_p A_{p0}\exp\!\left({{ip\over \tau_2} \tilde \zeta_1}\right)=\sum_p A_{p0}\exp({ip\, \zeta_1})$. This is perfectly consistent with the setup that $\zeta_2$-circle should cap smoothly at the south pole, implying that $\zeta_2$ dimension is absent there.
	\item Branch points are present on the Riemann surface of each mode $\Phi_{pq}$ at $\chi^{[k]}=ik\pi$, which can be studied by expanding Eq.(\ref{eq127}) around them. 
	\item The perturbative computation of saddle-point actions in Sec.\ref{OSactPert}, as well as the holographic renormalization for homogeneous scalar perturbation, can be easily extended to the inhomogeneous case (holographic renormalization is perturbative). Now instead of having only one pair of $(\alpha,\beta)$, we will have an infinite number of them, each from the asymptotic expansion of $\Phi_{pq}$ given by Eq.(\ref{eq127}). Thus they can be denoted by $(\alpha_{pq},\beta_{pq})$, where $(\alpha_{00},\beta_{00})$ is just $(\alpha,\beta)$ introduced in Eq.(\ref{Phi*40}). The imaginary part of the saddle-point actions are ${\cal I}=\sum_{(p,q)}B_{pq}\alpha_{pq}^2$, and $\tilde {\cal I}=\sum_{(p,q)}\tilde B_{pq}\alpha_{pq}^2$ where $B_{pq}=\sqrt{1-m^2}\,\tilde B_{pq}$. Further, for $q\neq0$, $B_{pq}$ and $\tilde B_{pq}$ depend explicitly on $\tau_1$.
	\item With the presence of explicit $\tau_1$-dependence just mentioned, we can show, taking into  account the coordinate change and boundary condition change, that the resulting on-shell action is invariant under the modular transform with $c=d=0$ and $e=1$ in Eq.(\ref{TDUAL}), so that when summing up all the inequivalent configurations on the $SL(2,\mathbb Z)$ orbit, we only sum over different $(c,d)$, same as for homogeneous scalar perturbation.
\end{itemize}

\begin{figure}
\begin{center}
	\includegraphics[width=0.65\textwidth]{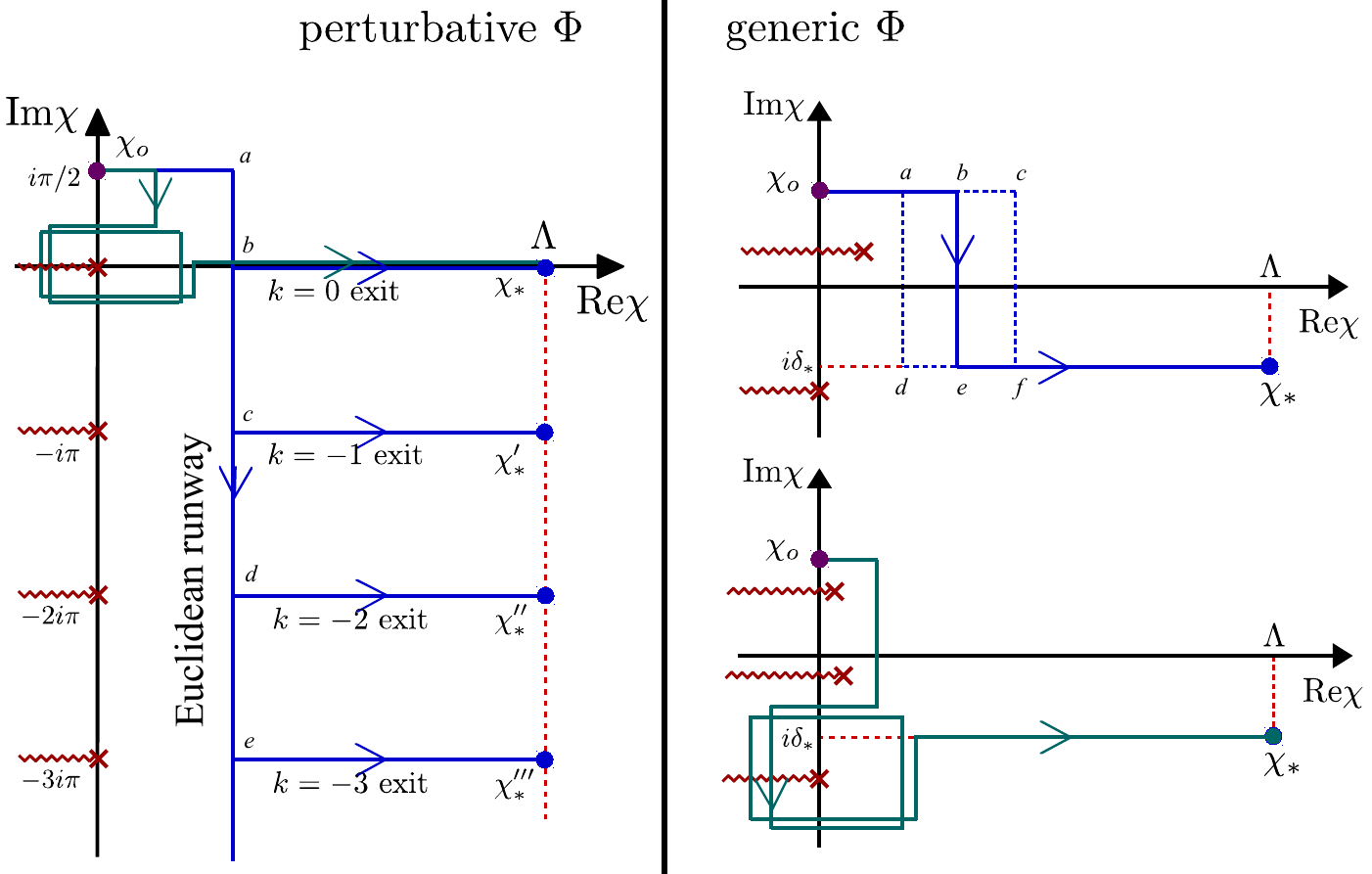}
\end{center}
	\caption{\footnotesize Complex time contours used for the saddle points studied in Sec.\ref{NUMT2}, along which Eqs (\ref{Eaph})--(\ref{Ephi}) are numerically integrated. 
{\bf Left:}  the form of the contours when $\Phi$ is perturbatively small. The $\chi$-plane is just the Riemann surface in Fig.\ref{RiemannSurfaces}. Two types of contours will be considered colored in blue and green. The blue contours stay always in the same layer of the Riemann surface. They start from $\chi_o$ along the Lorentzian direction $\chi_o\rightarrow a$, embark on a Euclidean evolution along $a\rightarrow e$ and can take any Lorentzian exit at $\im \chi=k\pi$ ($k\in\mathbb Z$). The green contour circles around the branch point $\chi=0$ and finishes up taking the $k=0$ exit on another Riemann sheet. The example shown here circles clockwisely twice the branch point, that is $p=2$ in the language of Sec.\ref{MVT2}.
{\bf Right:} the generic pattern of deformation of the Riemann surfaces and the contours when $\Phi$ is continuously tuned up. The Riemann surfaces have their singular points moving away from $ik\pi$ ($k\in\mathbb Z$); the contours need to be deformed accordingly. The blue contour may have its Euclidean part adjusted to the left or right to avoid singularities, for example $a\rightarrow d$ or $c\rightarrow f$. 
The green contour should have its loops continuously re-adapted in order not to let the branch point inside escape out nor outside in. For both contours the Euclidean shift $\im\chi_*=\delta_*$ will divert from $k\pi$ and should be determined by boundary conditions at $\Sigma_*$. 
} \label{ctNPafPaf} 
\end{figure}

\section{Beyond scalar  perturbation by numerics} \label{NUMT2}





Based on the perturbative result obtained previously, this section will set out into the realm of finite scalar deformation. Concretely, 
we will start from a certain saddle point found in Sec.\ref{PERTT2}, tuning up the scalar field, measured by $\Phi_o$ and $\alpha$, and meanwhile tracing its resulting deformation continuously by numerical means. The focus of this study are

\begin{itemize}

\item First the quantities worked out perturbatively in Sec.\ref{PERTT2} ($\beta$, $\cal S$, $\tilde {\cal S}$, $b_o$, $\Phi_o$, $\delta_*$), are now traced deep into the non-perturbative realm of scalar deformation, and their diversions from the perturbative behaviors are to be displayed.

\item Second the various saddles will provide a vast test field for verifying the relation Eq.(\ref{OPFunc}). This is done to all the saddle points covered in this section. This confirms that Eq.(\ref{HolRen2}) is the right result of holographic renormalization on the non-perturbative level. 


\item Third, the saddle-point actions before ($\cal S$) and after ($\tilde {\cal S}$) holographic renormalization will be compared. Especially according Eq.(\ref{II54}), holographic renormalization introduces a non-trivial imaginary part to the saddle-point action. And it will turn out that $\tilde {\cal I}$ can be drastically different from $\cal I$ for some saddle points. 

\end{itemize}

The study will be carried out in the following steps:

Sec.\ref{PRLIM4} will explain the key elements of the numerical scheme employed to trace the saddle points in the non-perturbative regime of scalar deformation. Also some useful properties of the physical quantities will be given, which decide the way that they are presented in the later subsections. The most important aspect is that we only need to trace various quantities as a function of $\alpha$ with $\tau_2$ fixed (the shape of $T^2$ fixed), because the results as such already contain all the information of the $\tau_2$-dependence. 

Sec.\ref{T2NCS} numerically studies the saddle points at finite scalar deformation in the order in which the perturbative cases were investigated in Sec.\ref{PERTT2}, where the scalar potential are taken to be quadratic potentials. First presented are the results of the saddle points whose time contours do not circle around any branch points but contain different amount of Euclidean history. Next presented are the results for those whose time contours contain the minimum amount of positive Euclidean history but circle around one branch point several times. It will especially be shown that the renormalized imaginary part of saddle-point action $\tilde {\cal I}$ can have severely different behavior and diverge to $\pm \infty$ with $\alpha$ for non-fundamental saddle points.

Whereas Sec.\ref{T2NCS} is concentrated on the physical outcomes of the different types of saddle points, Sec.\ref{INTC} studies the intrinsic characteristics of them, completing the full account of the properties of saddle points. These include the south pole values $\Phi_o$ and $b_o$, the Euclidean shift $\delta_*$ of the boundary $\Sigma_*$ (see Eq.(\ref{SigmatoI})), and the structure of the Riemann surfaces. Some characteristic pattern of Riemann surface deformation are empirically observed in case $\tilde {\cal I}$ presents scalar divergence.


\subsection{Numerical scheme in a nutshell} \label{PRLIM4}

In the numerical study of the saddle points, the general logic to follow stays the same: find out the saddle point that meets with the boundary data $(\tau, \alpha)$ assigned on $\Sigma_*$, and then compute its various physical quantities. This time all steps are to be done numerically, that is, the equations (\ref{Eaph})--(\ref{Ephi}) will be integrated numerically to obtain $a(\chi)$, $b(\chi)$ and $\Phi(\chi)$. The integration starts from the south pole\footnote{
Note that it is not possible when concretely doing numerics, to start integrating the equations (\ref{Eabp})---(\ref{Ephi}) exactly from the south pole $\chi_o$, since $a(\chi_o)=0$, which produces $0$ in the denominator. Instead, the initial conditions will in practice be assigned infinitesimally close to the south pole $\chi_o+\epsilon$ with $\epsilon\sim 0$. Referring to appendix \ref{ACEQGEN}, the initial conditions to be actually used are
\begin{align}
	& a(\chi_o+\epsilon)=i\, \epsilon,\ \ b(\chi_o+\epsilon)=b_o=|b_o|\, e^{i\gamma},\ \ \Phi(\chi_o+\epsilon)=\Phi_o=|\Phi_o| e^{i\theta} ; \nonumber
	\\
	&  \dot a(\chi_o+\epsilon)=i,\ \ \dot b(\chi_o+\epsilon) = \Big[ 1+V\! \left(\Phi_o \right) \Big] b_o \, \epsilon\,, \ \ \dot \Phi(\chi_o+\epsilon)=-{V'\! \left( \Phi_o \right)\over 2}\, \epsilon\, , \nonumber
	\\
	& \ \ {\rm where} \ \ \epsilon\in \mathbb C,\ |\epsilon|\ll1; \ \ \ b_o,\Phi_o \in \mathbb C,\ \ {\rm and}\ \  \gamma, \theta \in\mathbb R, \nonumber
\end{align}
and since Eq.(\ref{Eabp}) is used in deriving these conditions, when integrating the equations we only need to integrate the second order ones (\ref{Eaph})---(\ref{Ephi}) and the solutions will automatically satisfy the constraint Eq.(\ref{Eabp}).
}
$\chi_o$ to the final moment $\chi_*=i\delta_*+\Lambda$ (see Eq.(\ref{SigmatoI})), with $\Lambda$ fixed to a large number to approximate $I^+$ and $\delta_*$ to be determined. By adjusting the south pole data $(b_o,\Phi_o)$ as well as $\delta_*$, we can fit the numerical solutions to the boundary data $(\tau,\alpha)$ assigned on $\Sigma_*$. 
In this procedure there are five real conditions $\im (a_*b_*)=0$, ${b_*\over a_*}=\tau_2$ and $(a_*b_*)^{\Delta_-/ 2}\Phi_*=\alpha$, exactly what is needed to determine the five real defining parameters of the saddle points: $(b_o,\Phi_o)$ and $\im \chi_*=\delta_*$.\footnote{Here the story of saddle point searching is told with the narrative ``given $(\alpha,\tau)$, we determine $(b_o,\Phi_o,\delta_*)$ to find out the saddle point'', rather for pedagogical reason, while the actual steps are more subtile with the numerical code used here.
Instead of leaving $(b_o,\Phi_o,\delta_*)$ totally free and have them determined by the boundary conditions, we need to pre-determine values for $|\Phi_o|$ and $|b_o|$, only leaving the phases $\theta=\arg \Phi_o$ and $\gamma=\arg b_o$, and also the Euclidean shift $\delta_*$ free to vary. The field values at $\chi_*$: ($a_*, b_*,\Phi_*$), obtained from integrating the equations of motion, are thus functions of $\delta_*$, $\gamma$ and $\theta$, recalling that $\re \chi_*=\Lambda$ is fixed once for all. The next step is to numerically solve the equations ${\im a_* \over \re a_*}={\im b_* \over \re b_*}={\im \Phi_* \over \re \Phi_*}=0$ which are due to the requirement that all quantities should be real on $\Sigma_*$, and this determines the three real parameters $\delta_*,\theta,\gamma$ that have been left free in the beginning. When this is done, with the pre-determined $|\Phi_o|$ and $|b_o|$, all the characterizing parameters of the saddle point are obtained.  Then from the numerical functions of $\{a(\chi),b(\chi),\Phi(\chi)\}$ describing the saddle point, we read off the boundary data $(\tau,\alpha)$ that we should have imposed to obtain this very saddle point.  All these steps are realized with Mathematica for the work in this paper. \label{FN2}} 
We can do this continuously over a domain in the $(\tau,\alpha)$-space and thus 
establish the continuous mappings $b_o=b_o(\tau,\alpha)$, $\Phi_o=\Phi_o(\tau,\alpha)$ and $\delta_*=\delta_*(\tau,\alpha)$, which represent a continuous family of saddle points. 

However for a fixed set of boundary data $(\tau,\alpha)$, there can be infinitely many results of $(b_o,\Phi_o,\delta_*)$, which we have already seen in the previous section where the scalar field is perturbative ($\Phi_o\propto \alpha\sim 0$). There for the same boundary condition $(\tau, \alpha)$, we have infinitely many Euclidean shifts $\delta_*=k\pi$ ($k\in \mathbb Z$), further accompanied by infinitely many ways for the complex time contour to circle around the singularities of the scalar field, leading to infinitely many values of $\Phi_o$. When the boundary data sweeps through a region in the $(\tau,\alpha)$-space starting out from each distinct result of $(b_o,\Phi_o,\delta_*)$, we obtain infinitely many different families of saddle points, each gives rise to distinct saddle-point action as function of $(\tau,\alpha)$. For instance in the case of scalar perturbation, we have  Eqs (\ref{TreeBLKPtb}) and (\ref{TreeBDRPtb}) with different family of saddle points resulting in different $\rho$ given in Eq.(\ref{RHO122}). 

The strategy of this section is to obtain the different families of non-perturbatively deformed saddle points based on the perturbatively deformed ones obtained in the previous section. 
Concretely, in the perturbative domain where $\alpha \sim 0$, the approximative south pole data to be suggested to the numerical code are $\Phi_o$ according to Eq.(\ref{afdetPhi0}) and $b_o\approx i$ according to Eq.(\ref{AtauMatch}). Once the exact numerical result for small $\alpha$ is established, we can tune up $\alpha$ gradually with fixed $\tau_2$,\footnote{As explained in footnote \ref{FN2}, in the numerical code that is actually used here, we have no direct control on $\alpha$ and $\tau_2$ but on $|\Phi_o|$ and $|b_o|$. Tuning up $\alpha$ is realized by increasing $|\Phi_o|$. For each $|\Phi_o|$, we fix $|b_o|$ to $1$, and we search for the saddle point. Once the saddle point is found for the $|\Phi_o|$ and $|b_o|$ we have chosen, we read off $\tau_2$, $\alpha$ and $\beta$ and we compute the on-shell action. After having covered a certain range of $|\Phi_o|$, we would obtain various quantities as functions of $|\Phi_o|$, especially $\alpha$, $\beta$, $\tau_2$, $\cal I$ and $\tilde {\cal I}$, which we then convert into functions of $\alpha$ using the $\alpha$-$|\Phi_o|$ relation. However for each value of $|\Phi_o|$, $\tau_2$ takes the value of what we read off, and is generically not a constant. Then we can simply use the relations to be introduced shortly after, Eqs (\ref{BTAgentau2})  (\ref{Igentau2}) and (\ref{tSgentau2}), to obtain the value of various quantities as functions of $\alpha$ for a fixed value of $\tau_2$, which is set to $1$ in this paper.} and meanwhile, the continuous deformation of the time contour, as well as the change in south pole data $(b_o,\Phi_o)$, are carefully traced. However we should be aware that it is possible that we miss saddle points in this way. Actually it can happen that there are families of saddle points that do not contain a perturbative regime, that is, when $\alpha \sim 0$, $\beta$ and $\Phi_o$ stay finite. An example of this will be given in Sec.\ref{S3BrchAB} for 5d models, while I have not yet found such family of saddle points in 3d models.

Fig.\ref{ctNPafPaf} shows schematically the generic pattern of Riemann surface deformation and complex time contour deformation with the increase of scalar field backreaction. The search for saddle points starts with some perturbative saddle point whose contour is as shown in the left half of the figure, blue or green, and then when the scalar field is tuned up, the Riemann surfaces and the contours are deformed to something else, generically like what is shown in the right half of the figure. Especially $\delta_*$ is no longer $k\pi$ as in the perturbative case, and, since the singularities move with the increase of scalar deformation, the contours need to be carefully re-configured in order to avoid running over them.

\vspace*{4mm}

Before setting out to present the numerical results, there are some properties of the model to mention, which will more or less directly determine the way that the numerical results are plotted.

\vspace*{3mm}

\noindent {\bf {\it i})} {\bf The infrared divergences in the bulk saddle-point action ${\cal S}$ are real}

\vspace{3mm}

\noindent This discussion is already carried out in Sec.\ref{T2INFD}, leading to Eqs (\ref{ImOSFin}) and (\ref{splitIm}), due to which in the following subsection $\cal I$ will be plotted against $\alpha$. We should keep in mind that this does not mean that $\cal I$ is in the $\alpha$-representation but rather this is a reparameterization of boundary condition. As explained in the beginning of Sec.\ref{HPES}, when $\cal A$ is very large and fixed, $\varphi\approx \alpha\, {\cal A}^{- \Delta_-/2}$.

An additional remark is that using the integral expression for the saddle-point action Eq.(\ref{IabpOnSh}) can also lead to the same conclusion. If we use the asymptotic expansion Eqs (\ref{Phi152})--(\ref{b154}) we find that the integrand evaluated at $\chi_*=i\delta_*+\Lambda$ ($\Lambda\rightarrow \infty$) contains only exponentially suppressed imaginary terms. Therefore increasing $\re \chi_*=\Lambda$ in Eq.(\ref{IabpOnSh}) does not introduce any contribution to the imaginary part of the result. The same logic also applies to the models with boundary topology $S^d$ ($d=2,3,4\dots$)

\vspace{3mm}

\noindent {\bf {\it ii})} {\bf Due to the rescaling invariance of the equations of motion, the two variable problem of $\tau$ and $\alpha$ can be reduced to a one variable problem of only $\alpha$.}

\vspace{3mm}

\noindent The set of equations (\ref{Eabp})--(\ref{Ephi}) are invariant under the rescaling of the scale factors $a(\chi)$ or $b(\chi)$.  It turns out that this property can suppress the variable $\tau_2$ of our problem. For the quantities of our interest: $b_o$, $\Phi_o$, $\beta$, ${\cal I}$ and $\tilde{\cal S}$ which are initially functions of $\tau$ and $\alpha$, their $\tau_2$-dependences can be derived from their $\alpha$-dependence at a fixed $\tau_2$ without solving the equations of motion. Here below are some more details.


Suppose we have integrated the equations from the south pole $\chi_o$ with the south pole values $(b_o,\Phi_o)$, to the final moment $\chi_*=i\delta_*+\Lambda$ with very large $\Lambda$, and that we have found the solutions $\{a(\chi),b(\chi),\Phi(\chi)\}$ of boundary values $({\cal A},\tau_1,\tau_2,\varphi)$. Then we do the same procedure again only with $b_o$ rescaled by $c$, so that the south pole values are $(c\,b_o,\Phi_o)$. By the rescaling-invariance of the equations, the solutions should be $\{a(\chi),c\, b(\chi),\Phi(\chi)\}$, whose boundary values become $\left({\cal A}', \tau_1',\tau_2', \varphi' \right)=\left(c\, {\cal A}, \tau_1,c\, \tau_2, \varphi \right)$. Then we can write down the boundary data of the scalar field before and after rescaling as:
\begin{align}
	&\varphi=\alpha {\cal A}^{-{\Delta_-\over 2}}+\beta(\tau_1,\tau_2,\alpha) {\cal A}^{-{\Delta_+\over 2}};\\
	&\varphi'=\alpha' {\cal A}'{}^{-{\Delta_-\over 2}}+\beta(\tau_1',\tau_2',\alpha') {\cal A}'^{-{\Delta_+\over 2}}=\alpha' (c{\cal A})^{-{\Delta_-\over 2}}+\beta(\tau_1,c\tau_2,\alpha') (c{\cal A})^{-{\Delta_+\over 2}}.
\end{align}
Equating the above two since the rescaling has no effect on $\Phi$, we obtain
\begin{align}
	\beta\big(\tau_1,\tau_2,\alpha \big)=c^{-{\Delta_+\over 2}}\beta\big(\tau_1,c\tau_2,\alpha \, c^{{\Delta_-\over 2}}\big),
\end{align}
from which we have the following relations which suppress the $\tau_2$-dependence of $\beta$:
\begin{align}
	\beta \big(\tau_1,\tau_2,\alpha \big)=\tau_2^{\Delta_+ \over 2}\beta\big( \tau_1,1,\alpha \, \tau_2^{-{\Delta_- \over 2}} \big); \ \  \beta \big(\tau_1,1,\alpha \big)=\tau_2^{-{\Delta_+ \over 2}}\beta\big( \tau_1,\tau_2,\alpha \, \tau_2^{\Delta_- \over 2} \big). \label{BTAgentau2}
\end{align}
Then let us examine the property of the saddle-point actions inferred from the rescaling invariance. Suppose before the rescaling, the solutions $\{a(\chi),b(\chi),\Phi(\chi)\}$ lead to the bulk saddle-point action ${\cal S}\!\left({\cal A},\tau_1,\tau_2,\varphi \right)$. Then it follows from Eq.(\ref{Itotabp}) that after the rescaling $b_o\rightarrow c\, b_o$, $\cal S$ is rescaled by the same amount, and becomes $c{\cal S}\!\left({\cal A},\tau_1,\tau_2,\varphi \right)$. On the other hand, from the point of view of the boundary data on $\Sigma_*$, which now becomes $\left(c\, {\cal A}, \tau_1,c\, \tau_2, \varphi \right)$, the action after rescaling is also ${\cal S}\! \left(c\, {\cal A},\tau_1,c\,\tau_2,\varphi \right)$. Thus we establish the equality
\begin{align}
	{\cal S}\!\left(c\, {\cal A},\tau_1,c\,\tau_2,\varphi \right)=c\, {\cal S}\! \left({\cal A},\tau_1,\tau_2,\varphi \right). \label{Sresc}
\end{align}
For the imaginary part ${\cal I}=\im{\cal S}$, since it does not have $\cal A$ dependence when ${\cal A}\rightarrow \infty$, Eq.(\ref{Sresc}) implies
\begin{align}
	{\cal I}\!\left(\tau_1,c\,\tau_2,\alpha \, c^{\Delta_-\over 2} \right)=c\, {\cal I}\!\left(\tau_1,\tau_2,\alpha  \right)
\end{align}
from which by properly choosing $c$ and rescaling $\alpha$, we obtain the relations
\begin{align}
	 {\cal I}\! \left(\tau_1,\tau_2,\alpha \right)=\tau_2\,{\cal I}\big(\tau_1,1,\alpha \, \tau_2^{-{\Delta_- \over2}} \big),\ \ \  {\cal I}\! \left(\tau_1,1,\alpha \right)=\tau_2^{-1}\,{\cal I}\big(\tau_1,\tau_2,\alpha\, \tau_2^{\Delta_-\over 2} \big). \label{Igentau2}
\end{align}
We can further on do the same discussion on Eq.(\ref{HolRen2}), where we take into account the rescaling of $b_o$ and $\beta$ which have already been explained above, we will find the same relation for $\tilde {\cal S}$:
\begin{align}
	 \tilde {\cal S}\! \left(\tau_1,\tau_2,\alpha \right)=\tau_2\,\tilde {\cal S}\big(\tau_1,1,\alpha \, \tau_2^{-{\Delta_- \over2}} \big),\ \ \  \tilde {\cal S}\! \left(\tau_1,1,\alpha \right)=\tau_2^{-1}\, \tilde {\cal S}\big(\tau_1,\tau_2,\alpha\, \tau_2^{\Delta_-\over 2} \big), \label{tSgentau2}
\end{align}
and automatically the same relation holds for $\tilde{\cal S}_{\rm R}$ and $\tilde {\cal I}$. 

The relations Eqs (\ref{BTAgentau2}) (\ref{Igentau2}) and (\ref{tSgentau2}) allow us to obtain the value of the various physical quantities of interest for arbitrary $\tau_2$ from its value at $\tau_2=1$ (at some other $\alpha$) and vice versa. Therefore in presenting the numerical result of those quantities, only the curves against $\alpha$ with $\tau_2$ fixed to $1$ will be plotted.

\vspace{4mm}

\noindent {\bf {\it iii}) Large $\tau_2$ (``high temperature'') behavior of the on-shell actions}

\vspace{3mm}

\noindent The large $\tau_2$ behaviors of the saddle-point actions are determined by their perturbative forms. This is immediately seen when we take the $\tau_2\rightarrow \infty$ limit in the first equalities in Eqs (\ref{Igentau2}) and (\ref{tSgentau2}) where the third argument $\alpha\, \tau_2^{-{\Delta_- \over2}}$ on the righthand side becomes small with the limit, so that their perturbative results apply and we obtain Eqs (\ref{ImcalS}) and (\ref{tItot90}). With this observation, now we can finish up the suspended discussion at the very end of Sec.\ref{OSactPert}: at large $\tau_2$, corresponding to the very stretched $T^2$ or very ``high temperature'', the saddle point contributions to the Hartle-Hawking wave function takes the perturbative form whatever the value of $\alpha$. Therefore whether a certain saddle point leads to temperature divergence depends on the sign of $\im \rho$, regardless of the extent of scalar deformation.

\vspace{2mm}

\subsection{On the facade: numerical results of physical quantities} \label{T2NCS}

In this subsection, the saddle points are studied in the order that perturbative saddle points are studied in Sec.\ref{PERTT2}. Only the quadratic potentials are used, while non-quadratic potentials will be briefly studied in Sec.\ref{NoQuPsec}. The saddle points involved are now non-perturbative in scalar deformation; however since they are always obtained by giving finite scalar deformation to the perturbative saddle points, it makes sense to have them classified according to their characteristics when they are still perturbative. The results to be presented in the following are the physical quantity outputs of the saddle points: $\beta$, $\cal S$ and $\tilde {\cal S}$. The focus points are as announced in the introduction of this section, briefly: comparing the perturbative/non-perturbative results; testing Eq.(\ref{OPFunc}); comparing $\cal I$ and $\tilde {\cal I}$ mainly to check scalar divergence.

\subsubsection*{Fundamental saddle points; results in Fig.\ref{k0grid}; contour in Fig.\ref{ctNPafPaf} $\chi_o \rightarrow a\rightarrow b\rightarrow \chi_*$}

The first case computed are fundamental saddle points of several different masses. The notion of ``fundamental saddle point'' was introduced in the perturbative regime by the end of Sec.\ref{HPES}, referring to the saddle points with time contours containing a minimum amount of positive Euclidean history and having no loop around branch points. Here in the non-perturbative context, it refers to those continuously deformed out of the perturbative fundamental saddle points. Therefore their contours are $\chi_o \rightarrow a\rightarrow b\rightarrow \chi_*$ in Fig.\ref{ctNPafPaf} when $\Phi$ is perturbative. The masses investigated are $m=0.93,\, 0.96,\, 0.98$ and $0.99$. The features worth attention (see immediately relevant details in the figure and in the caption) are, regarding Fig.\ref{k0grid}:

\begin{itemize}

\item With the increase of $\alpha$, the imaginary parts of $\beta$   (first lower frame) increase from $0$, reach a peak, and then fall off and asymptote to 0, in contrast to the families of non-fundamental saddle points to be shown in the next figures, where $\im \beta$ generally diverges.

\item The imaginary part of saddle-point actions in the bulk field representation (second column upper frame), $\cal I$, are bounded as function of scalar deformation ($\alpha$). They differ slightly from $\tilde {\cal I}$, their counterpart in the boundary data representation, which is the consequence that $\im \beta$ asymptotes to $0$ with large $\alpha$.

\item The ``one-point function'' relation Eq.(\ref{OPFunc}) holds as shown in the right two frames in the lower row. More detail is shown in Fig.\ref{SSMatchDetail} for the saddle points of scalar mass $0.96$.

\item Perturbative results are shown with dotted red lines when they apply. Here they are shown in the inset frames for the sake of visibility. We see that they are in excellent match with the numerical results when $\alpha\sim 0$.

\end{itemize}

\subsubsection*{Longer Euclidean history for several masses; results in Fig.\ref{k1grid}; contour in Fig.\ref{ctNPafPaf} $\chi_o \rightarrow a\rightarrow c\rightarrow \chi_*'$}

The second case studied is when the saddle points have just next-to-minimum amount of Euclidean history, corresponding to the blue contour in Fig.\ref{ctNPafPaf} which ends up at the $k=-1$ Lorentzian exit. Saddle points of several masses are studied, which are $m=0.93,\, 0.94,\, \sqrt{8/9},\,0.95,\, 0.96$. Below are the noticeable features of these results, in contrast to those of the fundamental saddle points:

\begin{itemize}
\item For $\alpha\sim 0$, $\im \beta$ starts off increasing or decreasing from $0$, following $\im \rho$ being positive or negative, with $\rho=\rho(m,k)$ given in Eq.(\ref{rhomk}) where we put $k=-1$. For $m=0.93,\, 0.94$, $\im \rho>0$, and $\im \beta$ grows, while for $m=0.95,\, 0.96$, $\im \rho<0$, so that $\im \beta$ decreases. When $m=\sqrt{8/9}$, $\im \rho=0$, $\im \beta=0$ for all $\alpha$ covered by numerics. 

\item As $\alpha$ grows larger,  $\im \beta$ keeps increasing or decreasing in the same way as when $\alpha$ is small. The monotony is not always the case as we will see in the next category of saddle points studied. However it is general for saddle point containing non-minimum amount of Euclidean history that $\im \beta$ diverge to $\pm \infty$ when $\alpha$ increases. This is contrary to the previous case with fundamental saddle points, where $\im \beta$ asymptotes to $0$ when $\alpha$ increases.

\item The imaginary parts of the saddle point-actions in the bulk field representation $\cal I$ are bounded (the second upper frame) but their boundary counterparts $\tilde {\cal I}$ tend to diverge to $\pm \infty$ with $\alpha$, in the opposite directions to the divergences of $\im \beta$. The special case is $m=\sqrt{8/9}$, where $\im \!\rho=0$ and we obtain ${\cal I}=\tilde{\cal I}=0$ over the range of $\alpha$ covered by numerical computation. It is interesting to see whether this property holds for other forms of potential.

\end{itemize}


\subsubsection*{Several lengths of Euclidean history; results in Fig.\ref{M0p95Floors}; contour in Fig.\ref{ctNPafPaf} the blue line taking the $k=0$ till the $k=-5$ Lorentzian exit}

\begin{figure}
\begin{center}
	\includegraphics[width=\textwidth]{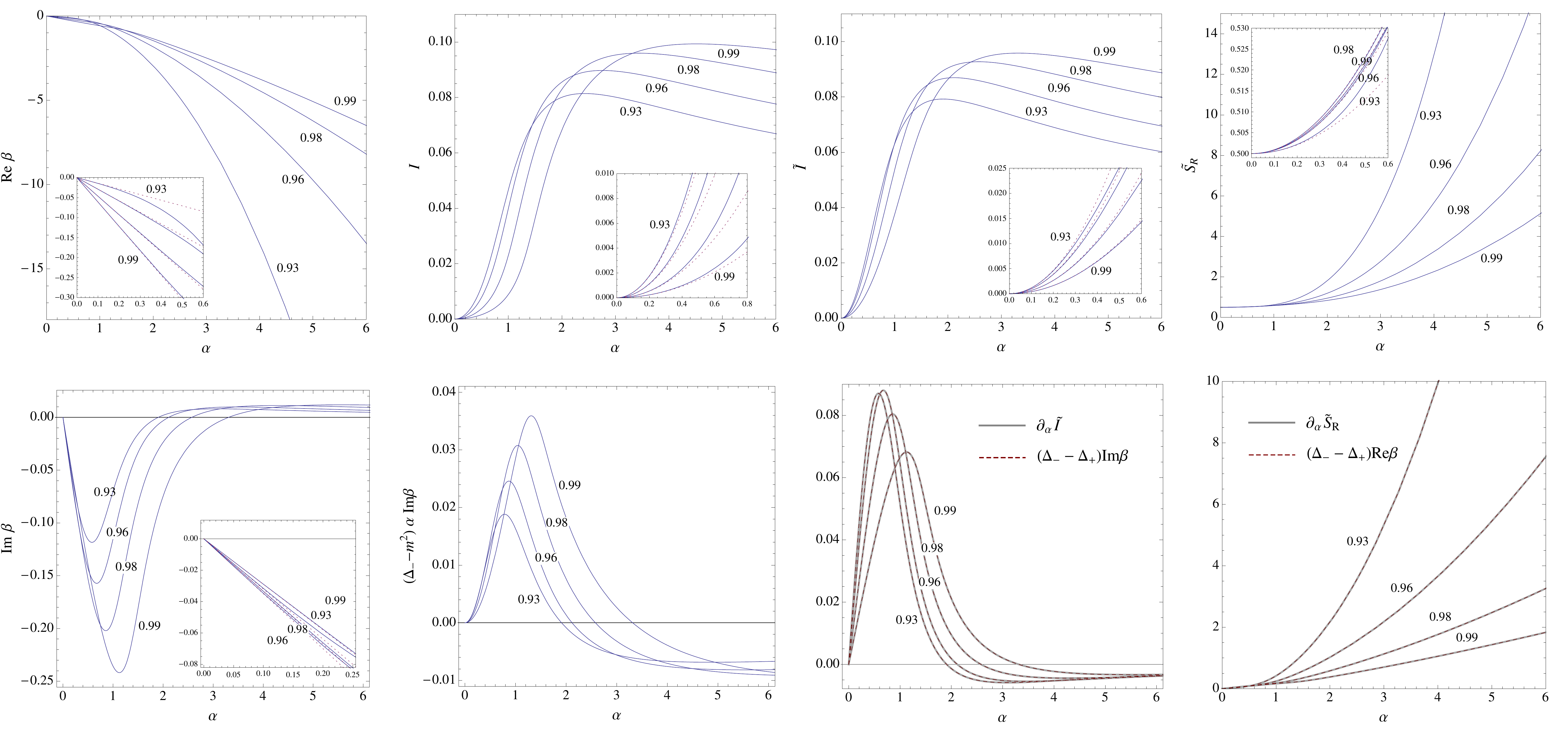}
\end{center}
\vspace*{-5mm}
 \caption{\footnotesize Results of fundamental saddle points of boundary topology $T^2$. The scalar potential is quadratic, with scalar masses $m=0.93,0.96,0.98,0.99$, marked beside the corresponding curves. The $\chi$-contours of these saddle points are $\chi_o\rightarrow a \rightarrow b \rightarrow \chi_*$ in Fig.\ref{ctNPafPaf} when the scalar field is perturbative. The curves plotted are obtained with the ``Interpolation'' command of Mathematica on discrete data with default interpolation order $3$ (same for all other results). {\bf Column 1:} the $\alpha$-$\re \beta$ and $\alpha$-$\im \beta$ relations, where with the increase of $\alpha$, $|\re \beta|$ grows monotonically, while $\im \beta$ peaks and then falls back and asymptotes to $0$. {\bf Column 2:} the upper frame shows the imaginary part of the Lorentzian bulk on-shell action ${\cal I}$ against $\alpha$. The lower frame shows the boundary term induced by the holographic renormalization, which when added to ${\cal I}$ leads to the imaginary part of boundary on-shell action $\tilde {\cal I}$. {\bf Column 3:} the upper frame shows the imaginary part of the boundary Lorentzian saddle-point action, and the lower frame shows their derivatives $\partial_{\alpha}\tilde {\cal I}$, together with the imaginary part of the conjugate momentum of $\alpha$, i.e., the $\alpha$-$(\Delta_--\Delta_+)\im\beta$ curves, where the two sets of curves match precisely each other. This verifies the imaginary part of Eq.(\ref{OPFunc}). {\bf Column 4:} the $\tilde {\cal S}_{\rm R}$ counterpart of the third column, where we see again that the $(\Delta_--\Delta_+)\re\beta$ curves run exactly over the $\partial_\alpha \tilde {\cal S}_{\rm R}$ curves, thus verifying the real part of Eq.(\ref{OPFunc}). The thin red dotted lines in the inset windows are the perturbative results obtained in Sec.\ref{HPES} and Sec.\ref{OSactPert}.}\label{k0grid}
\end{figure}

The third case studied consists of saddle points of the same scalar mass $m=0.95$ but living through different amount of Euclidean histories. These saddle points have, when the scalar deformation is perturbative, the blue contours in Fig.\ref{ctNPafPaf} starting from $\chi_o$ and ending up in the branches of $k=0,-1,\dots,-5$. The content of the figures are arranged in the same way as the previous two cases, and the features relevant to our purpose are:
\begin{itemize}
\item  $k=0$ corresponds to the fundamental saddle points, and their results have similar features as those presented in Fig.\ref{k0grid}: at large $\alpha$, $\im \beta$ tend to $0$, $\tilde {\cal I}$ differ slightly from $\cal I$ where both are bounded.

\item Saddle points of $k\neq 0$ show similar features as those of the previous case in Fig.\ref{k1grid}: at small $\alpha$, $\im \beta$ increase or degreases following $\im \rho>0$ or $\im \rho <0$; for large $\alpha$, $\im \beta$ show tendency to diverge to infinities, ${\cal I}$ are bounded while $\tilde {\cal I}$ diverge to infinity in the opposite direction that $\im \beta$ diverges.

\item Different from the previous case however, $\im \beta$ have turning point, for instantce for $k=-3$: it starts off decreasing but then turns up and shows tendency of diverging to $+ \infty$. As a result, $\tilde {\cal I}$ starts off increasing when $\alpha \sim0$ but then it starts to decrease and shows tendency of diverging to $-\infty$.

\end{itemize}

\subsubsection*{With contours circling a branch point; results in Fig.\ref{M0p95loops}; contour in Fig.\ref{ctNPafPaf} in green}  

This category of saddle points of boundary topology $T^2$ are those with contours circling around a branch point as shown in Fig.\ref{ctNPafPaf} in green. When the scalar field is perturbative, the $\chi$-contours circle the first branch point below the south pole from counterclockwisely twice to clockwisely tree times, i.e., $p=-2,-1,\dots,3$. The result in Fig.\ref{M0p95loops} shows on the non-perturbative level the different physical outcomes of these different saddle points, where the features worth mentioning are 

\begin{itemize}
\item Similar as for the fundamental saddle points, $\im \beta$ asymptotes to $0$ when $\alpha$ is large. In the whole range of $\alpha$ investigated each keeps the same sign, which is identical to that of $\im \rho$, with $\rho$ given by Eq.(\ref{RHO122}). See the right frame of Fig.\ref{ImRFLp} where we can obtain the signs of $\im\rho$.

\item The imaginary parts of the saddle-point actions in the bulk field and boundary data representations, $\cal I$ and $\tilde {\cal I}$, differ slightly from each other. However different from the case of fundamental saddle points, they both show mild tendency of diverging to $\pm \infty$. 
\end{itemize}

%
%
%
%
%
%

\begin{figure}
\begin{center}
	\includegraphics[width=0.75\textwidth]{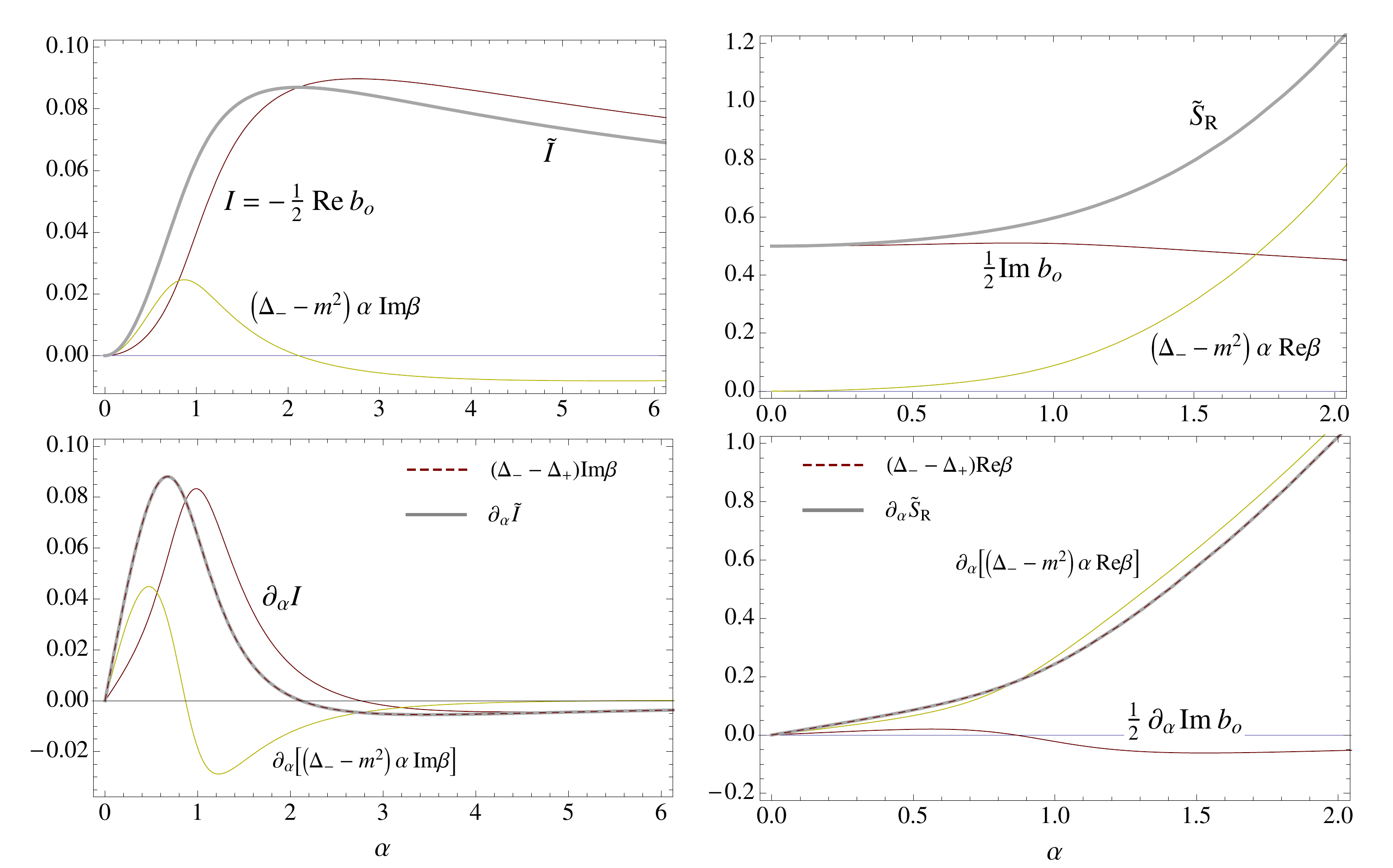}
\end{center}
\vspace*{-5mm}
 \caption{\footnotesize  Details of the matching $\partial_\alpha \tilde {\cal S}=(\Delta_- -\Delta_+)\beta$ for the fundamental saddle points of scalar mass $m=0.96$. The left and the right columns show respectively the real and imaginary parts of this relation, which correspond to the third and fourth columns in Fig.\ref{k0grid}. The upper row visualizes the detail of the addition in Eqs (\ref{SS53}) and (\ref{II54}) for obtaining $\tilde {\cal S}_{\rm R}$ and $\tilde {\cal I}$, and the lower row shows the derivative of the curves being added in the upper row to show how they add up to a $\partial_\alpha \tilde {\cal S}$ that matches exactly $(\Delta_- -\Delta_+)\beta$.}\label{SSMatchDetail}
\end{figure}

\begin{figure}
\begin{center}
	\includegraphics[width=\textwidth]{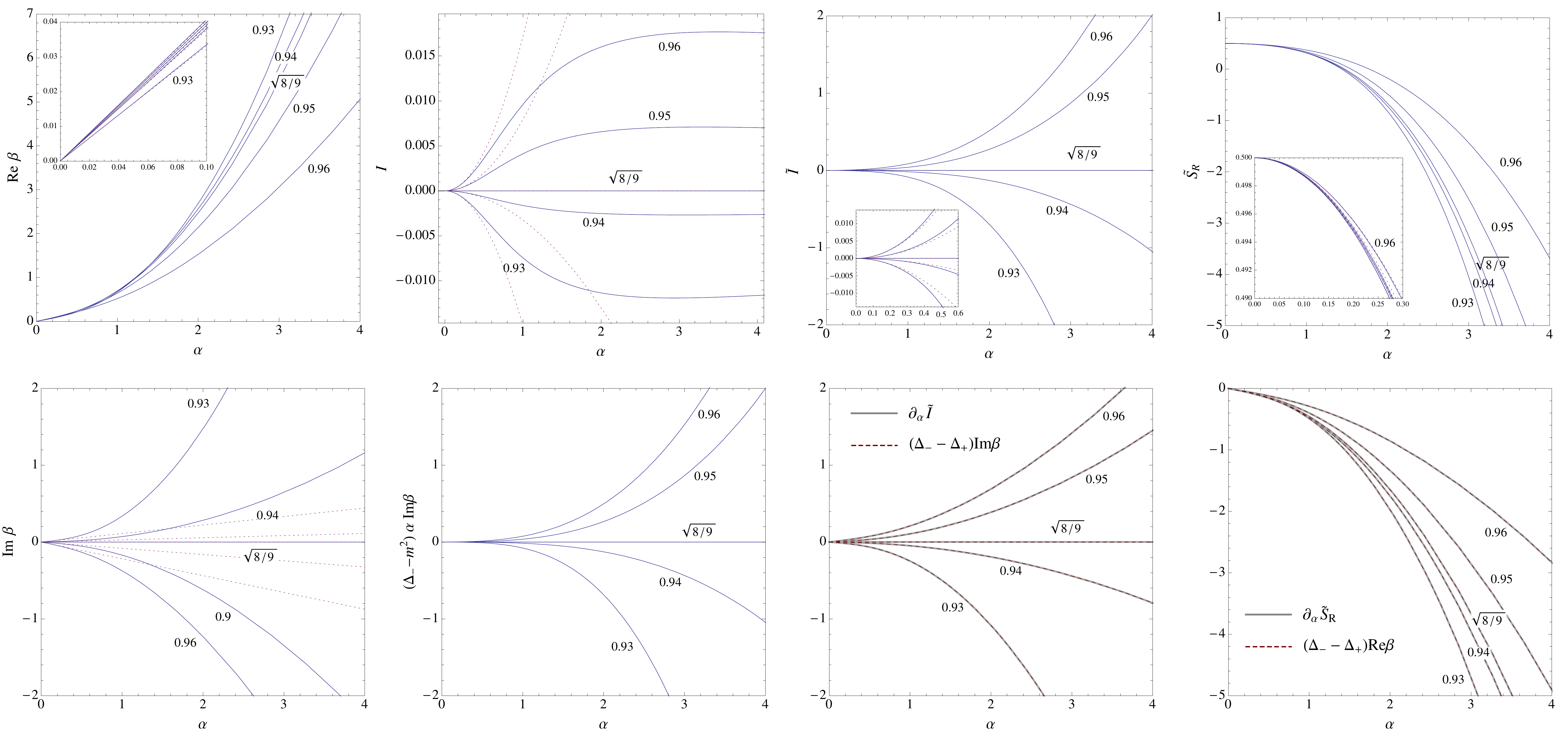}
\end{center}
\vspace*{-5mm}
 \caption{\footnotesize Results of saddle points of boundary topology $T^2$ whose $\chi$-contrours are $\chi_o\rightarrow a\rightarrow c\rightarrow \chi_*'$ ($k=-1$) in Fig.\ref{ctNPafPaf}. The scalar  masses covered are $m=0.93,0.94,\sqrt{8/9},0.95,0.96$. The contents of the figures are arranged in the same way as in Fig.\ref{k0grid}. Note the difference between the fundamental saddle points in Fig.\ref{k0grid}. Here $\im \beta$ diverge with $\alpha$ and resulting in  divergent $\tilde {\cal I}$, although $\cal I$ is bounded. The thin red dotted lines are perturbative results.}\label{k1grid}
\end{figure}


\begin{figure}
\begin{center}
	\includegraphics[width=\textwidth]{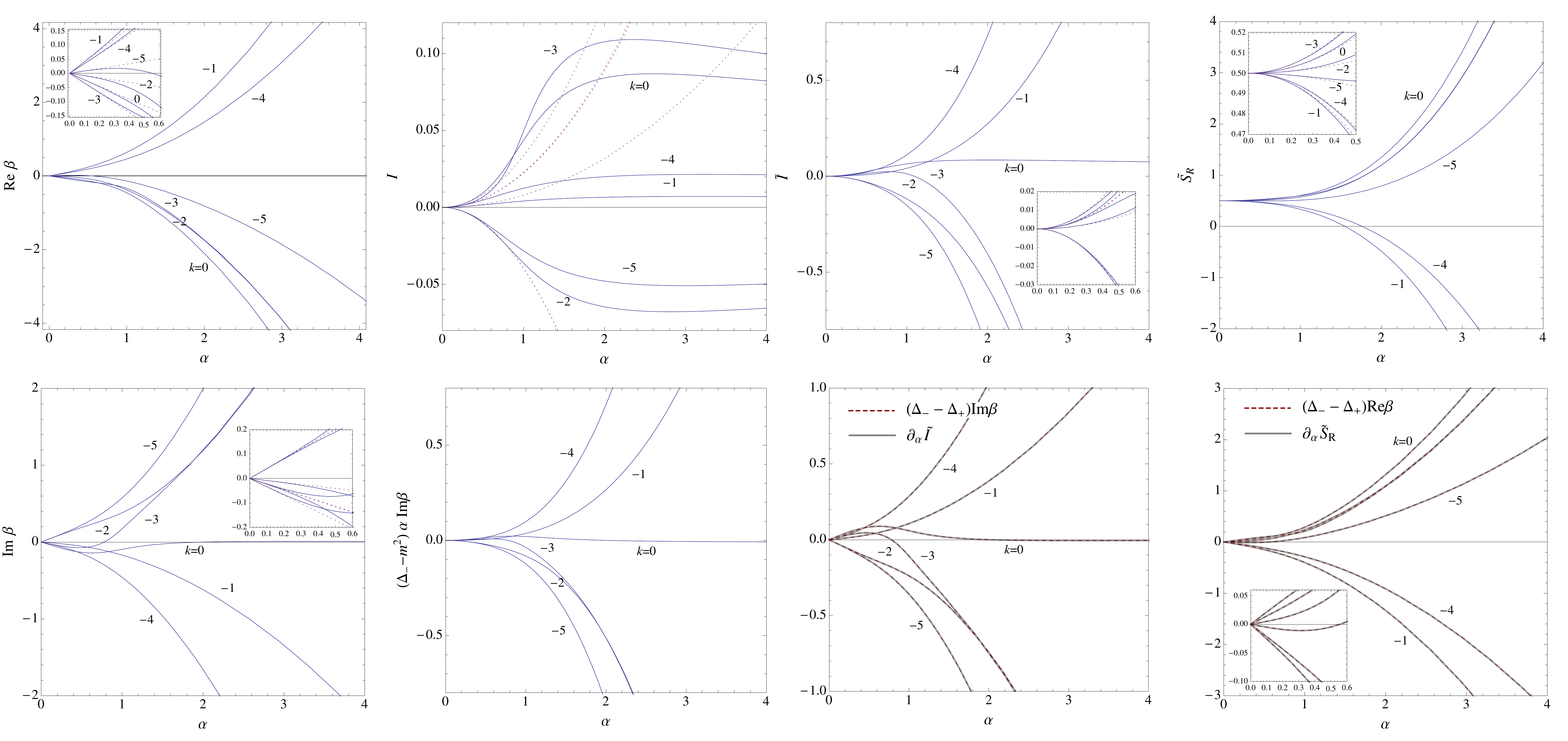}
\end{center}
\vspace*{-5mm}
 \caption{\footnotesize Results of saddles points with scalar mass $m=0.95$ whose contours contain different amount of positive Euclidean time labelled by the values of $k$ attached to the corresponding curve. The contents of the figures are arranged in the same way as in Fig.\ref{k0grid}. Note that the general pattern of these curves is similar with those of the non-fundamental saddle points shown in Fig.\ref{k1grid}, but with the increase of Euclidean time length, more varied details show up. For example the $\alpha$-$\im \beta$ curve has turning point when $k=-3$.}\label{M0p95Floors}
\end{figure}

\begin{figure}
\begin{center}
	\includegraphics[width=\textwidth]{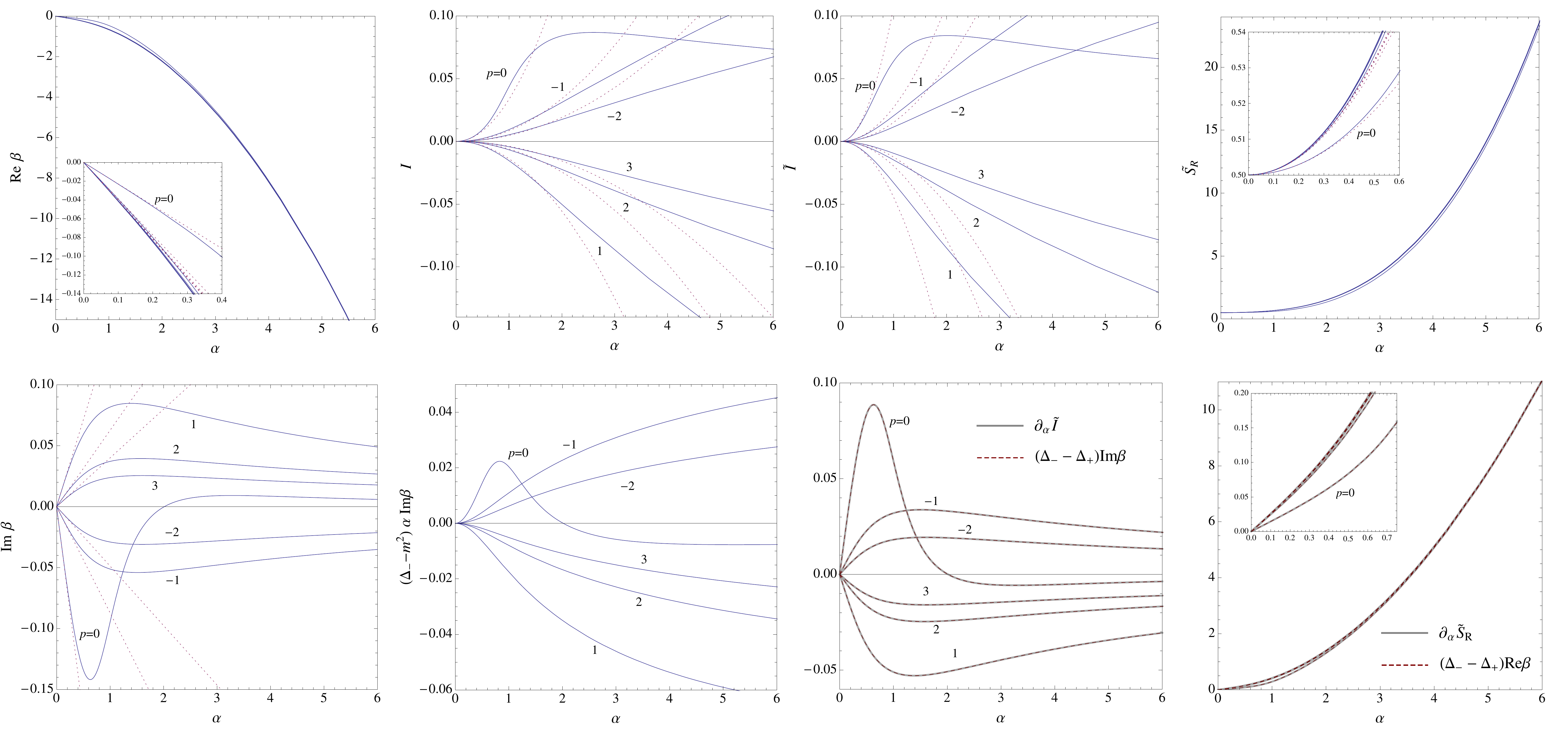}
\end{center}
\vspace*{-5mm}
 \caption{\footnotesize Results of saddle points of boundary topology $T^2$ and with scalar mass fixed at $0.95$. The time contours circle around a branch point which, when the scalar field is perturbative, is located at $\chi=0$. The integers $p=-2,-1,\dots,3$ attached to each curve are the times that the contour circles around the branch point. The contents of the figures are arranged in the same way as in Fig.\ref{k0grid}. The red dotted lines are perturbative results derived from Eqs (\ref{AFA})--(\ref{RHO122}), where $n=k=0$ in Eq.(\ref{RHO122}).}\label{M0p95loops}
\end{figure}

\begin{figure}
\begin{center}
	\includegraphics[height=8.4cm]{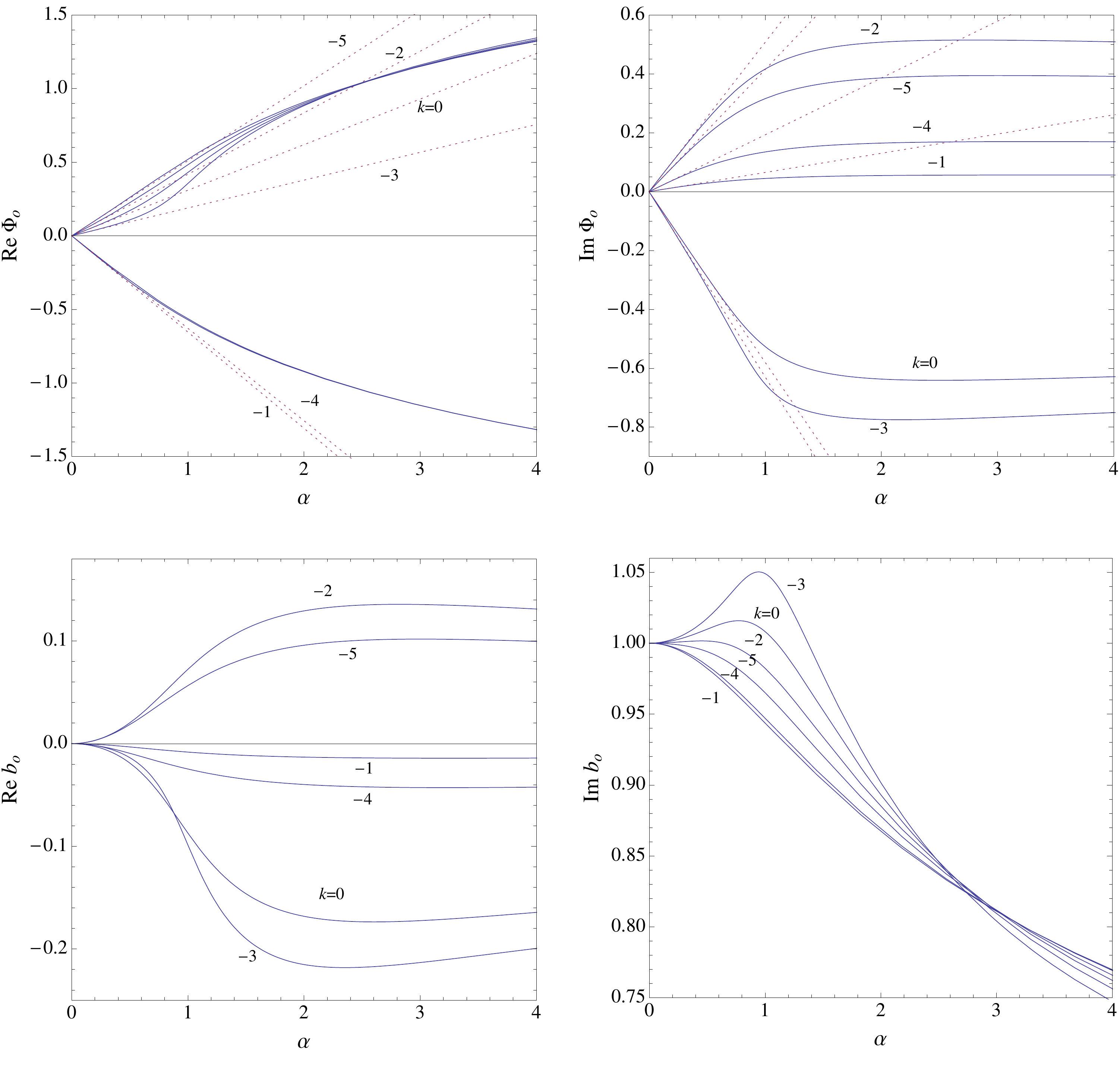}\hspace{9mm}\includegraphics[height=8.35cm]{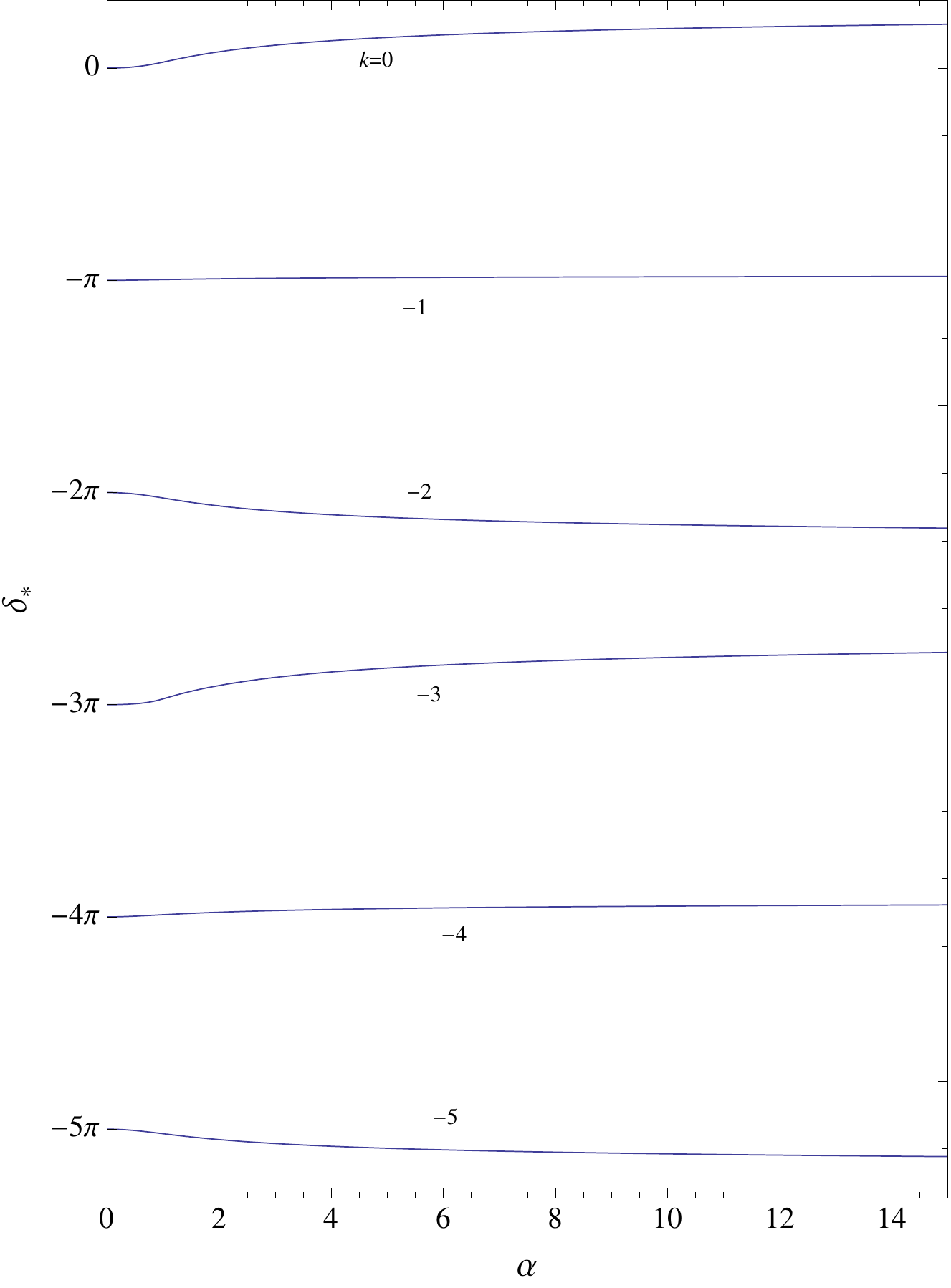}
\end{center}
\vspace*{-5mm}
 \caption{\footnotesize Defining parameters $(b_o,\Phi_o,\delta_*)$ of the saddle points studied in Fig.\ref{M0p95Floors}, traced as a function of scalar boundary data $\alpha$. The index $k=0,-1,\dots,-5$ are used in the same way as in Fig.\ref{M0p95Floors} and are shown next to the corresponding curves. {\bf Left:} The four frames are plots of $\Phi_o$ and $b_o$. The red dotted lines are perturbative results presented in Eq.(\ref{afdetPhi0}). {\bf Right:} Plots of the Euclidean shift $\delta_*$. It turns out that $\delta_*$ can be increasing or decreasing with the increase of $\alpha$, according to whether $\im\! \rho<0$ or $\im\! \rho>0$.
}\label{SPTPFloors}
\end{figure}

\begin{figure}
\begin{center}
	\includegraphics[height=8.8cm]{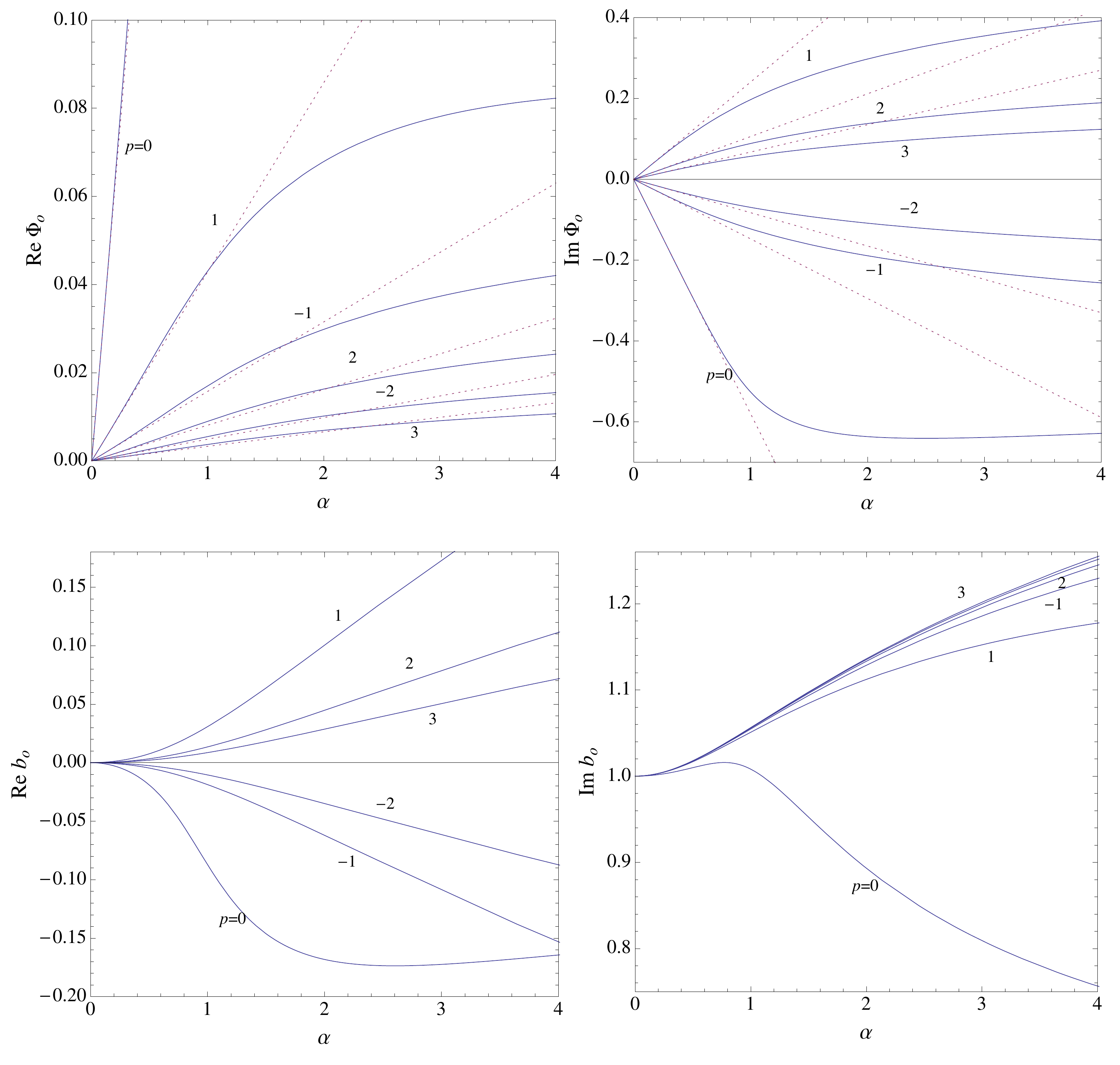}\hspace{8mm}\includegraphics[height=8.7cm]{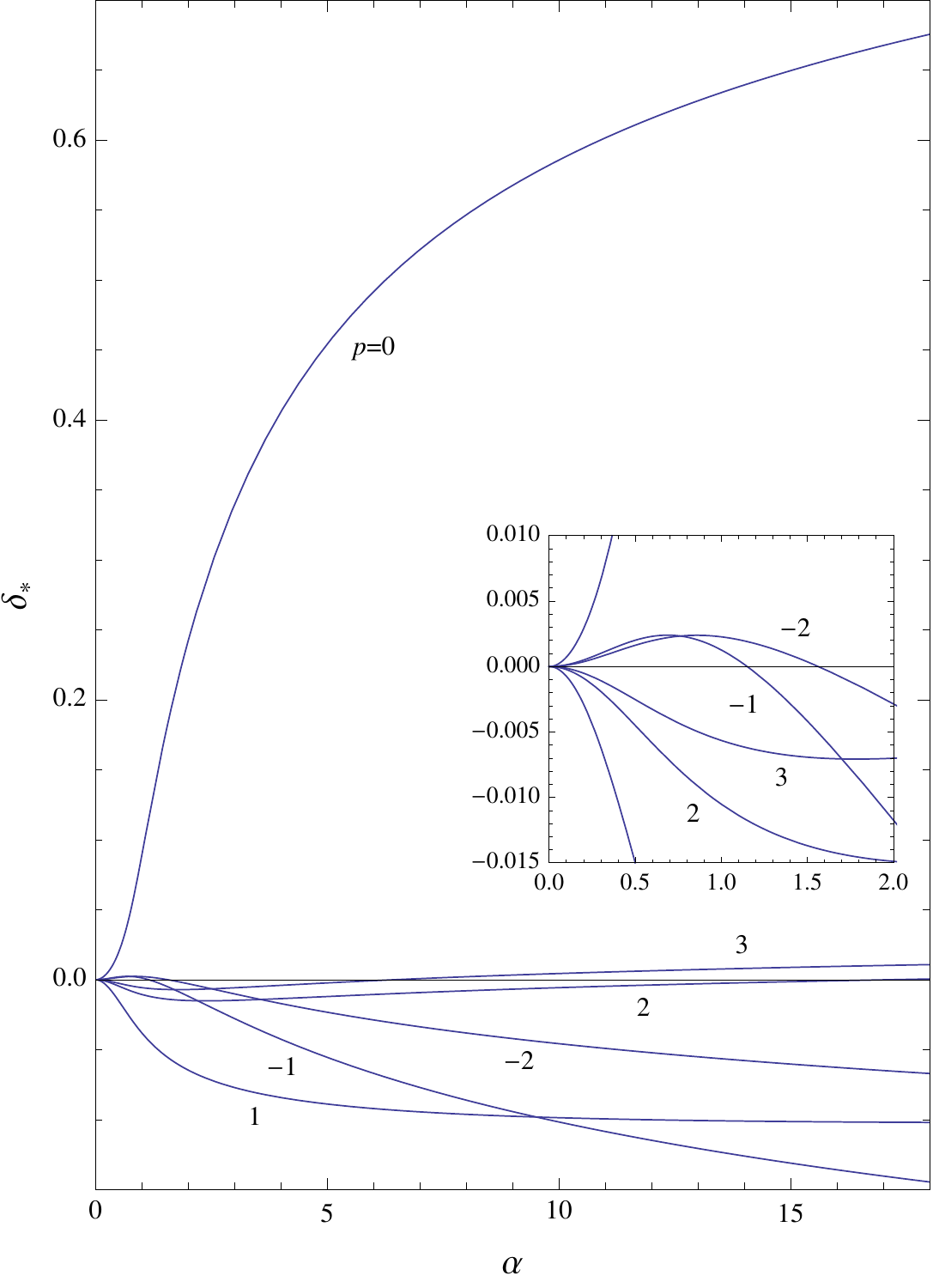}
\end{center}
\vspace*{-5mm}
 \caption{\footnotesize Defining parameters $(b_o,\Phi_o,\delta_*)$ of the saddle points studied in Fig.\ref{M0p95loops}. The curves are labeled $p=-2,-1,\dots,3$ in the same way as results are labeled in Fig.\ref{M0p95loops}. The red dotted lines represent perturbative results of Eq.(\ref{Phioaf124}). The general behavior of the quantities plotted behave in similar manner as those presented in Fig.\ref{SPTPFloors}. The right frame shows that $\delta_*$ are not necessarily monotonic with the growth of $\alpha$.}\label{SPTPLoops}
\end{figure}

%
%

%
%
\begin{figure}
\begin{center}
	\hspace*{-0cm}\includegraphics[width=\textwidth]{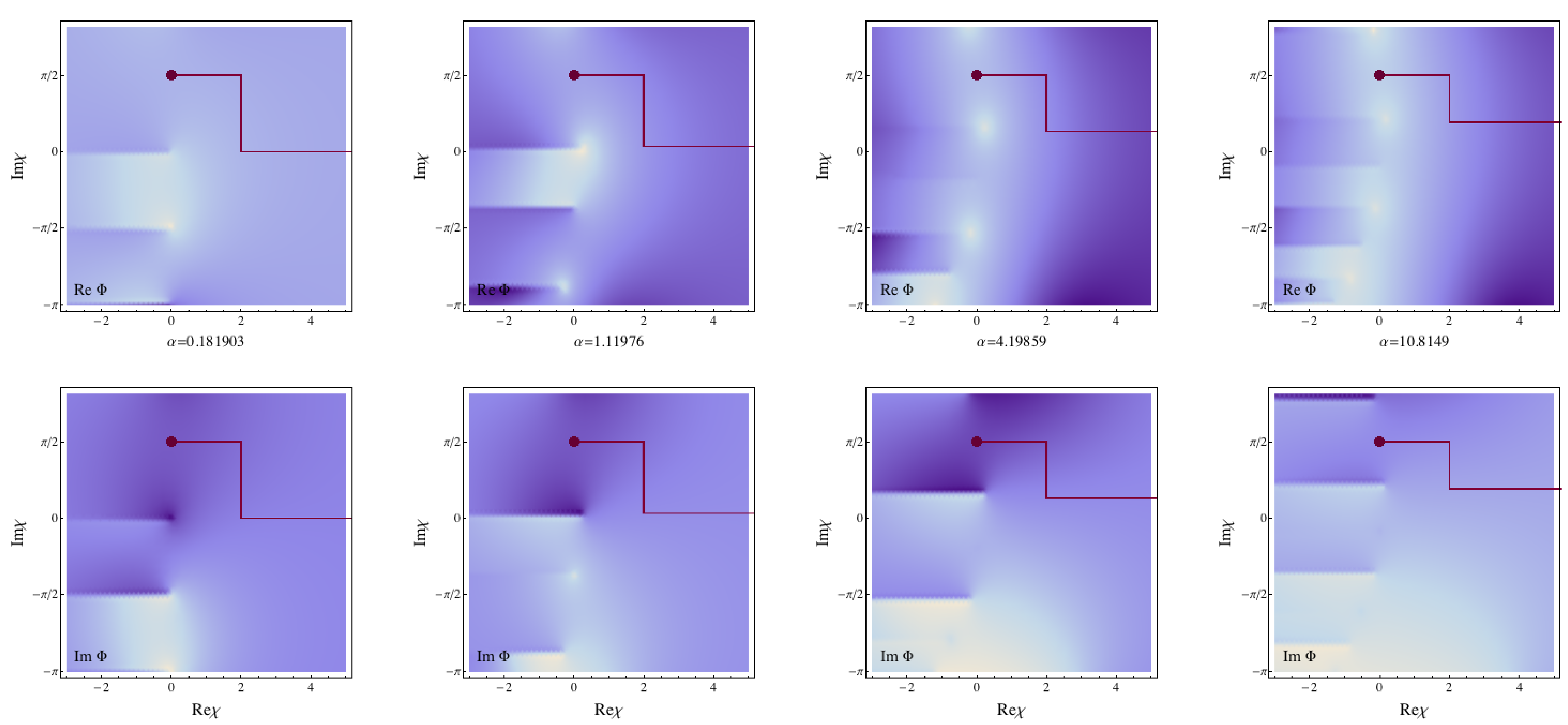}  
\end{center}
\vspace*{-5mm}
\caption{\footnotesize Deformation of the Riemann surface of $\Phi$ with the increase of $\alpha$, for the fundamental saddle points with quadratic scalar potential of scalar mass $m=0.95$. The physical quantities of this case are presented in Fig.\ref{M0p95Floors} by the $k=0$ curves. The red solid lines are the $\chi$-contours along which the equations of motion are integrated. In the perturbative regime $\alpha\sim 0$ (first column) the Riemann surface is well approximated by that in Fig.\ref{RiemannSurfaces} with branch points manifestly seen at $\chi=0$ and $-\pi$, and the integral contour is just the $\chi_o\rightarrow a\rightarrow b\rightarrow \chi_*$ in Fig.\ref{ctNPafPaf}. However the branch point at $-i\pi/2$ cannot be explained by perturbative calculation. When $\alpha$ increases, we observe the singular points clustering towards the south pole, and they are no longer in the line of $\re \chi=0$ as they do in the first column. Meanwhile the length of Euclidean history shortens, indicated by the decreasing length of the vertical segment of the contour; this last behavior is also shown in the right half of Fig.\ref{SPTPFloors} by the increasing $\delta_*$ for the $k=0$ curve.} \label{DPk0}
\end{figure}

\subsection{Inside story: tracing the intrinsic characteristics} \label{INTC}

The previous subsection has focused on the physical outcomes of the saddle points concerned. To give a complete account of these saddle points, it is important to also show how their intrinsic structures evolve with the scalar deformation. This subsection will show the results of this respect and try to identify some connection between these results and the features of the physical quantities observed in the previous subsection. Especially an interesting correlation is observed between the way that singularities move in the scalar field's Riemann surfaces and the divergence properties of $\im \beta$.

\subsubsection*{South pole data $(b_o,\Phi_o)$ and Euclidean shifts $\delta_*$}

Let us first examine the behavior of the defining parameters of the mini-superspace saddle points. They are presented in Fig.\ref{SPTPFloors} and Fig.\ref{SPTPLoops}, which are from exactly the same saddle points whose physical quantities are presented in Fig.\ref{M0p95Floors} and Fig.\ref{M0p95loops} respectively. The aspects worth mentioning are mainly about the comparison to their perturbative behaviors:

\begin{itemize}

\item $\Phi_o$ is proportional to $\alpha$ just as Eq.(\ref{afdetPhi0}) for $\alpha\sim 0$ as shown in the $\alpha$-$\Phi_o$ plots in Fig.\ref{SPTPFloors} and Fig.\ref{SPTPLoops}. Indeed, Eq.(\ref{afdetPhi0}) is plotted with red dotted lines, which are in very good match with the non-perturbative results, plotted with solid lines, when $\alpha \sim0$. When $\alpha$ increases, $|\Phi_o|$ increases the more and more slowly. However in general $|\Phi_o|$ does not necessarily grow with $\alpha$, but such example have not yet been found for the 3d models studied here, while in Sec.\ref{S3BrchAB} such a case will be shown in the model of scalar-deformed dS$_5$.

\item For different values of $\alpha$, $b_o$ needs to be adjusted differently to keep the value of $\tau_2$ to be $1$, making $b_o$ a non-trivial function of $\alpha$. Especially we have already seen that ${\cal I}=-{1\over 2}\re b_o$ as in Eq.(\ref{splitIm}).

\item The Euclidean shifts $\delta_*$ start off from the values found in perturbative theory when $\alpha$ increases from $0$. In Fig.\ref{SPTPFloors} they start from $0,\,-\pi,\dots,\,-5\pi$ for the saddle points of $k=0,-1,\dots,-5$, while in Fig.\ref{SPTPLoops} they all start from $0$. Then with the increase of $\alpha$ for very small $\alpha$, $\delta_*$ increases or decreases according to whether $\im \rho$ is negative or positive, but $\delta_*$ is not necessarily monotonic when $\alpha$ increases all the way to large values.

\end{itemize}


\subsubsection*{Riemann surfaces of the scalar field}

There is another intriguing aspect that seems to have the potential of providing more insight into the property of the saddle points. That is the analytical structure of the scalar field $\Phi$ as a function of the complex time $\chi$. To expose this feature in the following contents, the deformation as a function of $\alpha$ of the scalar field's Riemann surfaces will be shown by density plots. Note that they have already been studied in Sec.\ref{PERTT2} in the cases where $\Phi$ is perturbative, and are shown in Fig.\ref{RiemannSurfaces}. A rough illustration of how they get deformed when $\Phi$ becomes non-perturbative has been given in Fig.\ref{ctNPafPaf}, while now a precise account will be provided.

In the following, three cases are given attention to, which can well summarize all the cases encountered so far, all of them are of scalar mass $m=0.95$.

\begin{figure}

%
%
\begin{center}
	\hspace*{-0cm}\includegraphics[width=\textwidth]{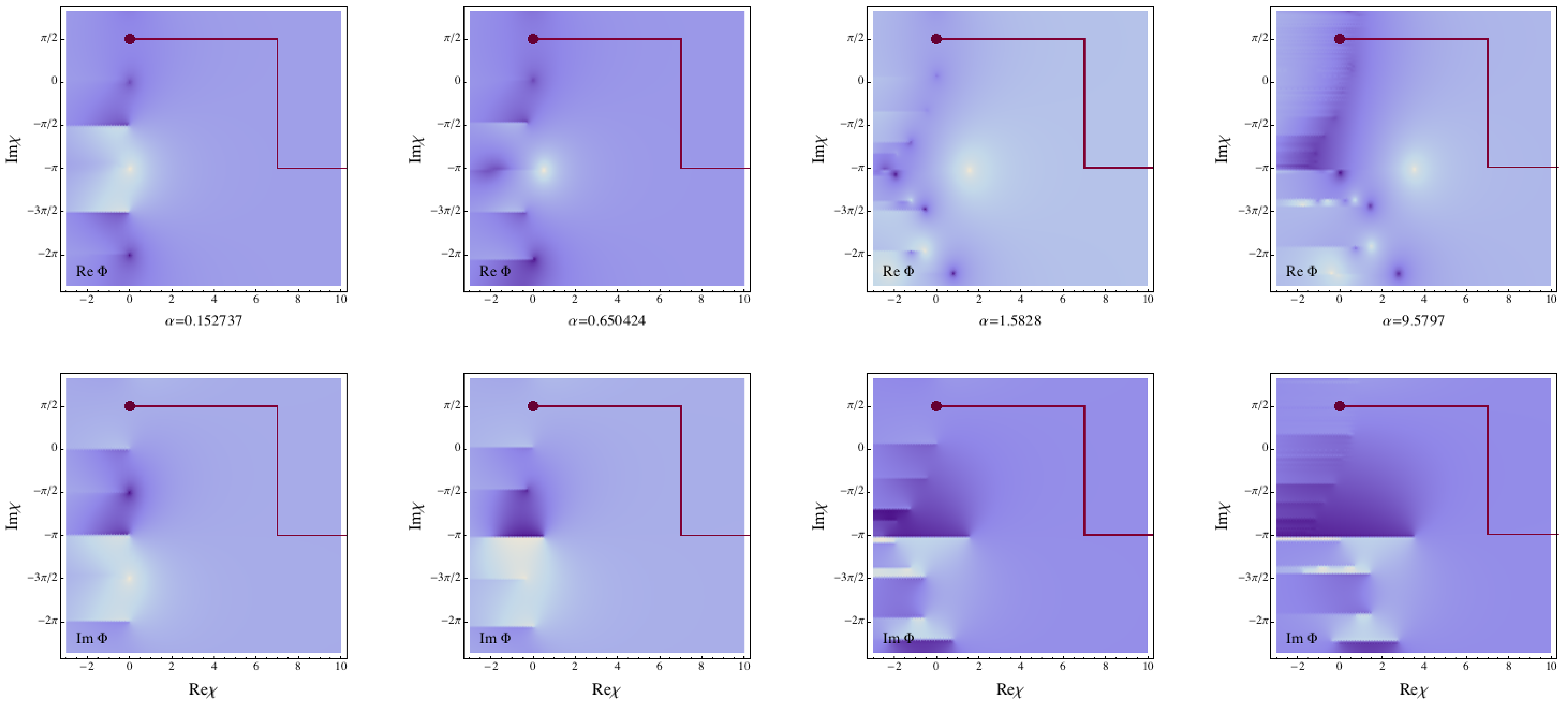}  
\end{center}
\vspace*{-5mm}
\caption{\footnotesize Density plots of $\Phi$ showing the deformation of Riemann surface of $\Phi(\chi)$ of $m=0.95$ with the increase of the boundary data $\alpha$. The saddle points concerned are the non-fundamental ones whose time contour is  given by $\chi_o\rightarrow a \rightarrow c \rightarrow \chi_*'$ in Fig.\ref{ctNPafPaf} when the scalar field is perturbative, as shown in the first column. The physical quantities are plotted in Fig.\ref{M0p95Floors} with the $k=-1$ curves. As $\alpha$ increases, there is one singular point taking off from the $\im \chi$-axis and tend to collide into the Euclidean part of the contour. This is not observed in Fig.\ref{DPk0} and seems to be related with the divergence of $\im \beta$ seen in the first column in Fig.\ref{M0p95Floors}.}\label{DPk1}
\end{figure}

\underline{\it The first two cases} are the saddle points whose contours do not circle around singularities of $\Phi(\chi)$ but cover different amounts of Euclidean time. They are the fundamental and the next-to-fundamental saddle points, whose physical quantities are presented in Fig.\ref{M0p95Floors} by the $k=0$ and $k=-1$ curves. When the scalar field is perturbative their time contours are given by the blue ones ending at $k=0$ and $k=-1$ levels in Fig.\ref{ctNPafPaf}. The Riemann surfaces of $\Phi$ of these two families of saddle points are shown in Fig.\ref{DPk0} and Fig.\ref{DPk1}. Immediately relevant descriptions are given in the captions, while I would like to emphasize the following features:

\begin{itemize}
\item When $\alpha\sim0$ (in the first columns) the Riemann surfaces for both cases are well approximated by that obtained in Sec.\ref{PERTT2}, shown in Fig.\ref{RiemannSurfaces}. Also, the Euclidean segments (vertical part) of the two saddle points terminate at $0$ and $-\pi$, which are just the blue contours $\chi_o\rightarrow a\rightarrow b\rightarrow \chi_*$ and $\chi_o\rightarrow a\rightarrow c\rightarrow \chi_*'$ in Fig.\ref{ctNPafPaf}.

\item As $\alpha$ increases, the Riemann surfaces for the two cases deform differently, manifestly in the way that the singular points behave. In Fig.\ref{DPk1} depicting the $k=-1$ saddle points, we observe a singular point leaving off the $\im \chi$-axis towards the positive Lorentzian time direction. If we keep on increasing $\alpha$, it will collide with the Euclidean part of the contour. Then we will have to retreat the latter to the right to avoid the collision, in order to ensure the continuous tracing of the saddle points. However for the $k=0$ saddle points as shown in Fig.\ref{DPk0}, the singular points also leave off the $\im \chi$-axis but towards the negative Lorentzian time direction, and no ``aggressive behavior'' to the integral contour is observed. 

\item In regard of the different ways that singular points behave, and the fact that $k=-1$ saddle points have divergent $\im \beta$ with the increase of $\alpha$ but not the $k=0$ saddles, it is tempting to associate the presence of a singularity tending to collide into the integral contour, with the divergence of $\im \beta$. Among all the model studied in this paper, this association holds. We will see more examples in the following studies.

\end{itemize}

\begin{figure}
\begin{center}
	\includegraphics[width=\textwidth]{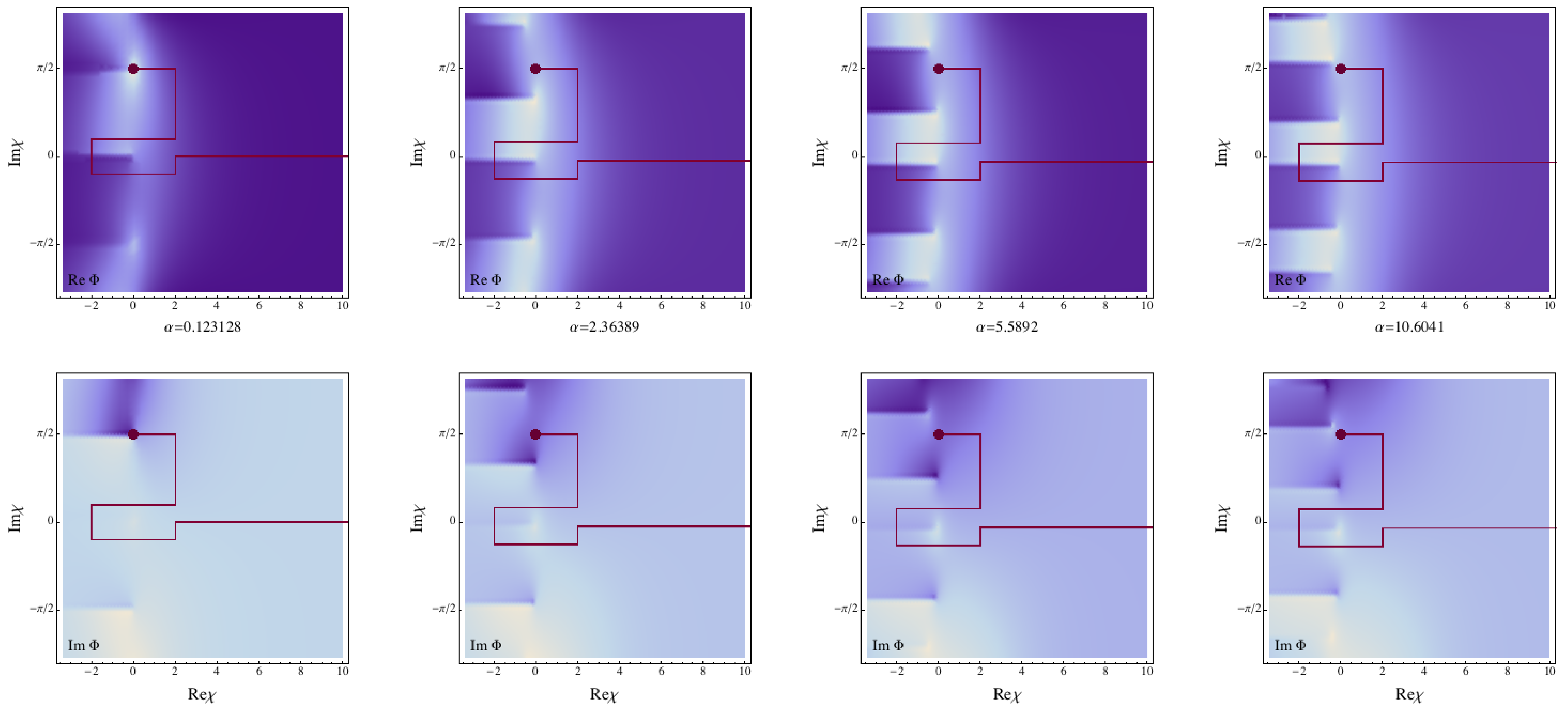}  
\end{center}
\vspace*{-5mm}
\caption{\footnotesize Deformation of Riemann surfaces of $\Phi$ with $m=0.95$. The complex time contours of the saddle points,  when $\Phi$ is perturbative, are as the green contour in Fig.\ref{ctNPafPaf}, but circle clockwisely once the branch point $\chi=0$. In the first column where $\Phi$ is perturbative, the south pole should be regular, but we see a branch cut leaving precisely from the south pole. This is due to the fact, referring to the discussion at the end of Sec.\ref{MVT2}, that the Riemann sheet shown in the plots is that of the end point $\chi_*$ and is not the same sheet as that of the south pole. On the Riemann sheet of $\chi_*$, the perturbative scalar field is given by the second line of Eq.(\ref{eq76}) with $p=1$ and $n=0$, which clearly has branch points at $\chi=ik\pi/2$ ($k\in \mathbb Z$), among which there is the south pole.} \label{DPp1}
\end{figure}

\underline{\it The third case} shown in Fig.\ref{DPp1} involves the saddle points, when $\Phi$ is perturbative, having their contours circling around once ($p=1$) the branch point $\chi=0$, as shown by the solid red contour in the first column. As $\alpha$ increases, the deformation of the Riemann surface and the contour are evident, while here I would like to mention that the singular points do not leave away from the $\im \chi$-axis towards the right, which corresponds well to the non-divergent behavior of $\im \beta$ in terms of $\alpha$ shown in Fig.\ref{M0p95loops}. This provides another example of the guess in the last paragraph, that the divergence of $\im\beta$ is always accompanied with a singular point aggressive to the integral contour.

%
%
%
%
%
%

%
%
%


\section{A parallel study: boundary topology $S^2$} \label{COSMOS2}

All the previous sections are devoted to the no-boundary quantum cosmology with boundary topology $T^2$, from which many results are drawn. It is thus interesting to carry out a study in parallel on another model with different boundary topology to see the effect of topology on the results. A very obvious candidate is $S^2$. This section will be dedicated to this model. 

All the conceptual elements and general formalisms developed for the case of $T^2$ in Sec.\ref{FFR}, including the general method of addressing the problem, spacetime ansatz, the definition of the Hartle-Hawking wave function, computation of its saddle-point contribution, can be adapted to the $S^2$ case here, and they will not be constructed in detail. What will be given below are the immediately useful formulae and equations. 

Again for the sake of feasibility of quantitative computation, the search of saddle points will be restricted within the range that mini-superspace formalism can reach. Therefore we will only be dealing with the saddle points whose spacetime can be sliced globally into homogeneous isotropic $S^2$-hypersurfaces against the coordinate time $\chi$. The latter ranges from the south pole $\chi_o$ to some final moment $\chi_*$, where $\chi_*-\chi_o$ is generically complex. This time at $\chi_o$, which will still be set at $i\pi/2$, it is an $S^2$ that is supposed to shrink smoothly to zero size. The final moment $\chi_*$ is mapped to the spacelike boundary $\Sigma_*$, and it is still parameterized as in Eq.(\ref{SigmatoI}), to formulate the limit $\Sigma_*\rightarrow I^+$.

The mini-superspace formalism is actually simpler to operate than in the case of $T^2$ boundary, since there is no detail in the boundary geometry to take into account. We can refer to the appendix \ref{ACEQGEN} for the action and the equations of motion, where we put $d_1=d$, $d_2=0$ in Eqs (\ref{abAction})--(\ref{EqVphi}). The action for no-boundary saddle points is
\begin{align}
	  \kappa S=&\, {1\over 2 }\int_{\chi_o}^{\chi_{{}_*}} d\chi \, a^d\Bigg[ d(d-1) \left(-{\dot a^2\over a^2}+{1\over a^2}\right) -d ( d-1)+\dot \Phi^2-2V(\Phi) \Bigg]-d\, a_o^{d-1} {\dot a_o}, \label{aAction}
\end{align}
where $a(\chi)$ is the scale factor or the radius of $S^2$, which can be complex but must be real at $\chi_*$; $\kappa={8 \pi G \over \ell^{d-1}\Omega_d}$, with $\Omega_d$ the volume of a $d$-dimensional sphere. We will set $\kappa=1$ for simplicity. Unlike the case of the previous case studied of $T^2$ spatial slices, the south pole term $-2 d\, a_o^{d-1} {\dot a_o}$ vanishes due to the south pole condition $a_o=0$. The equations derived from this action are
\begin{align}
	d(d -1)\left({\dot a^2 \over a^2}+ {1 \over a^2}\right) -d (d-1) -\dot\Phi^2 -2V(\Phi)=0, \label{EqVNd2}\\
	2(d-1){\ddot a\over a} -2(d-1)(d-2) \left({\dot a^2 \over a^2}+ {1 \over a^2}\right) - d(d -1)+\dot\Phi^2 -2V(\Phi)=0, \label{EqVad2} \\
	\ddot \Phi+d{\dot a\over a}\dot \Phi +V'(\Phi)=0.\label{EqVphid2}
\end{align}
These equations are to be integrated from $\chi_o$ to $\chi_*$ in the search for saddle points. At $\chi_o$ the fields should satisfy the conditions just as Eqs (\ref{NBCab}) and (\ref{NBCPhi}) but $b_o$ is not there. On the other hand at $\chi_*$, the boundary conditions to impose is simply the scalar boundary data $\alpha$, where the expressions of the asymptotic behaviors can be found in appendix \ref{ABAsydS}. The notation $\cal A$ will still be used, but now ${\cal A}=a_*^2\sim C_a^2e^{2\Lambda}$ measuring the surface area of the sphere $S^2$ at $\Sigma_*$.  Therefore the asymptotic behavior of the scalar field can still be expressed as Eq.(\ref{Phi*38}) and Eq.(\ref{Phi*40}), but now $\alpha=C_a^{\Delta_-}\hat \alpha$ and $\beta=C_a^{\Delta_+}\hat \beta$.


The work to be presented in this section will be based on the variation principle presented above. The following content will be covered. Sec.\ref{PertS2} studies the saddle points with perturbative scalar deformation, of boundary topology $S^d$ keeping $d$ generic. Just as for the case of $T^2$ in Sec.\ref{PERTT2}, the perturbative results will later serve as the starting points for numerically tracing the non-perturbative scalar deformation of the saddle points. The remarkable difference from the $T^2$ cases on the perturbative level, is the absence of branch cuts in the Riemann surfaces of all modes of $\Phi$ in the expansion against the $S^d$-harmonics. Sec.\ref{NPS2Act} studies the on-shell action of the saddle points of boundary topology $S^2$. Its holographic renormalization will be worked out for finite homogeneous scalar deformation. Sec.\ref{NUMS2} carries out the numerical calculation on the saddle points of boundary topology $S^2$, in the same manner as the studies have been presented in Sec.\ref{T2NCS}--Sec.\ref{INTC}.



%
%

%

\begin{figure}
\begin{center}
	\includegraphics[width=0.5\textwidth]{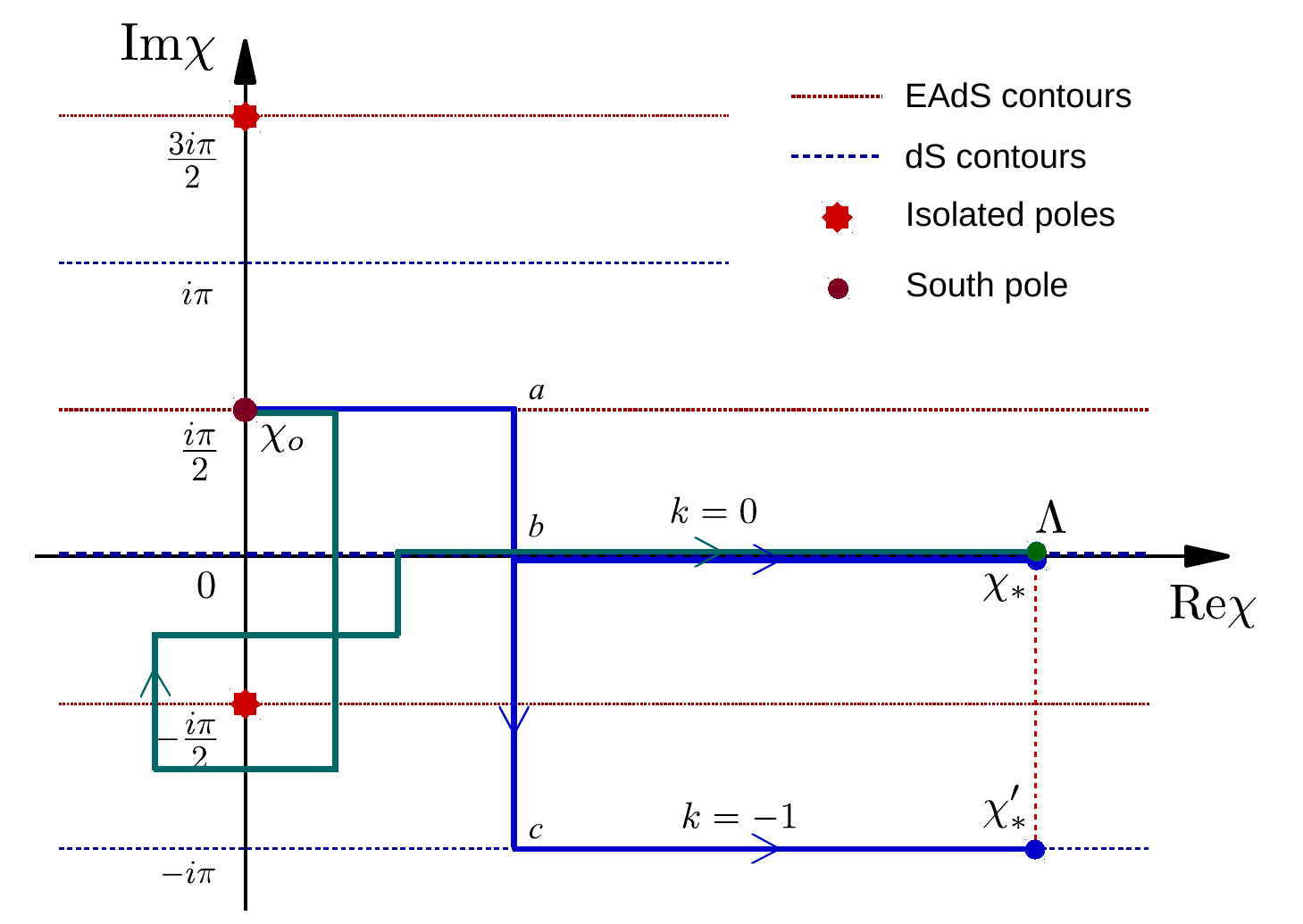}  
\end{center}
\vspace*{-5mm}
\caption{\footnotesize Riemann surface of homogeneous perturbative scalar field $\Phi$ in the background Eq.(\ref{dSdMetric}), as a function of the complexified time $\chi$. The south pole is set at $\chi_o=i\pi/2$ as usual. The expression of $\Phi$ is the $L=0$ component in Eq.(\ref{INHOMO146}), with which we locate singular points at $\chi=i(2k-{1\over 2})\pi$ ($k\in \mathbb Z$). They are simple poles of order $d-1$ and thus $\Phi$ when treated perturbatively is single-valued. Features of the background spacetime Eq.(\ref{dSdMetric}) are also shown in the figure: the metric describes a Lorentzian dS$_{d+1}$ space when $\im\chi=k\pi$ and an Euclidean AdS space when $\im \chi=\left(k +{1\over2}\right)\pi$ ($k\in \mathbb Z$). The defining contours of saddle points start from the south pole $\chi_o$ and terminate on one of the Lorentzian dS contours: $\chi_*=ik\pi+\Lambda$ with $k\in \mathbb Z$ and $\Lambda\rightarrow \infty$. The saddle points studied in Sec.\ref{COSMOS2} will have, in the perturbative regime of $\Phi$, the $\chi$-contours as shown by the solid blue and green lines.} \label{RiemannSdSd}
\end{figure}

\subsection{Scalar perturbation} \label{PertS2}

\subsubsection*{Solution of perturbative scalar fields}

Let us directly work out the inhomogeneous result for boundary topology $S^d$ ($d=2,3,\dots$). Again the method is already known in the literature (cf for example \cite{Bousso:2001mw}), and what is to be presented below is rather a reformulation using the notations and expressions adapted to our context.

First ignoring the scalar field, setting the south pole at $\chi_o=i\pi/2$, we solve from Eq.(\ref{EqVNd2}) and obtain the scale factor $a(\chi)=\cosh\chi$. Therefore the spacetime metric is that of an empty dS${}_{d+1}$ 
\begin{align}
	{ds^2\over \ell^2}=-d\chi^2+\cosh^2\! \chi \,d\Omega_d^2. \label{dSdMetric}
\end{align}
Just like the empty spacetime saddle points studied in Sec.\ref{EPTNBSD}, we can figure out the dS contours and EAdS contours for this metric, which are respectively $\im \chi=i k\pi$ and $\im \chi=i\big(k+{1\over 2}\big)\pi$ ($k\in \mathbb Z$). An illustration of these contours is shown in Fig.\ref{RiemannSdSd} (ignore the red spots for the moment which are the poles to be induced by the scalar field). Therefore we have another example of no-boundary saddle points which can be continued from asymptotic dS into asymptotic EAdS \cite{Hertog:2011ky}, in addition to the saddle points in Sec.\ref{EPTNBSD} which allow continuation between EBTZ and Kerr-dS$_3$ spacetimes. The no-boundary saddle points are hereby specified by the metric (\ref{dSdMetric}) supplemented with a complex $\chi$-contour. The latter starts off from the south pole $\chi_o$ and terminates on the asymptotic boundary of any dS contour: $\chi_*=i\delta_*+\Lambda$ where $\delta_*=k\pi$ ($k\in \mathbb Z$) and $\Lambda \rightarrow \infty$. Therefore we have an infinite number of saddle points labeled by the integer $k$. Examples of these contours are as shown in Fig.\ref{RiemannSdSd} by the blue and the green lines (again ignore the red spots).

Then let us further on switching on the scalar field perturbation. The Klein-Gorden equation against this background is
\begin{align}
	- {1\over \cosh^d\!\chi} {\partial \over \partial \chi}\left(\cosh^d\! \chi {\partial \Phi \over \partial \chi}\right)+{1\over \cosh^2\! \chi}\nabla_{\!\!S^d}^2 \Phi-m^2\Phi=0, \label{KGSd}
\end{align} 
where $\nabla_{\!\!S^d}^2$ is the Laplacian on $S^d$. We can expand the scalar field against the $d$-dimensional spherical harmonics $Y_L^j(\Omega)$ (c.f. \cite{Bousso:2001mw}):
\begin{align}
	\Phi(\chi,\Omega)=&\sum_{L,j} \Phi_{Lj}(\chi)Y_L^j(\Omega), \ \ {\rm with}\\  &\nabla_{\!\!S^d}^2 Y_L^j=-L(L+d-1)Y_L^j, \nonumber 
\end{align}
where $L=0,1,2,\dots$ and $j=-L,-L+1\dots,L$, and $\Omega$ denote collectively the coordinates of $S^d$. Inserting this expansion into the Klein-Gordan equation (\ref{KGSd}), we obtain the equations for each component $\Phi_{Lj}(\chi)$:
\begin{align}
	\ddot \Phi_{Lj} +d\, \tanh\chi \, \dot \Phi_{Lj} +\left[ m^2+{L(L+d-1) \over \cosh^2 \chi} \right]\Phi_{Lj}=0. \label{fL146}
\end{align}
To obtain an equation in a more obvious form we can do the following change of variables:
\begin{align}
	\Phi_{Lj}=z^{L\over 2} (1-z)^{L\over 2}y_{Lj},\ \ {\rm with}\ \ z={1+i \sinh \chi \over 2}.
\end{align}
Then we obtain from Eq.(\ref{fL146}) the equation for $y_{Lj}$:
\begin{align}
	z(1-z)(y_{Lj})''_{zz}+\left[{d+1+2L \over2} -(d+1+2L)z\right] (y_{Lj})'_z-\big[ m^2+L(d+L) \big] y_{Lj}=0,
\end{align}
which is a hypergeometric differential equation. We are interested in the solution regular at the south pole $z=0$. Therefore the branch of solution to pick up is
\begin{align}
	y_{Lj}=\, A_{Lj}\,  &{}_2F_1\!\! \left( \Delta_-+L, \Delta_++L,{d+1 \over 2}+L , {1+i\sinh\!\chi \over 2} \right)\!,\  {\rm where} \ \Delta_\pm ={d\pm \sqrt{d^2-4m^2} \over 2},
\end{align}
where $A_{Lj}$ are constants to be determined by boundary conditions at $\Sigma_*$. Thus we have the expression for the scalar field:
\begin{align}
	\Phi(\chi,\Omega)=\sum_{L,M}A_{Lj} \big[z(1-z)\big]^{L\over 2} {}_2F_1\! \! \left( \Delta_-+L, \Delta_++L,{d+1 \over 2}+L, z \right) Y_L^j(\Omega). \label{INHOMO146}
\end{align}
Note that the field value at the south pole $z=0$ is the coefficient of the homogeneous mode $A_{00}$, consistent with the fact that $z=0$ is not a boundary so that the dimensions represented by the angular coordinates $\Omega$ do not exist.

\subsubsection*{No branch cuts}

The solution, regarded as a function of $z$, is singular at $z=1$, where $\chi=\chi^{[n]}:=i \left(2k-{1\over 2} \right)\pi$. Expanding one single mode $\Phi_{Lj}$ of the scalar field around $z=1$, we have the asymptotic behavior
\begin{align}
 	\Phi_{Lj}\sim &\, \Big[C_1\, (1-z)^{L\over 2}+C_2\, (1-z)^{1-d-L\over 2} \Big] \times \Big[1+O(z-1)\Big] , \label{PoleOrd1}
\end{align}
where $C_1$ and $C_2$ are some coefficients depending on $m$, $d$ and $L$. Therefore we see that on the $z$-plane $z=1$ is a branch point when $L$ or $L+d-1$ is odd. However since $z={1+i\sinh\chi \over 2}$, therefore when $\chi\sim \chi^{[n]}:=i \left(2k-{1\over 2} \right)\pi$ the variable $z$ expands as
\begin{align}
	z\sim {1\over 4} \left(\chi - \chi^{[n]}\right)^2+O\left( \chi-\chi^{[n]}\right)^4 \label{PoleOrd2}
\end{align}
whose leading term is quadratic. Therefore $z=1$ or $\chi=\chi^{[n]}$ are not branch points but isolated poles in the $\chi$-plane of order $L+d-1$. Therefore examined perturbatively, the scalar fields are single-valued.  An illustration of the Riemann surface of $\Phi_{00}(\chi)$ is given in Fig.\ref{RiemannSdSd}. In the presence of a perturbative scalar field the no-boundary saddle points are specified by the metric Eq.(\ref{dSdMetric}), the scalar field Eq.(\ref{INHOMO146}) and a contour on the complex $\chi$-plane starting from the south pole and ending at the asymptotic boundary of a dS contour. Some examples of such contours are shown in Fig.\ref{RiemannSdSd}. Since the singular points in the Riemann surface (red spots) are isolated poles, the green contour and the $k=0$ blue contour represent the same saddle point. However later we will see that they will become different saddle points when the scalar field becomes non-perturbative.


\subsubsection*{Perturbative saddle-point actions}

We can work out the perturbative result of saddle-point action for inhomogeneous perturbations. However to be relevant with the numerical calculation, let us only focus on the homogeneous case, while the extension to the inhomogeneous case is straightforward. Thus the scalar field is given by the $L=0$ component of Eq.(\ref{INHOMO146}) and its asymptotic behavior near the dS boundary is:
\begin{align}
	\Phi_*=\varphi \sim \Phi_o \left[ {\Gamma\!\left({d+1 \over 2}\right)\Gamma(\Delta_+-\Delta_-) \over \Gamma(\Delta_+) \Gamma\!\left({d+1 \over 2}-\Delta_-\right)}\left( -{i\over 4}\, e^{\chi_*} \right)^{-\Delta_-} + {\Gamma\!\left({d+1\over 2} \right)\Gamma(\Delta_--\Delta_+) \over \Gamma(\Delta_-) \Gamma\!\left({d+1 \over 2}-\Delta_+ \right)}\left( -{i\over 4}\, e^{\chi_*} \right)^{-\Delta_+}\right], \label{expPhieq149}
\end{align}
where $\Phi_o=A_{00}$, and $\chi_*=ik\pi +\Lambda$ ($k\in \mathbb Z$ and $\Lambda \rightarrow \infty$). Therefore if on the other hand we require the standard asymptotic formula at future boundary:
\begin{align}
	\Phi_*=\varphi \sim \hat \alpha \,e^{-\Delta_-\Lambda}+\hat \beta \,e^{-\Delta_+\Lambda}\sim \alpha \,a_*^{-\Delta_-}+ \beta \,a_*^{-\Delta_+}, \label{afbt137}
\end{align}
where $\alpha=C_a^{\Delta_-} \hat \alpha$, $\beta=C_a^{\Delta_+} \hat \beta$ and $a_*=a(\chi_*)\sim {1\over 2}e^{\Lambda}$ ($C_a={1\over 2}$), we can read off the coefficients 
\begin{align}
	\alpha=2^{-\Delta_-}\hat \alpha=\Phi_o\, 2^{\Delta_-}\exp\!\left[i\left({1\over 2}-k\right) \pi \Delta_-\right]  {\Gamma\!\left({d+1 \over 2}\right)\Gamma(\Delta_+-\Delta_-) \over \Gamma(\Delta_+) \Gamma\!\left({d+1 \over 2}-\Delta_-\right)}, \label{afaPhiodS}\\
	\beta=2^{-\Delta_+}\hat \beta =\Phi_o\, 2^{\Delta_+}\exp\!\left[i\left({1\over 2}-k\right) \pi \Delta_+\right]{\Gamma\!\left({d+1\over 2} \right)\Gamma(\Delta_--\Delta_+) \over \Gamma(\Delta_-) \Gamma\!\left({d+1 \over 2}-\Delta_+ \right)}, \label{btaPhiodS}
\end{align}
where $-i$ in Eq.(\ref{expPhieq149}) should be understood as $e^{-i\pi/2}$. The ratio between $\alpha$ and $\beta$ will be useful in the expressions of perturbative on-shell action, and we denote it again by $\rho$ as previously for the $T^2$ case:
\begin{align}
	\rho:=&\, {\beta\over \alpha}= \left(-{ie^{ik\pi}\over 2} \right)^{\Delta_--\Delta_+}{\Gamma(\Delta_+)\Gamma \!\left({d+1\over 2}-\Delta_- \right)\Gamma(\Delta_--\Delta_+)\over \Gamma(\Delta_-) \Gamma\! \left({d+1\over 2}-\Delta_+ \right)\Gamma(\Delta_+-\Delta_-)}\nonumber \\
	=&\, {e^{i\pi\left({1\over2}-k\right) \sqrt{d^2-4m^2}} \over 2^{-\sqrt{d^2-4m^2}}}\,{\Gamma\left({d+\sqrt{d^2-4m^2}\over 2}\right)\Gamma\left({1+\sqrt{d^2-4m^2}\over 2}\right) \Gamma \left(-\sqrt{d^2-4m^2}\right)\over \Gamma \left({d-\sqrt{d^2-4m^2}\over 2}\right) \Gamma\left({1 -\sqrt{d^2-4m^2}\over 2}\right) \Gamma \left(\sqrt{d^2-4m^2}\right)}. \label{rhosp}
\end{align}
The results of the saddle-point actions in the bulk field representation ${\cal S}(a_*,\varphi)$  and boundary data representation $\tilde {\cal S}(\alpha)$ in terms of power series of $\alpha$, are respectively:
\begin{align}
	{\cal S}_{\rm R}(a_*, \alpha)=&\, [\text{IR divergences}]-{d^2-4m^2 \over 2d}\, \re \rho\, \alpha^2 + O(\alpha^4),\ \ {\cal I}(\alpha)= {\cal I}(0)-{d^2-4m^2 \over 2d}\, \im \rho\, \alpha^2 + O(\alpha^4), \label{PT145}\\
	\tilde {\cal S}_{\rm R}(\alpha)= & \, \tilde {\cal S}_{\rm R}(0)-{1\over 2} \sqrt{d^2-4m^2 }\, \re \rho\, \alpha^2+ O(\alpha^4),\ \ \ \ \tilde {\cal I}(\alpha)=  \tilde {\cal I}(0)-{1\over 2} \sqrt{d^2-4m^2 }\, \im \rho\, \alpha^2+ O(\alpha^4).\label{PT146}
\end{align}
Here ${\cal I}(0)=\tilde {\cal I}(0)$ is the value corresponding to empty spacetime, which depends on spacetime dimension and can be obtained by integrating Eq.(\ref{OSA183}) with $d_1=d$, $\Phi=0$ and $a(\chi)=\cosh\chi$. For example,
\begin{align}
	{\cal I}(0)=\tilde {\cal I}(0)=\left\{ \begin{array}{ll}  \displaystyle \Big(k-{1\over 2}\Big)\pi\, , & S^2\,, \\ \displaystyle -2\,, & S^3\, , \\ \displaystyle {3\over 2}\Big(k-{1\over 2}\Big)\pi\, , & S^4\, ,\end{array}\right.
\end{align}
where $k\in \mathbb Z$ is just the integer in the Euclidean shift $\delta_*$, appearing also in Eqs (\ref{afaPhiodS})--(\ref{rhosp}). The result of $S^4$ will be useful in Sec.\ref{S3BrchAB}.

\subsection{Saddle-point actions and holographic renormalization} \label{NPS2Act}

Starting from the action Eq.(\ref{aAction}) we can do as in the case of $T^2$, to have it converted into a form more amenable for the study of asymptotic behavior near the dS boundary. For this reason we need to do a partial integral to turn first order derivatives into second order ones, and use Eq.(\ref{EqVad2}) to eliminate the second-order derivative term in the resulting expression. Denoting the on-shell value of the action of $S$ using ${\cal S}$, we have
\begin{align}
	{\cal S}=&\, -{1\over 2}d(d-1)a^{d-1}\dot a\Big|_{\chi_{{}_o}}^{\chi_{{}_*}}+{1\over 2}\int_{\chi_o}^{\chi_*} d\chi \Big\{ 2d (d-1)(d-2) a^{d-2}\dot a^2 +d(d-1)^2 a^{d-2}\nonumber \\ &+{1\over 2} d(d-1)(d-2)a^d+{1 \over 2} (2-d) a^d \big[\dot \Phi^2 -2V(\Phi)\big]\Big\} -d \, a_o^{d-1}\dot a_o.
\end{align} 
When we go to 3d spacetime by setting $d=2$, the action becomes very simple:
\begin{align}
	 {\cal S}=&\, -a_*\dot a_* +\int_{\chi_o}^{\chi_*} d\chi = -a_*\dot a_* +\chi_*-\chi_o  = -a_*\dot a_* +i\left( \delta_*-{\pi \over 2}\right)+\Lambda \,. \label{SS2SimpAct}
\end{align} 
where the default setting of this paper is used: $\chi_o=i\pi/2$ and $\chi_*=\Lambda+i\delta_*$. The IR divergences are contained in the boundary term $-a_*\dot a_*$ and in the logarithmic term $\Lambda\sim \ln a_*$. Using the asymptotic behavior Eq.(\ref{asy213}), we can find that the IR divergent terms in  $-a_*\dot a_*$ are real, and that $-a_*\dot a_*$ does not contain finite imaginary terms. Thus the imaginary part of Eq.(\ref{SS2SimpAct}) is exclusively contained in the second term in its last step:
\begin{align}
	{\cal I}(\alpha)=\delta_*(\alpha)-{\pi\over 2},
\end{align}
which is just the amount of Euclidean history contained in the saddle point's whole complex history. The subtraction of the IR divergences in $-a_*\dot a_*$ proceeds essentially in the same way as that in the $T^2$ cosmology studied in Sec.\ref{T2INFD}, where the related counter terms are given by Eq.(\ref{CTTM}) with ${\cal A}=a_*^2$. This step yields the result
\begin{align}
	 \tilde {\cal S}=&\, -\Lambda+i \left( \delta_* -{\pi \over 2}\right) + \left( \Delta_- -m^2 \right) \alpha \beta +(\text{other counterterms}) \label{Substr1}
\end{align} 
Now we still have the logarithmic divergence $\Lambda$ to take care of. An immediate thinking is to mimic the AdS/CFT renormalization scheme by directly introducing a term of order $-\Lambda$, where a natural choice here is $-\ln a_*$. Indeed in AdS/CFT the same operation leads to the bulk computation of Weyl anomaly on the boundary field theory \cite{Henningson:1998gx}. Using the asymptotic behavior Eq.(\ref{asy213}) setting $u=e^{-\Lambda}$, we have
\begin{align}
	\ln a_*=\ln \left[C_a e^{ \Lambda}(1+\dots) \right]=\Lambda +\ln C_a+\dots \label{lna*}
\end{align}
When the scalar field is perturbatively small, we have $C_a=1/2$, while for finite $\Phi$, $C_a$ is generically a nontrivial function of $\alpha$. Including Eq.(\ref{lna*}) among the counter terms of Eq.(\ref{Substr1}) assuming that no other counter terms will be necessary, we arrive at a finite expression for the boundary on-shell action
\begin{align}
	 \tilde {\cal S}=&\, -\ln C_a+i \left( \delta_* -{\pi \over 2}\right) + \left( \Delta_- -m^2 \right) \alpha \beta. \label{Substr2}
\end{align} 
which is supposed to be a function of $\alpha$, and indeed on the righthand side, $C_a$, $\delta_*$ and $\beta$ are all functions of $\alpha$. Splitting the real and the imaginary parts, we have
\begin{align}
	 \tilde {\cal S}_{\rm R}= -\ln C_a + \left( \Delta_- -m^2 \right) \alpha \, \re\beta,\ \ \ \tilde {\cal I}= \delta_* -{\pi \over 2}+ \left( \Delta_- -m^2 \right) \alpha\, \im \beta. \label{Substr3}
\end{align} 
Next subsection will show numerically that this is the exact expression for the boundary on-shell action by showing the generation of ``one-point function'' Eq.(\ref{OPFunc}).

\begin{figure}
\begin{center}
	\includegraphics[width=\textwidth]{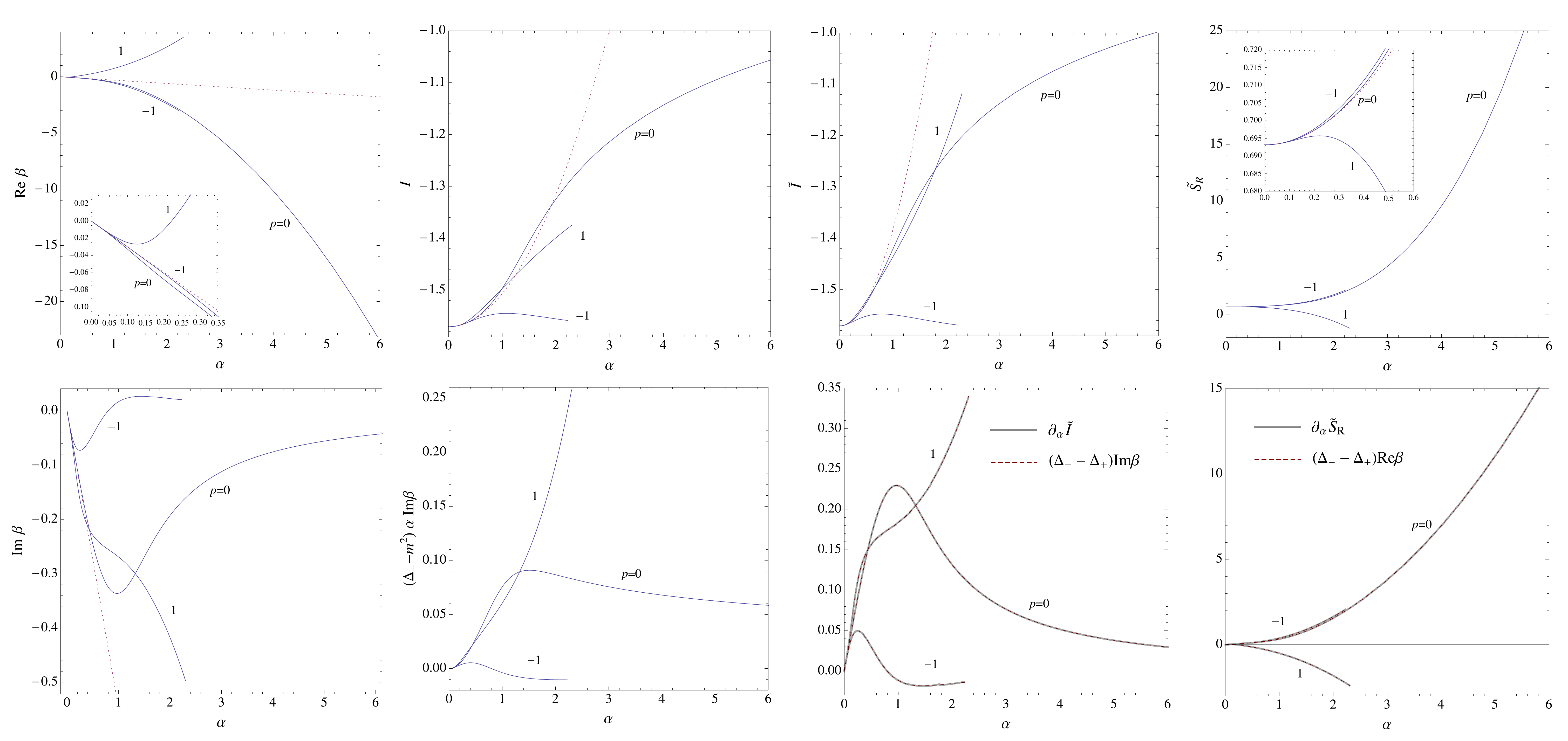}  
\end{center}
\vspace*{-5mm}
\caption{\footnotesize Results from saddle points of boundary topology $S^2$ with quadratic scalar potential and scalar mass $m=0.94$. The perturbative-regime contours are as the green one in Fig.\ref{RiemannSdSd}, where the cases studied are those with the contour circling $\pm 1$ and $0$ times around the singular point $\chi=-i\pi/2$. The corresponding curves are marked by $p=0,\pm1$ in the plots. The $p=0$ case is just the fundamental saddle point. Compared to the previous cases with boundary topology $T^2$ presented in Fig.\ref{M0p95loops}, here the adjustment of the contour for $p\neq 0$ is much more difficult because the movement of singularities is complicated. For this reason, the tracing of $p=\pm 1$ saddles did not go very far, but is enough to show that the curves divert from each other when $\alpha$ increases. Therefore the perturbative result, showing no branch cuts existing in the Riemann surfaces, do not hold for finite scalar perturbation. 
} \label{LOOP_S2M0p94}
\end{figure}

\begin{figure}
\begin{center}
	\includegraphics[width=0.75\textwidth]{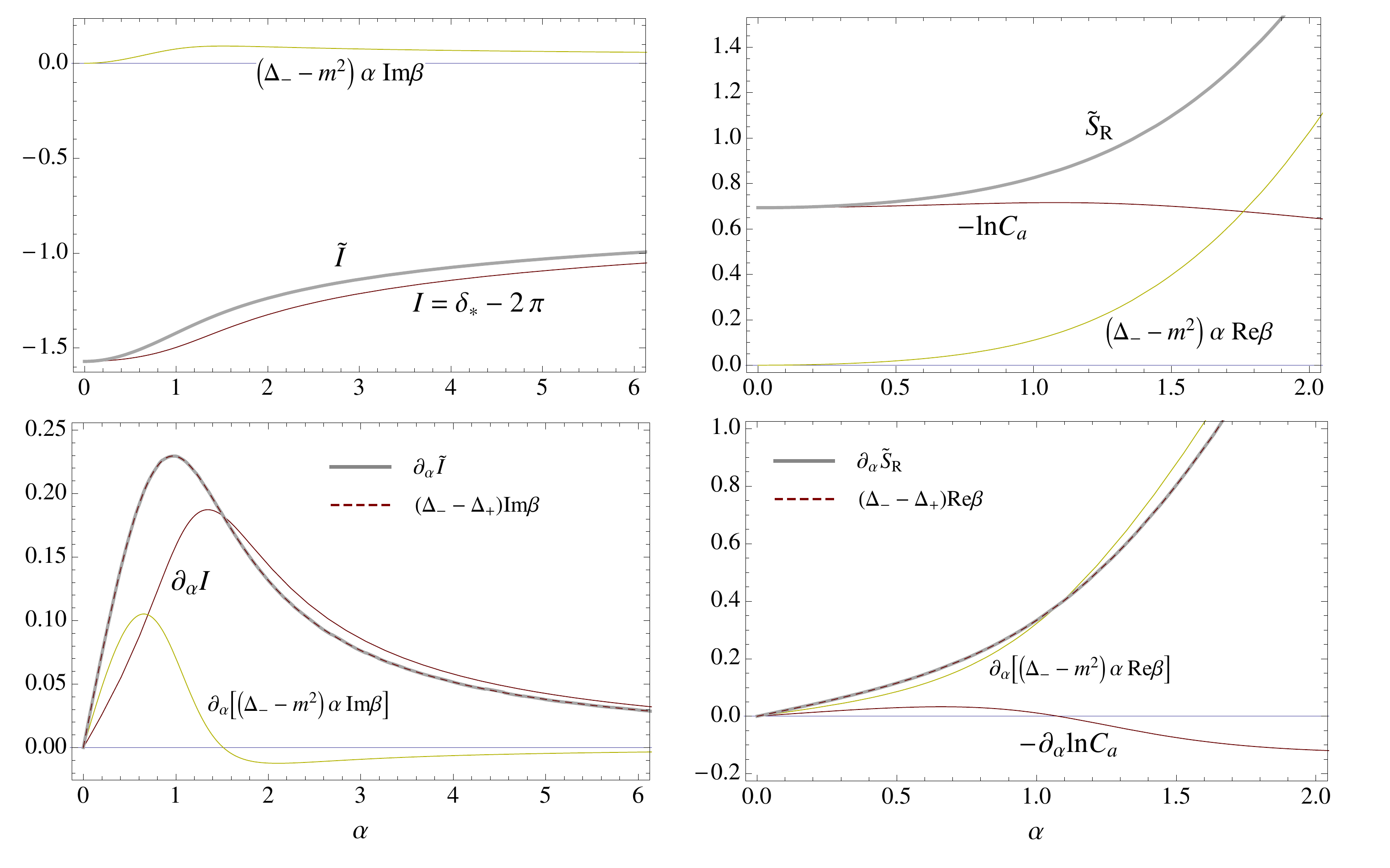}  
\end{center}
\vspace*{-5mm}
\caption{\footnotesize Details of the matching $\partial_\alpha \tilde {\cal S}=(\Delta_--\Delta_+)\beta$ for the fundamental saddle point, whose result has been presented with the $p=0$ curves in Fig.\ref{LOOP_S2M0p94}. The left column shows the matching of the imaginary part, and the right column real part, corresponding respectively to the third and fourth columns in Fig.\ref{LOOP_S2M0p94}. The upper frames show the detail of how terms on the righthand sides of Eq.(\ref{Substr3}) add up to the boundary on-shell action, and the lower frames shows the addition of the derivatives and the matching between the resulting $\partial_\alpha \tilde {\cal I}$ with $(\Delta_--\Delta_+)\beta$.} \label{DetMatS2}
\end{figure}

\begin{figure}
\begin{center}
	\includegraphics[height=6.2cm]{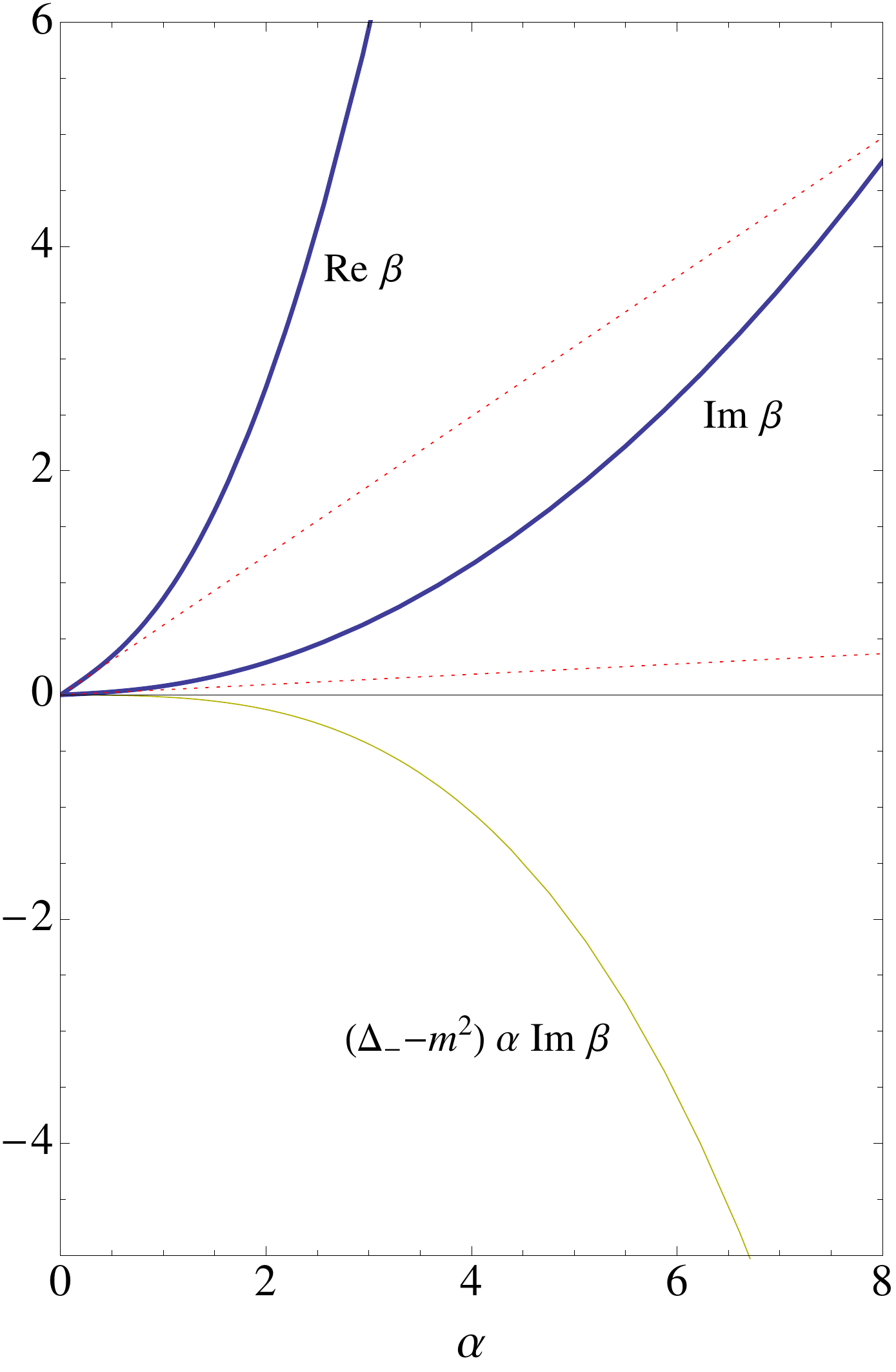}\hspace{2mm}\includegraphics[height=6.2cm]{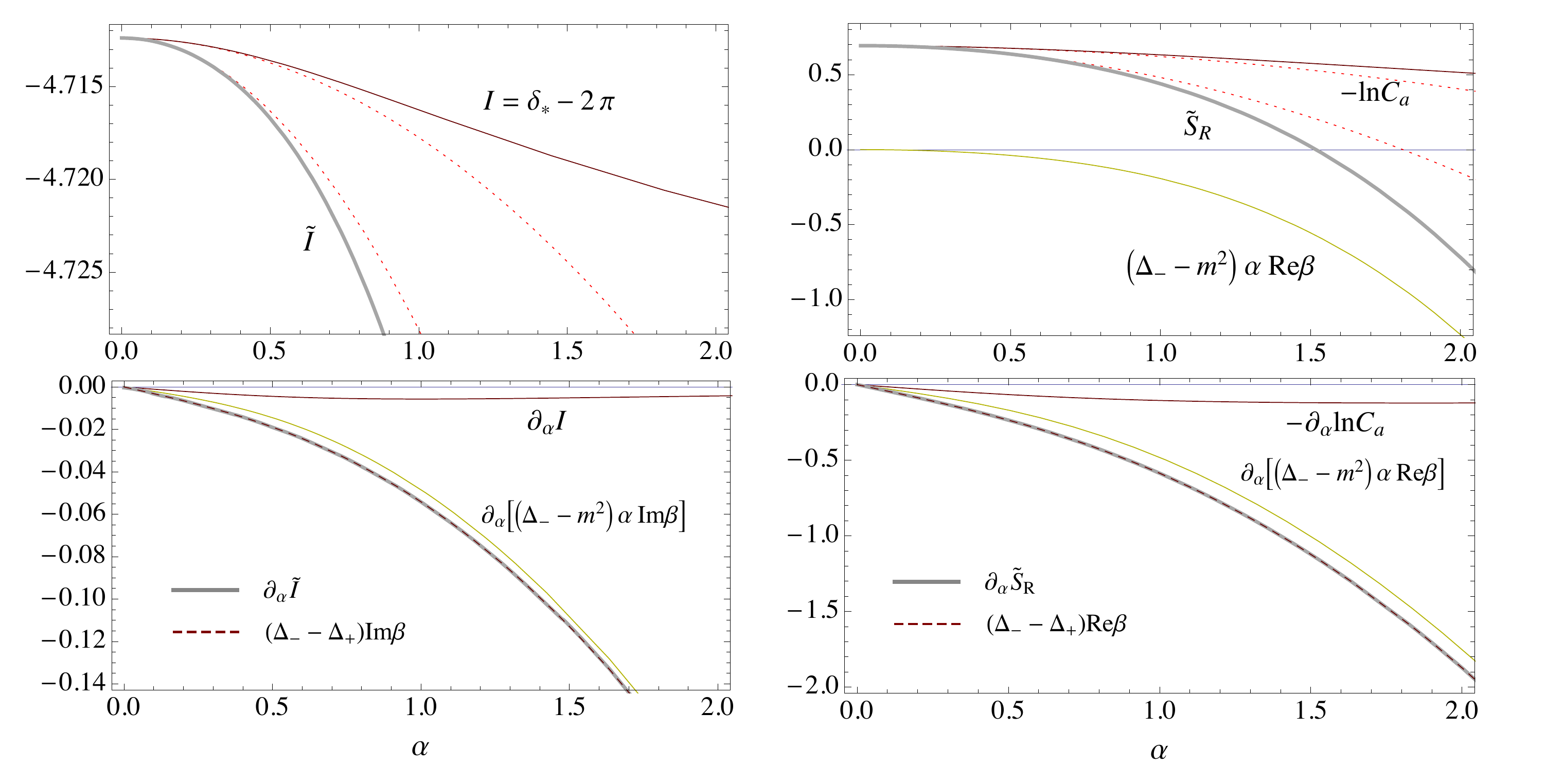}  
\end{center}
\vspace*{-5mm}
\caption{\footnotesize Results of the saddle points of boundary topology $S^2$, which have contours given by $\chi_o\rightarrow a\rightarrow c\rightarrow \chi_*'$ in the regime of perturbation scalar deformation. 
The left-most frame shows the $\alpha$-$\beta$ relation (blue) and the imaginary part of the boundary term produced by holographic renormalization  (yellow). The latter is to be added to the imaginary part of the bulk saddle-point action $\cal I$ to obtain its boundary counterpart $\tilde {\cal I}$. The middle column shows the plots related to the imaginary part of bulk on-shell action; the right column the real part. The dotted red lines are perturbative results. Since there is only one family of saddle points presented, the ``one-point function'' generation is shown in detail as Fig.\ref{DetMatS2}.} \label{FIG24}
\end{figure}

\begin{figure}
\begin{center}
\begin{minipage}[t]{0.5\textwidth}
\begin{center}	
	\includegraphics[width=\textwidth]{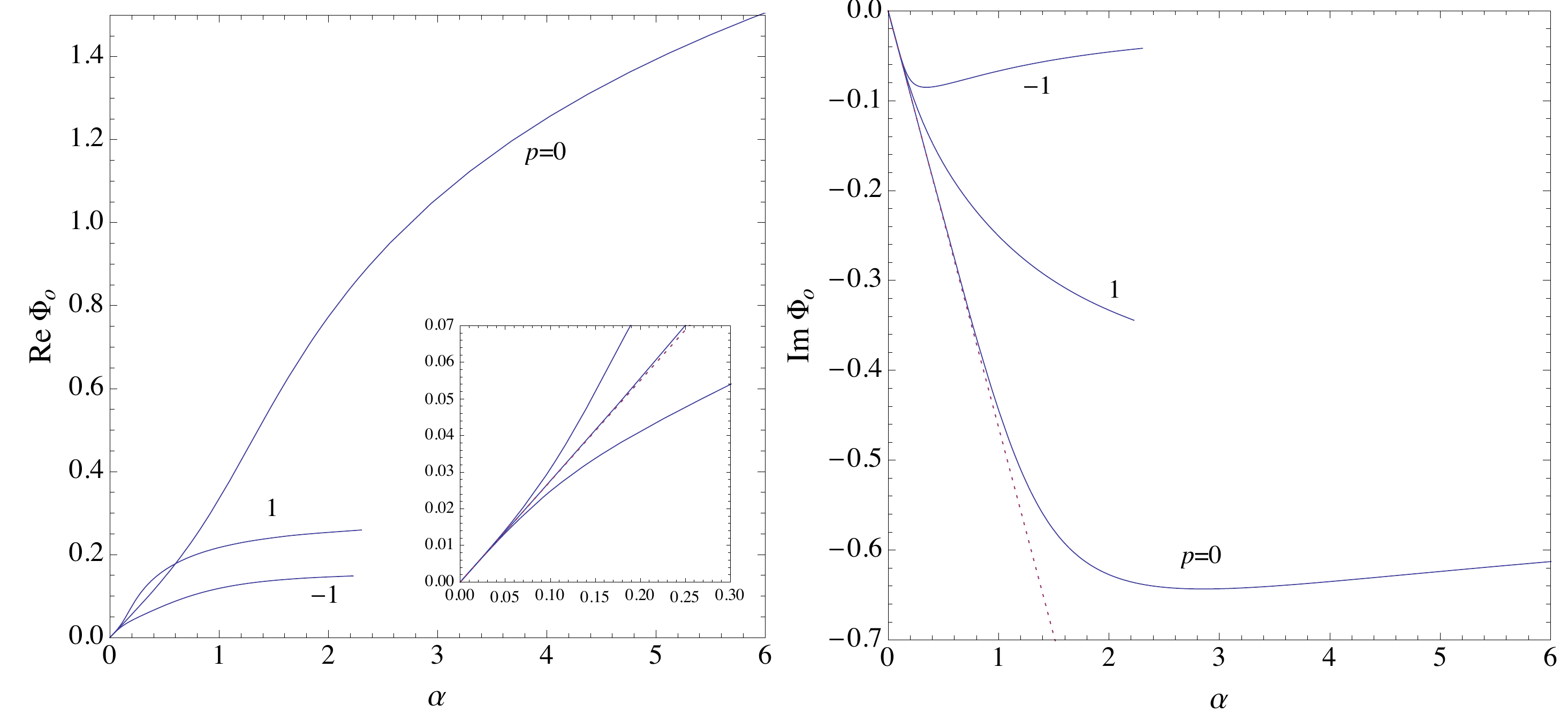}
	\caption{\footnotesize The south pole value $\Phi_o$ of the saddle points in Fig.\ref{LOOP_S2M0p94}, where the saddle points have time contours circling the first singular point under the south pole $0$ and $\pm 1$ times, marked beside the curves. The dotted lines are perturbative results from Eq.(\ref{afaPhiodS}).}  \label{S2PhioLoops}
\end{center}
\end{minipage}\hspace*{3mm}
\begin{minipage}[t]{0.45\textwidth}
\begin{center}	
	 \includegraphics[width=0.5\textwidth]{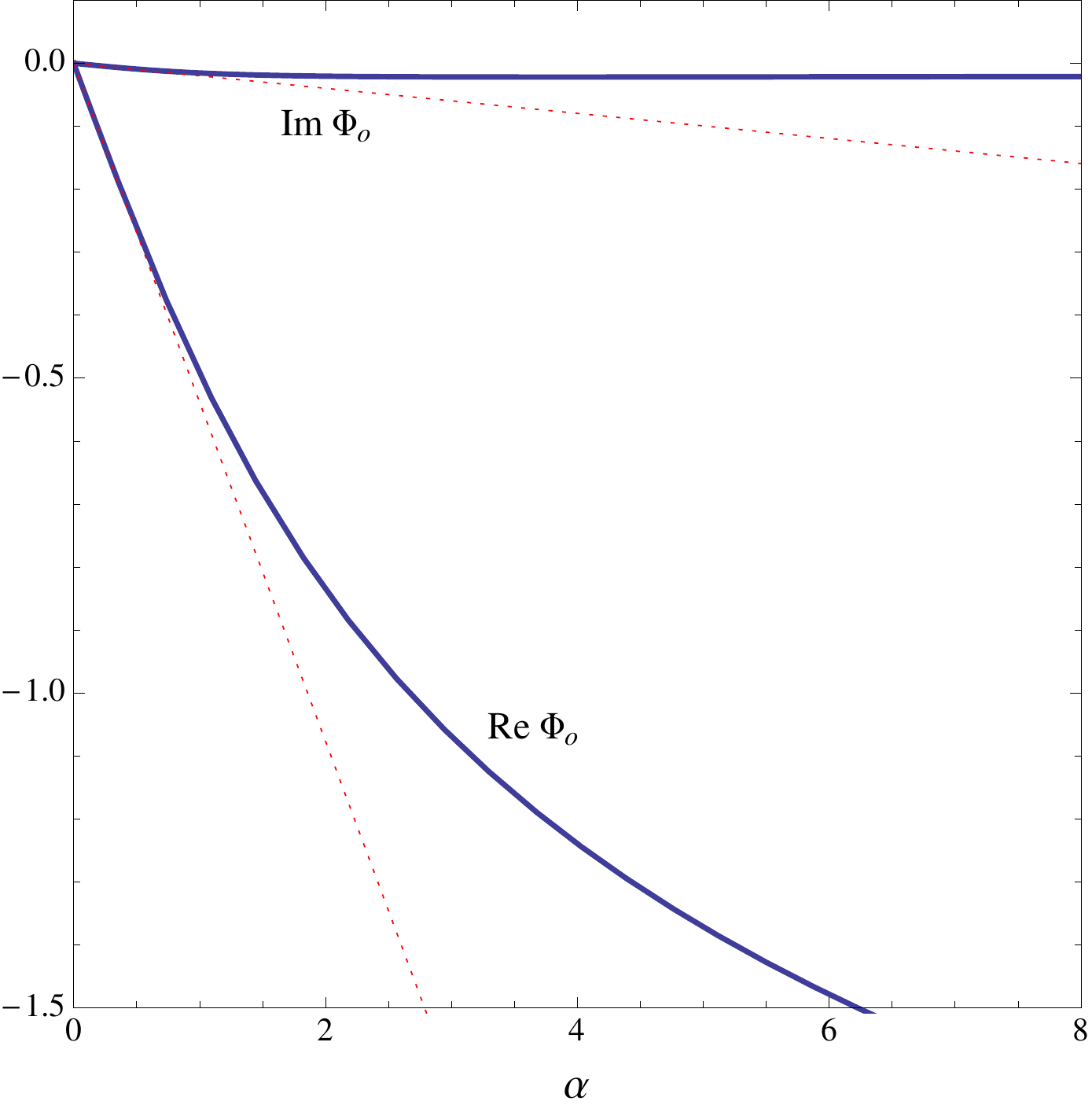}
	\caption{\footnotesize South pole value $\Phi_o$ of the saddle points studied in Fig.\ref{FIG24}. The defining time contour in the perturbative regime is as $\chi_o\rightarrow a\rightarrow c\rightarrow \chi_*'$ ($k=-1$) in Fig.\ref{RiemannSdSd}. The dotted red lines are perturbative results.} \label{S2PhioF2}
\end{center}
\end{minipage}
\end{center}
\end{figure}

\begin{figure}
\begin{center}
	\includegraphics[width=\textwidth]{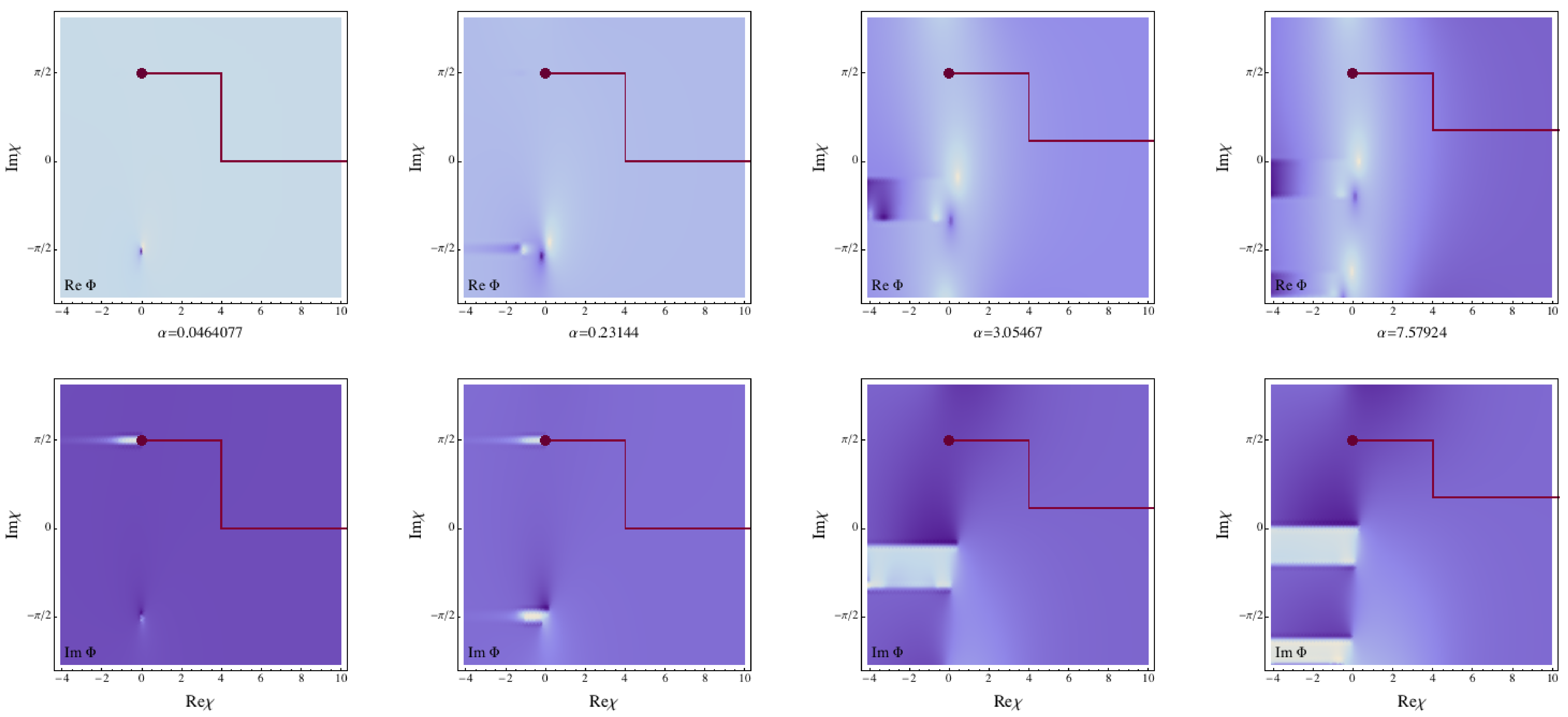}  
\end{center}
\vspace*{-5mm}
\caption{\footnotesize The scalar field profile of the fundamental saddle points of $S^2$ boundary and quadratic scalar potential with scalar mass $m=0.94$. Their time contour in the perturbative regime is given by $\chi_o\rightarrow a\rightarrow b \rightarrow \chi_*$ in Fig.\ref{RiemannSdSd}. In the perturbative regime $\Phi_o\sim \alpha\sim 0$ the singular point at $\chi=-i\pi/2$ is of the pattern of a $+\infty$ point closely attached to a $-\infty$ point, which is just the characteristic of a simple pole of order $1$. This corresponds exactly to the discussion under Eq.(\ref{PoleOrd1}). As $\alpha$ grows, the two opposite infinity points leave away from each other, and very likely there is a branch cut stretching between them. Since the numerical scheme solves the equations along horizontal lines, the branch cuts are all horizontal. However it seems that in the third and the fourth column in the lower figures, the branch cuts at the upper and lower edges of the white bands should be joining each other at infinity.} \label{FIG19}
\end{figure}

\begin{figure}
\begin{center}
	\includegraphics[width=\textwidth]{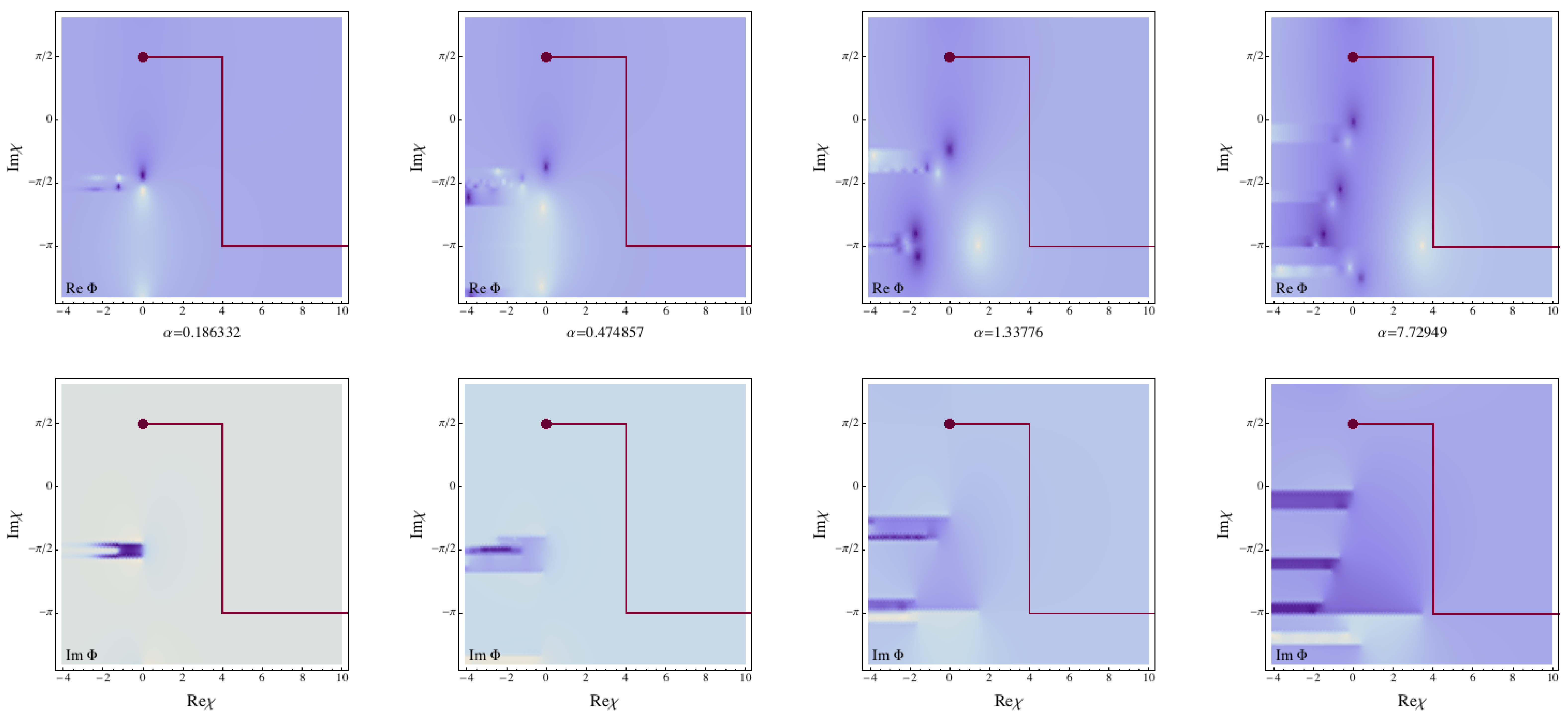}  
\end{center}
\vspace*{-5mm}
\caption{\footnotesize Scalar field profile of the saddle points whose contour is $\chi_o \rightarrow a\rightarrow c\rightarrow \chi_*'$ in the perturbative regime ($\Phi_o\sim \alpha\sim 0$). The scalar potential is quadratic with scalar mass $0.94$. The singular point at $\chi=-i\pi/2$ has the form of an order $1$ pole for small $\alpha$, containing closely neighboring $+\infty$ point and $-\infty$ point. With the increase of $\alpha$ or $|\Phi_o|$, the pole splits into two separate branch points, one of which moves toward the $\chi$-contour. This seems related to the divergence of $\im \beta$ and hence $\tilde {\cal I}$ in Fig.\ref{FIG24}.} \label{FIG25}
\end{figure}

\begin{figure}
\begin{center}
	\includegraphics[width=\textwidth]{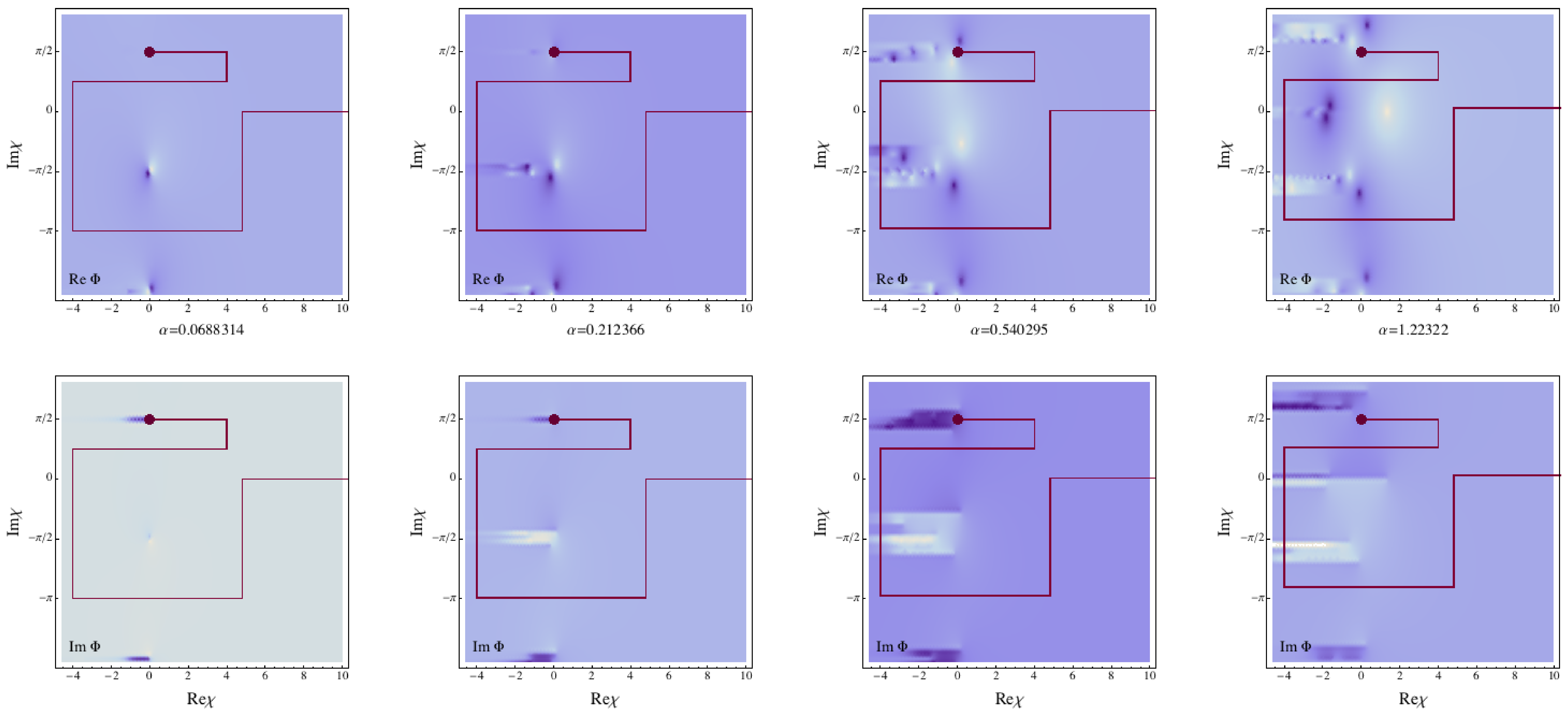}  
\end{center}
\vspace*{-5mm}
\caption{\footnotesize Riemann surfaces for the saddle points studied in Fig.\ref{LOOP_S2M0p94}, where the time contour circles clockwisely once ($p=1$) the first singular point under the south pole when the scalar field is perturbative. The singularity being circled around starts off as a pole of order $1$ (first column), while with the increase of $\alpha$, the $\infty$ point and the $-\infty$ point are teared away, seemingly having a branch cut linking them. One of them (the $\infty$ point) moves towards the positive direction of the Lorentzian time and tends to collide into the $\chi$-contour. This may be related to the tendency of divergence of $\im \beta$ (the $p=1$ curve) in Fig.\ref{LOOP_S2M0p94}.}\label{FIG26}
\end{figure}

\subsection{Numerical results} \label{NUMS2}

The numerical computations follow the same method and logic as those in Sec.\ref{T2NCS} and Sec.\ref{INTC}. That is, they will carry out the search for saddle points subjected to non-perturbative scalar deformation, starting out from the perturbative results obtained in Sec.\ref{PertS2}. When the scalar field is tuned up from perturbative to finite, the deformations caused to the saddle point are traced continuously. Once the family of saddle points is found, further computation will be done to obtain their physical quantities and intrinsic features as function of scalar deformation measured by the boundary data $\alpha$. In the following content, the results for physical quantities of the saddle points, their south pole data, and the Riemann surfaces will be presented. The saddle points involved all have quadratic scalar potential with mass $m=0.94$.

\subsubsection*{Physical quantities}

The numerical calculation cover two situations in parallel with the situations studied in Sec.\ref{T2NCS} for the case of boundary topology $T^2$: saddle points with time contours circling differently around singularities of $\Phi$, and those with time contours containing different amount of Euclidean time. The corresponding contours in the regime of perturbative scalar deformation are shown in Fig.\ref{RiemannSdSd} in green and blue respectively.  

\underline{\it The first situation} involves saddle points whose complex time contours, when scalar deformation is perturbative, takes the form of the green contour shown in Fig.\ref{RiemannSdSd}, with the whole complex history containing minimum amount of positive Euclidean time: $\delta_*=0$. The contour circles the singularity $\chi=-i\pi/2$ from $-1$ time to $+1$ time ($p=-1,0,1$). The case of circling $0$ time corresponds to the fundamental saddle point. The motivation of this computation comes from the perturbative result showing that no branch cut is present on the perturbative level in the Riemann surface of the scalar field; therefore the numerical study can tell whether this property holds when scalar field becomes non-perturbative. The result of this situation is summarized in Fig.\ref{LOOP_S2M0p94}, where I would like to mention here two things worth attention.

\begin{itemize}

\item The absence of branch points in the Riemann surfaces in the perturbative regime, stated below Eq.(\ref{PoleOrd2}), clearly does not persist in the realm non-perturbative scalar perturbation. In the figure it is shown that the results for $p=0,\pm 1$ start off from $\alpha=0$ in the same direction, sharing the same perturbative results (dotted red lines), but soon they divert from one another as $\alpha$ increases. 

\item The ``one-point function'' generation Eq.(\ref{OPFunc}) is verified as shown in the third and fourth frames in the lower row. Fig.\ref{DetMatS2} shows the detail of the verification of Eq.(\ref{OPFunc}) for the fundamental saddle points of $p=0$.

\end{itemize}

\vspace{2mm}

\underline{\it The second situation} involves saddle points with no circling contours around singularity but with more Euclidean time length covered by its whole complex history than the fundamental saddle point. The results are shown in Fig.\ref{FIG24} where the saddle points studied have complex time contours in the regime of scalar perturbation given by the $k=-1$ contour $\chi_o\rightarrow a\rightarrow c \rightarrow \chi_*'$ in Fig.\ref{RiemannSdSd}.  Again as the non-fundamental saddle points studied in Fig.\ref{k1grid} and Fig.\ref{M0p95Floors}, we observe the divergence of $\im \beta$ with the increase of $\alpha$. As a result, ${\cal I}(\alpha)$ is bounded as a function of $\alpha$ while $\tilde {\cal I}$ diverges to $-\infty$ when $\alpha$ increases. Therefore the $k=-1$ saddle points lead to bounded tree-level contribution to the bulk Hartle-Hawking wave function but exponentially divergent contribution to the boundary Hartle-Hawking wave function.

%


\subsubsection*{South pole data $\Phi_o$}

The south pole data here for the case of $S^2$ contains only $\Phi_o$. Fig.\ref{S2PhioLoops} and Fig.\ref{S2PhioF2} show the $\alpha$-$\Phi_o$ relations for the saddle points covered in Fig.\ref{LOOP_S2M0p94} and Fig.\ref{FIG24} respectively. With the relation ${\cal I}=\delta_*-{\pi\over 2}$, the Euclidean shift is not especially plotted since they are already given in Fig.\ref{LOOP_S2M0p94} and Fig.\ref{FIG24}. The noticeable properties is that the small $\alpha$ behavior match the perturbative result Eq.(\ref{afaPhiodS}) with $\rho$ given by Eq.(\ref{rhosp}). Also for saddle points differing in the way the time contours circle around the singularity (Fig.\ref{S2PhioLoops}) different curves share the same perturbative approximation but start to divert when $\alpha$ increases.

\subsubsection*{Riemann surfaces}

The results of tracing Riemann surface deformations are summarized in Fig.\ref{FIG19}, Fig.\ref{FIG25}, and Fig.\ref{FIG26}, which cover the case of fundamental saddle points (Fig.\ref{LOOP_S2M0p94}), $k=-1$ saddle points (Fig.\ref{FIG24}) and $p=-1$ saddle points (Fig.\ref{LOOP_S2M0p94}) respectively. Summarizing the three cases, we have the following points worth attention:
\begin{itemize}

\item For very small $\alpha$ (first columns) we notice the singularity at $\chi=-i\pi/2$, has exactly the look of an isolated pole of order $1$. There is a $+\infty$ spot closely attached to a $-\infty$ spot. This corresponds perfectly to the discussion around Eq.(\ref{PoleOrd2}) in the perturbative context. 

\item Comparing Fig.\ref{FIG19} and Fig.\ref{FIG26}, when $\alpha$ increases we see clearly on the level of Riemann surfaces that the two families of saddle points ($p=0$ and $p=-1$) are after all different, although perturbatively they are the same. 

\item In Fig.\ref{FIG25} and Fig.\ref{FIG26}, when $\alpha$ increases we observe a singular point moves towards the positive direction of Lorentzian time with the tendency of collision with the $\chi$-contour. Meanwhile we notice that $\im \beta$ for these two cases tend to diverge. On the other hand in Fig.\ref{FIG19} for the fundamental saddle points, all singularities stick well close to the $\im \chi$-axis, and meanwhile the corresponding $\im\beta$ curve in Fig.\ref{LOOP_S2M0p94} ($p=0$) asymptotes to $0$ when $\alpha$ increases. This  conforms to the speculation in Sec.\ref{INTC} that divergent $\im \beta$ is accompanied by a singular point of $\Phi$ that moves towards the positive direction of Lorentzian time with the increase of $\alpha$, and which tend to collide into the integral contour of the equations of motion. 

\end{itemize}

\section{Further issues in brief} \label{FTIS}

%

This section browses through two topics which are in the immediate continuation of the line of thinking of the work being presented so far, in order to give a preview of the possible direction that future work can be oriented to as well as making up for some important aspect that the previous sections did not cover. 


Sec.\ref{NoQuPsec} will show numerically that the holographic renormalization results Eqs (\ref{HolRen2}) and (\ref{Substr2}) are valid for non-quadratic potentials. Sec.\ref{S3BrchAB} will try to extend the holographic renormalization in Sec.\ref{NPS2Act} for boundary topology $S^2$ to higher dimensions $S^d$ ($d=3,4,\dots$), where attention will be focused on the renormalization of the imaginary part of the saddle-point action. An application of the result to the case of $S^4$ will be shown.

\subsection{Potentials containing quartic or higher order terms} \label{NoQuPsec}

The numerical verification of the holographic renormalization via Eq.(\ref{OPFunc}) has so far been carried out for quadratic scalar potentials only, while the derivation only required the potential to take the form $\ell^2V(\Phi)={1\over 2}m^2\Phi^2+O(\Phi^4)$ as in Eq.(\ref{Vph2O}).  This subsection shows with two examples of non-quadratic scalar potentials that the holographic renormalization works beyond quadratic level. 
The potentials tested are: the quadratic potential (for reference); the $\Phi^4$ potential, containing a quartic term; and also the $\cosh$ potential, containing an infinite series of even order terms in $\Phi$. They will be denoted by $V_0$, $V_1$ and $V_2$ respectively:
\begin{align}
	\ell^2V_0(\Phi)={m^2\over 2} \Phi^2;\ \ \ \ \ell^2V_1(\Phi)={m^2\over 2} \Phi^2+{m^2 \over 4!} \Phi^4;\ \ \ \ \ell^2V_2(\Phi)=m^2(\cosh\Phi -1). \label{VVV}
\end{align}
The potentials are chosen as such in order to let all three share the same quadratic term, and let $V_1$ and $V_2$ have the same quartic term.

\begin{figure}
\begin{center}
	\includegraphics[width=\textwidth]{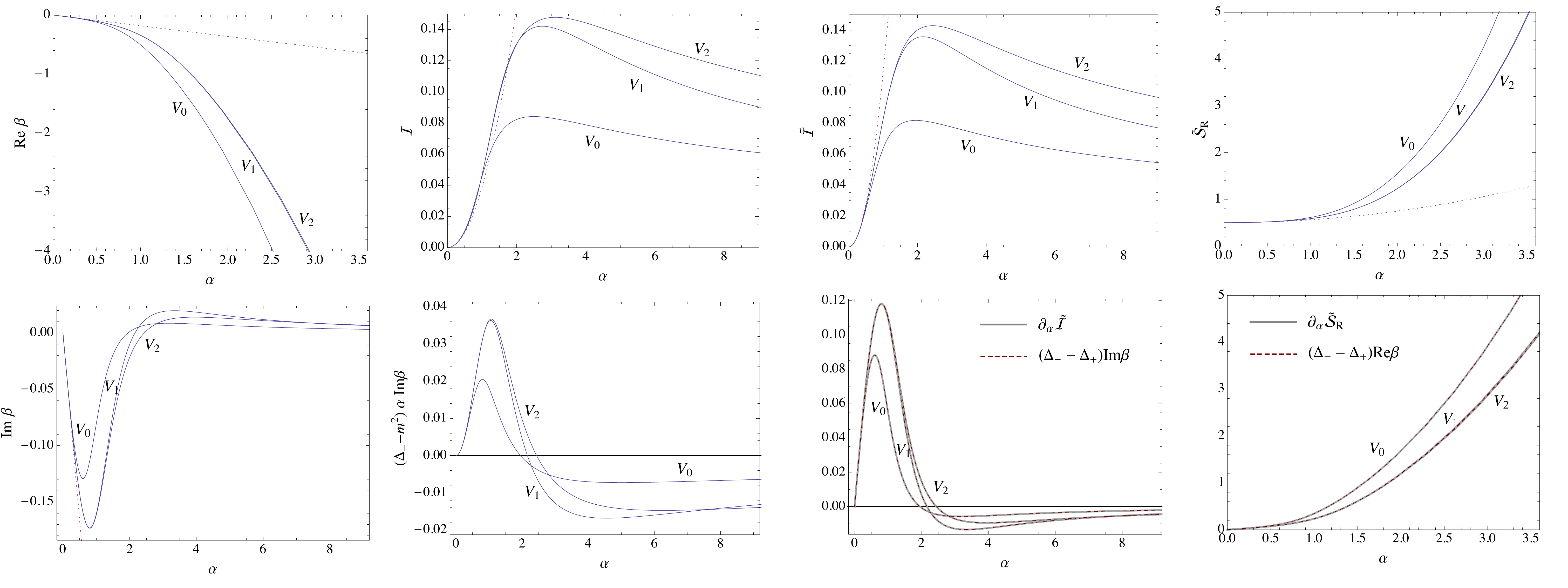} 
\end{center}
\vspace*{-5mm}
\caption{\footnotesize Results of saddle points of boundary topology $T^2$, comparing quadratic and non-quadratic scalar potentials. The contents are organized in the same way as in Fig.\ref{k0grid}. The potentials $V_0$,  $V_1$ and $V_2$ are given by Eq.(\ref{VVV}), and the scalar mass is $m=0.94$. In each frame, the three curves start off from $\alpha=0$ with almost the same behavior. This is because their potentials have the same quadratic term, and the leading order approximation of the curves are determined by the quadratic term. Therfore the perturbative results, represented by the dotted red lines, fit all three curves in the frames where they apply. When $\alpha$ becomes larger, the differences begin to show up. First it is $V_0$-curves that split from the $V_1$ and $V_2$-curves when the effect of quartic term becomes important, and then $V_1$ and $V_2$-curves begin to divert from each other when the terms beyond quartic terms in $V_2$ become important. The general pattern is that larger potential tends to enhance the value of $\cal I$ as well as $\tilde {\cal I}$. The result also shows that the relation of ``one-point function'' generation Eq.(\ref{OPFunc}) is satisfied.} \label{T2NoQuM0p94}
\end{figure}

\begin{figure}
\begin{center}
	\includegraphics[width=\textwidth]{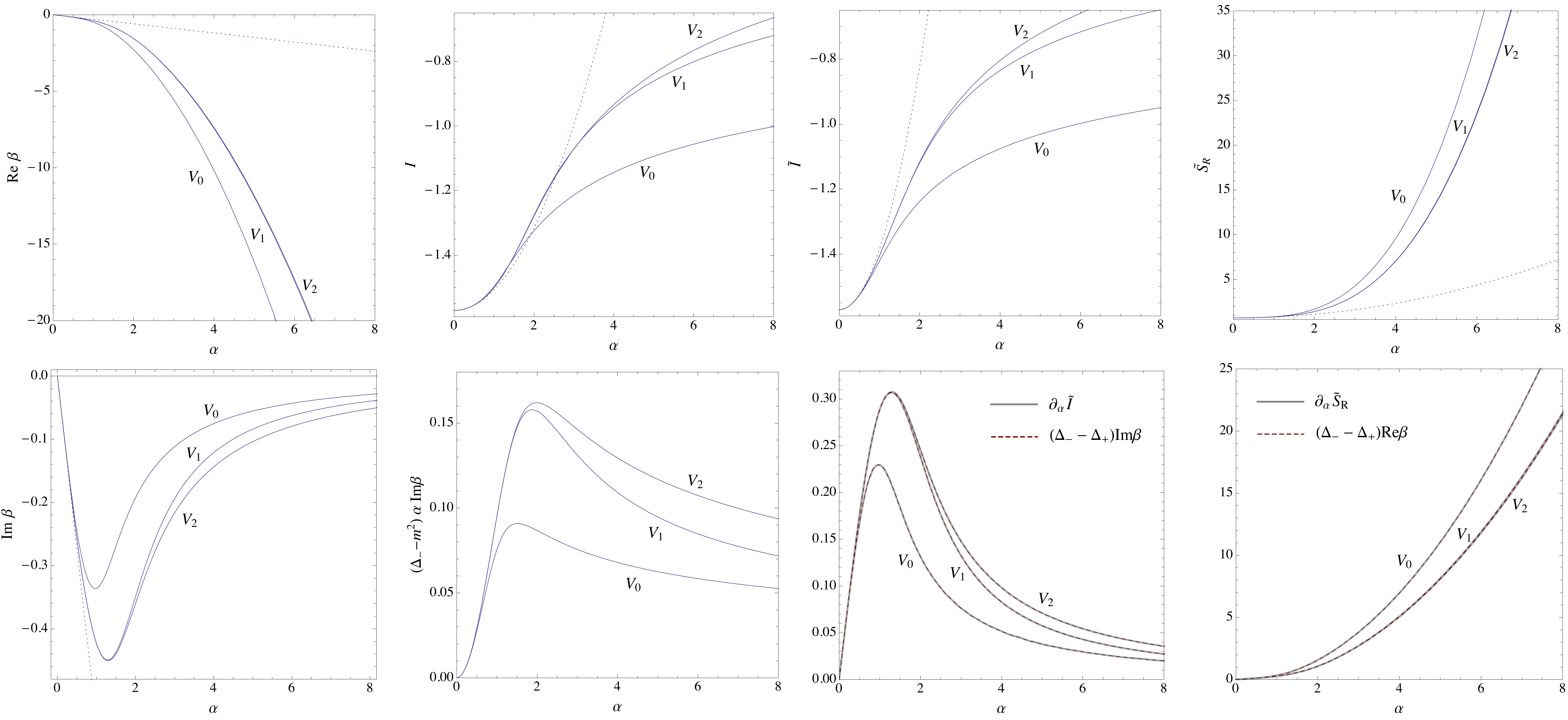} 
\end{center}
\vspace*{-5mm}
\caption{\footnotesize Results of saddle points of spatial topology $S^2$, comparing quadratic and non-quadratic scalar potentials. Except for the boundary topology, all other settings are the same as in Fig.\ref{T2NoQuM0p94}. Here we observe the similar feature of diversion of the three curves when $\alpha$ increases. Also we have Eq.(\ref{OPFunc}) verified.} \label{S2NoQuM0p94}
\end{figure}

The numerics have been done to both boundary topologies of $T^2$ and $S^2$, but only for the fundamental saddle points for simplicity. The results of physical quantities are presented in Fig.\ref{T2NoQuM0p94} and Fig.\ref{S2NoQuM0p94} respectively. The content are organized in the same way as results are presented in Sec.\ref{T2NCS}. The plots show that all three curves start off from $\alpha=0$ in much the same way, sharing the same leading order approximation as shown by the dotted red lines. Then as $\alpha$ increases, the higher and higher order terms in the potentials start to enact, and so we see first $V_0$ curves start to divert from $V_1$ and $V_2$ curves and then $V_1$ and $V_2$ curves start to divert from each other.

\subsection{Holographic renormalization extended to higher dimensions} \label{S3BrchAB}

In Sec.\ref{PertS2} the perturbative results of the saddle-point actions were derived for boundary of topologies $S^d$ ($d=2,3,\dots$) presented in Eqs (\ref{PT145}) and (\ref{PT146}), while later in Sec.\ref{NPS2Act} the holographic renormalization for non-perturbative scalar deformation was worked out only for $S^2$ in Eqs (\ref{Substr2}) and (\ref{Substr3}). It seems that a generalization to the boundary topologies $S^d$ ($d=3,4,\dots$) can be very ready worked out, but there is actually the difficulty that analyzing the IR divergence becomes much more complicated than for $d=2$. However in the context of no-boundary quantum cosmology where saddle-point actions are generally complex, a partial generalization is still possible: it is possible to work out $\im\! {\cal S}_{\rm ct}$, and hence we can obtain $\tilde {\cal I}$ from $\cal I$. 

The derivation is to a great extent a repeating of the steps for obtaining Eq.(\ref{Substr2}) while the important difference is that when applying the formula Eq.(\ref{CTTM}) for counter terms, $\cal A$ is understood as the volume of the boundary $a_*^d$ and the leading counter term $\cal A$ should acquire a coefficient $d-1$: ${\cal A}\rightarrow (d-1)a_*^d$ \cite{Balasubramanian:1999re,Balasubramanian:2001nb}. The next-to-leading term is still ${\Delta_-\over 2} {\cal A} \varphi^2= {\Delta_-\over 2}a_*^d \varphi^2$. The further subleasing terms are more difficult to obtain, but in fact to work out $\tilde {\cal I}$ does not need the knowledge about these terms, since they do not contribute to $\tilde  {\cal I}$. Therefore $\tilde {\cal I}$ and ${\cal I}$ differ by the imaginary part of $(d-1)a_*^d+{\Delta_-\over 2}a_*^d \varphi^2$. Using the asymptotic expansions Eqs (\ref{asy196}) and (\ref{asy197}), we obtain\footnote{The result in the $\beta$-representation is $\tilde {\cal I}(\beta)={\cal I}(\beta)+\left( \Delta_+ -{2m^2\over d}\right) \beta\, \im\alpha$.}
\begin{align}
	\tilde {\cal I}(\alpha)={\cal I}(\alpha)+\left( \Delta_- -{2m^2\over d}\right) \alpha\, \im\beta. \label{IdIdtd}
\end{align}
This formula is valid for scalar potentials of the form $V(\Phi)={1\over 2} \ell^{-2}m^2 \Phi^2+O(\Phi^4)$, and scalar mass within the range ${d^2-1 \over 4}<m^2 <{d^2 \over 4}$, where the formalisms in appendix \ref{ABAsydS} are valid.

\subsubsection*{A model of boundary topology $S^4$}

In the following I show an application of this result on the model of scalar-deformed dS$_5$ cosmology, where the scalar field is minimally coupled and has quadratic potential. The scalar mass is $m=1.96$, which is within the range where Eq.(\ref{IdIdtd}) is valid. All the technicalities used in Sec.\ref{COSMOS2} can be very easily transplanted here. The most important difference however, is that this time we have to do the bulk saddle-point action $\cal I$ directly using the integral formula: Eq.(\ref{OSA183}) with $d_1=4$ and $d_2=0$. This sometimes makes the numerical error difficult to control. 

An intriguing aspect of this model is that we can very easily see that there are families of saddle points which do not allow the scalar field to be switched off. That is, when $\alpha$ vanishes, $\beta$ and $\Phi_o$ are still finite, unlike the saddle points studied in the previous sections all of which have $\alpha \propto \beta \propto \Phi_o \sim 0$ when $\alpha$ approaches $0$. This feature is also found in scalar-deformed dS$_4$ models \cite{RMYV}. Following the same pattern that the results are presented in Sec.\ref{T2NCS} and Sec.\ref{INTC}, I show the results of two families of saddle points of boundary topology $S^4$, where one family contains and the other does not contain a perturbative regime, and they are labeled by A or B respectively. 

\begin{figure}
\begin{center}
	\includegraphics[width=\textwidth]{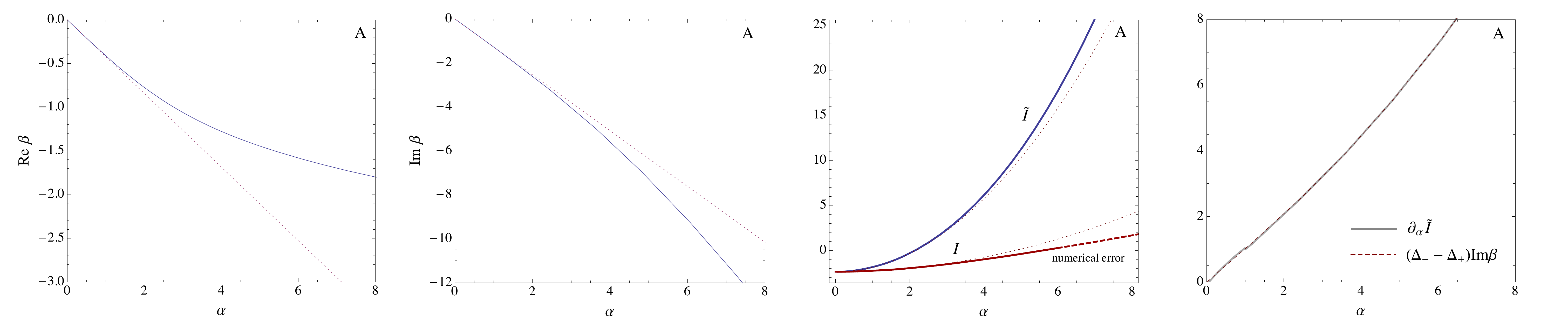}  	\includegraphics[width=\textwidth]{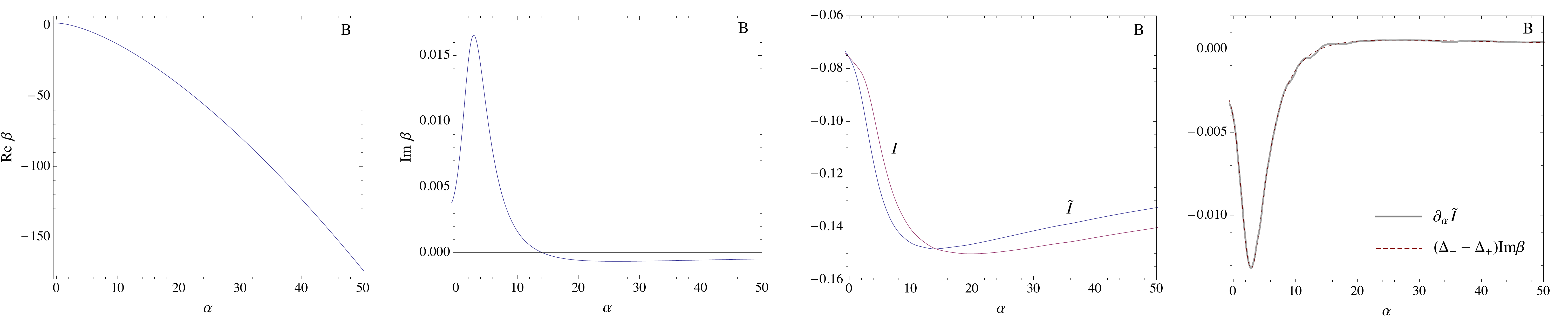}  
\end{center}
\vspace*{-5mm}
\caption{\footnotesize Results of two distinct families, labeled by A and B, of saddle points of boundary topology $S^4$. They are minimally coupled to a scalar field of quadratic potential and mass $m=1.96$. The family A (upper row) contains a perturbative regime where its complex time contour is $\chi_o\rightarrow a\rightarrow b\rightarrow \chi_*$ ($k=0$) in Fig.\ref{RiemannSdSd}. The family B (lower row) does not contain a perturbative regime. The four columns show respectively the $\alpha$-$\beta$ curves in the first two columns, the imaginary part of the saddle-point actions in the bulk ($\cal I$) and on the boundary ($\tilde {\cal I}$) related by Eq.(\ref{IdIdtd}) in the third column, and the generation of 1-point function by the boundary saddle-point action, i.e., the imaginary part of Eq.(\ref{OPFunc}). In the fourth column there are some wiggles in the $\partial_\alpha \tilde{\cal I}$ curves, and this is very probably because the bulk on-shell action $\cal I$ is computed by direct integration, which is numerically less stable than the indirect way used in $T^2$ and $S^2$ cases. The find dotted red lines in the upper row are perturbative results Eqs (\ref{afaPhiodS})--(\ref{PT146}).} \label{BRCHAB}
\end{figure}

\begin{figure}
\begin{center}
	\includegraphics[width=\textwidth]{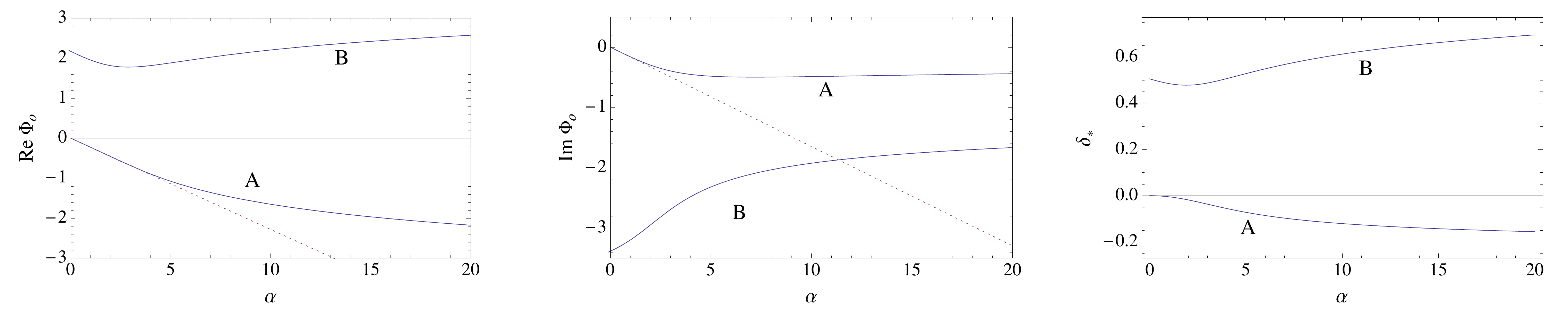} 
\end{center}
\vspace*{-5mm}
\caption{\footnotesize Defining parameters of the no-boundary saddle points of spatial topology $S^4$ with scalar mass $m=1.96$. The physical quantities are shown in Fig.\ref{BRCHAB}. The family B does not cover a perturbative regime, which is seen from the fact that when $\alpha\sim0$, $\Phi_o$ is finite. For about $0<\alpha<2$ we observe $|\Phi_o|$ decreasing with the increase of $\alpha$ for branch B, which is not seen for other cases studied so far. The red dotted lines are results of scalar perturbation Eq.(\ref{afaPhiodS}), applying only to family A. We see in the third frame that the family B, which cannot go perturbative, contains less Euclidean history and in fact it is the family of fundamental saddle points.} \label{PHIoDT*}
\end{figure}

\begin{figure}
\begin{center}
	\includegraphics[width=\textwidth]{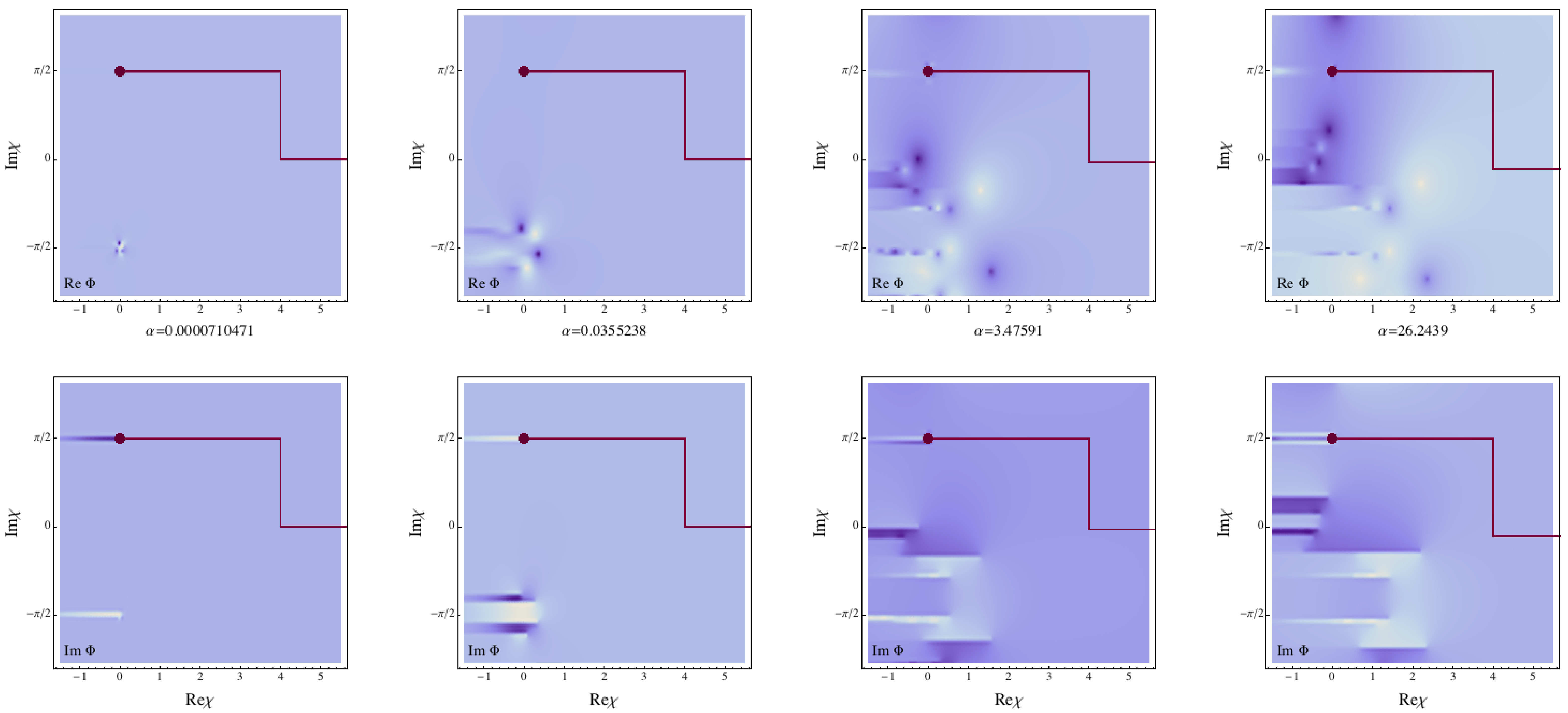}  
\end{center}
\vspace*{-5mm}
\caption{\footnotesize The scalar profile of the saddle points of boundary topology $S^4$ in the family A, whose physical quantities are presented in the upper row in Fig.\ref{BRCHAB}. In the perturbative realm as shown in the first, the singular point at $\chi=-i\pi/2$ has the characteristics of a pole of order $3$, where we see thee $-\infty$ points clustered with three $+\infty$ points. Then with the growing of $\alpha$ the system becomes non-perturbative, the $\pm \infty$ points are separated away, each becomes a branch point. Meanwhile one of them is attracted the more and more towards the integral contour of the equations of motion, which seems to be related to the divergent behavior of $\im \beta$ shown in Fig.\ref{BRCHAB}. } \label{RSBchA}
\end{figure}

\begin{figure}
\begin{center}
	\includegraphics[width=\textwidth]{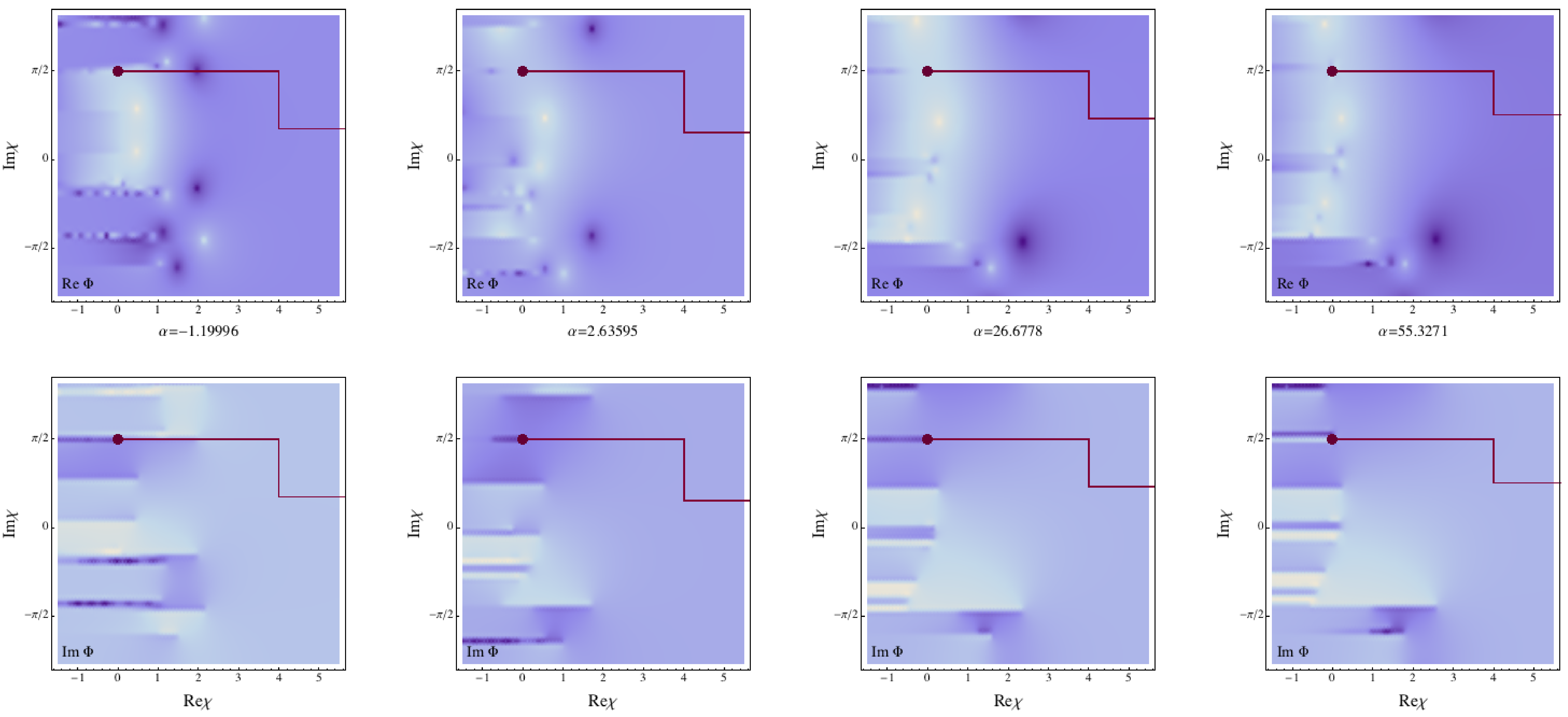}  
\end{center}
\vspace*{-5mm}
\caption{\footnotesize The scalar profile of the saddle points of boundary topology $S^4$ in family B. The physical qualities of these saddle points are presented in the lower row of Fig.\ref{BRCHAB}. 
With the $\chi$-contour configured as in these plots, $\alpha$ cannot decrease below about $0$ since a singularity (dark spot) collides into the first Lorentzian segment of the contour from above. Then when $\alpha$ increases, there is a branch point (bright spot) to the left of the Euclidean part of the contour that moves towards the right  as shown in the second column, and then falls back to the $\im \chi$ axis. This seems to correspond to the presence of the peak of the $\alpha$-$\im \beta$ curve at $\alpha \sim 3$ in Fig.\ref{BRCHAB}. As $\alpha$ continues to grow, no singular point tends to collide into the $\chi$-contour from the left, and this well corresponds to the behavior of $\im \beta$ for large $\alpha$, which asymptotes to $0$ as shown in the lower row of Fig.\ref{BRCHAB}. }\label{RSBchB}
\end{figure}

The physical quantities are shown in Fig.\ref{BRCHAB}, the south pole data $\Phi_o$ and the Euclidean shifts $\delta_*$ are shown in Fig.\ref{PHIoDT*} and the Riemann surfaces of the scalar field of the two families of saddle points are shown in Fig.\ref{RSBchA} and Fig.\ref{RSBchB}. Here are the features relevant to our purpose:

\begin{itemize}

\item The  saddle points of family A are obtained by augmenting the scalar deformation in the perturbative saddle points with contour $\chi_o\rightarrow a\rightarrow b\rightarrow \chi_*$ ($k=0$) shown in Fig.\ref{RiemannSdSd}. In 3d when we do so, we obtain fundamental saddle points, but here they are not fundamental. From the behavior of the Euclidean shifts $\delta_*$ in Fig.\ref{PHIoDT*}, we see that saddle points of family B actually experience less Euclidean time, and in fact they are the fundamental saddle points of the $S^4$ model. 

\item The behavior of $\Phi_o$ in Fig.\ref{PHIoDT*} shows that for the family A, when $\alpha\sim 0$ so does $\Phi_o$. However this is not the case for family B, showing that family B does not cover a perturbative regime of the scalar deformation where $\alpha\propto \Phi_o\sim 0$. This can also be seen from the $\alpha$-$\beta$ relation in Fig.\ref{BRCHAB} where we have $\alpha\propto \beta\sim 0$ for family A but not for family B.

\item The relation of ``one-point function'' generation Eq.(\ref{OPFunc}) is verified for the imaginary part, seen from the forth column of Fig.\ref{BRCHAB}. Therefore $\tilde {\cal I}$ obtained from Eq.(\ref{IdIdtd}) is indeed the imaginary part of the boundary saddle-point action. There are wiggles in the curve $\partial_{\alpha} \tilde{\cal I}$, which is very probably due to the numerical error introduced in the computation of $\cal I$ by direct integral Eq.(\ref{OSA183}) with $d_1=4$ and $d_2=0$. 

\item The behaviors of $\im\! \beta$ in Fig.\ref{BRCHAB} have the same feature as the previous models of boundary topologies $T^2$ and $S^2$. When $\alpha$ increases $\im \beta$ asymptotes to $0$ for family B the fundamental saddle points, while for family A of non-fundamental saddle points, $\im\! \beta$ diverges. 

\item As shown in Fig.\ref{BRCHAB}, for both families A and B, $\cal I$ are bounded. For family A, $\tilde {\cal I}$ diverges to $+\infty$, while for family B, $\tilde {\cal I}$  and $\cal I$ differ very little, as expected of the fundamental saddle points. Therefore in the boundary wave function, the contribution from saddles of family A is exponentially suppressed when $\alpha$ increases (in case higher-loop contributions are suppressed). 

\item There are also families of saddle points giving rise to $\tilde {\cal I}$ diverging to $-\infty$ when $\alpha$ increases. For example those that have $k=1$ contour (blue contour in Fig.\ref{RiemannSdSd}, with the final Lorentzian segment along $\im\chi=\pi$) when scalar deformation is perturbative. Such saddle points can lead to exponentially divergent contribution to the Hartle-Hawking wave function.  It is not necessary to show this case in detail because all the physical quantities are just the opposite of those of the family A in Fig.\ref{BRCHAB}. 

\item For the deformation of the Riemann surfaces of $\Phi$, Fig.\ref{RSBchA} shows that for saddle points of family A, with the increase of $\alpha$, a singular point moves the more and more to the right and tends to collide into the $\chi$-contour (it is really the case if we further increase $\alpha$), while it is not the case in Fig.\ref{RSBchB}. Meanwhile we already noticed in Fig.\ref{BRCHAB} that $\im \beta$ diverges for family A, while asymptotes to $0$ for family B, when $\alpha$ increases. This again suggests the connection between the divergence of $\im \beta$ and the presence of a singularity taking off from the $\im\! \chi$-axis and moving towards the $\chi$-contour. 

\item For both families A and B, we can extend $\alpha$ to negative values. For family A the extension is obvious: ${\cal I}(\alpha)$, $\tilde {\cal I}(\alpha)$ and $\delta_*(\alpha)$ are even functions, while $\beta(\alpha)$ and $\Phi_o(\alpha)$ are odd; but for family B these quantities do not have definite parity as function of $\alpha$. Indeed Fig.\ref{RSBchB} (especially the first two columns) shows that when $\alpha$ drops towards $0$, there is a singularity (the dark spot) colliding into the first Lorentzian segment of the $\chi$-contour from above, preventing $\alpha$ from decreasing to negative values. It is possible to further adjust the $\chi$-contour to let it avoid this singularity so as to further decrease $\alpha$. However, I hope to leave this to future work when the physical meaning of the result is better figured out.

\end{itemize}


\section{Summary and discussion} \label{DCS}

\subsubsection*{\underline{\it What is done}}

This paper studied several models of no-boundary quantum cosmology in a dS/CFT holographic setting, computing the Hartle-Hawking wave function on the asymptotic boundary based on bulk computations and holographic renormalization. The goal is to address the non-normalizability problem of the Hartle-Hawking wave functions obtained by direct dS boundary computations. The models studied are mainly 3d universes governed by Einstein gravity, with a positive cosmological constant and a minimally coupled scalar field, and they have either $T^2$ or $S^2$ as the topology of asymptotic spacelike boundary.

An extensive search for the probable contributing saddle points has been carried out in the bulk, where mini-superspace formalism is used for this search to be feasible (at the cost of losing saddles). The admission criteria for the saddle points is the no-boundary proposal: they should be compact spacetimes accommodating on-shell fields that satisfy the wanted boundary conditions. The search was first carried out for perturbative scalar deformation (Sec.\ref{PERTT2} and Sec.\ref{PertS2}) and then traced all the way to finite scalar deformation (Sec.\ref{NUMT2} and Sec.\ref{NUMS2}). 

For some selected saddle points, supposedly representative, their saddle-point actions are computed first directly in the bulk ${\cal S}={\cal S}_{\rm R}+i\, {\cal I}$ and then on the asymptotic boundary $\tilde {\cal S}=\tilde {\cal S}_{\rm R}+i \, \tilde {\cal I}$ [cf. Eqs (\ref{ReSImSSplit}) and (\ref{RItSSplit})], as functions of scalar deformation $\alpha$ as in Eqs (\ref{Phi*40}) and (\ref{afbt137}), and of boundary geometry $\tau_2$ as in Eq.(\ref{hijAT13}) in case the boundary geometry is $T^2$. Therefore the individual tree-level contribution of each saddle is $e^{i{\cal S}}=e^{i{\cal S}_{\rm R}}e^{-{\cal I}}$ to the bulk Hartle-Hawking wave function, and $e^{i\tilde {\cal S}}=e^{i\tilde {\cal S}_{\rm R}}e^{-\tilde {\cal I}}$ to the boundary one. I have resorted to the non-trivial amplitudes $e^{-\cal I}$ and $e^{-\tilde {\cal I}}$ for indications of the normalizability property of the Hartle-Hawking wave functions.

One key step in the above part of the work is the holographic renormalization which computes $\tilde {\cal S}$ from $\cal S$. Technically it is just a dS application of the standard AdS/CFT holographic renormalization, where one finds out the counter term action ${\cal S}_{\rm ct}$ which cancels the IR divergences in $\cal S$, so as to obtain $\tilde {\cal S}={\cal S}+{\cal S}_{\rm ct}$. This is done in detail in Sec.\ref{HR_OSA}, and less in detail in Sec.\ref{NPS2Act}. The results are applicable to the case of finite homogeneous scalar deformation.

Finally, a partial generalization of holographic renormalization to higher dimensional models of boundary topologies $S^d$ ($d=3,4,\dots$) is realized. In particular, the quantity explicitly worked out is the imaginary part of the counter term action, and therefore we are able to compute $\tilde {\cal I}={\cal I}+\im {\cal S}_{\rm ct}$, and hence to infer from $e^{-\cal I}$ and $e^{-\tilde {\cal I}}$ the normalizability properties of the Hartle-Hawking wave functions.

\subsubsection*{\underline{\it What is found}}

In the search for saddle points, the coordinate time in the mini-superspace is necessarily complex, so that the whole coordinate time history of a saddle point is depicted by a generic curve on the complex time plane. This is in fact already well-known in the literature.

For a given set of boundary condition, expressed in terms of scalar deformation and boundary geometry $(\alpha,\tau)$, an infinite number of saddle points exist which can be described by mini-superspace formalism. These saddle points found are distinguished according to their complex time contours. Different saddle points can have a contour covering different amount of Euclidean time (blue contours in Fig.\ref{ctNPafPaf} and Fig.\ref{RiemannSdSd}) and/or winding differently around the singularities (comparing blue and green contours in the same figures) in the Riemann surface of the scalar field. 

Among all saddle points figured out, we can single out one family which, in the literature where the Hartle-Hawking wave functions are computed in the bulk, are very commonly considered as the contributing saddle points. It is the one whose complex time contour covers a minimum amount of positive Euclidean history and does not circle around singularities of the scalar field (contour $\chi_o\rightarrow a\rightarrow b\rightarrow \chi_*$ in figures \ref{ctNPafPaf} and \ref{RiemannSdSd}). The second property means that the complex time contour always lies in the same layer of the Riemann surface. These saddle points are referred to as ``fundamental'' in this paper. All other saddle points are thus called non-fundamental.

When the actions are computed for the fundamental saddle points, it is found that as function of $\alpha$, $\cal I$ and $\tilde {\cal I}$ differ very little and are bounded as function of $\alpha$ (figures \ref{k0grid},\ref{SSMatchDetail},\ref{DetMatS2},\ref{T2NoQuM0p94},\ref{S2NoQuM0p94},\ref{BRCHAB}); and as function of boundary geometry $\tau_2$ when the boundary topology is $T^2$, they both tend to $+\infty$. Therefore no remarkable implication can be drawn from these results concerning the normalizability property of the Hartle-Hawking wave function.

For the actions of the non-fundamental saddle points with time contours not circling singularities in the Riemann surface of the scalar field (blue contours with $k\neq0$ in figures \ref{ctNPafPaf} and \ref{RiemannSdSd}), but containing non-minimum length of Euclidean history, it turns out that $\cal I$ and $\tilde {\cal I}$ can differ violently from each other.
For all such cases studied, the imaginary part of bulk action $\cal I$ is bounded while the boundary counterpart $\tilde {\cal I}$ diverges towards $\pm \infty$ as a function of scalar deformation $\alpha$ (figure \ref{k1grid},\ref{M0p95Floors},\ref{FIG24}). For boundary topology $T^2$, it is also shown that the ``high temperature'' ($\tau_2\rightarrow \infty$) behavior is the same as what the scalar perturbation results tell: $\cal I$ and $\tilde {\cal I}$ tend simultaneously to $+\infty$ or $-\infty$. Thus this class of non-fundamental saddle points can probably be responsible for both the scalar divergences and the high temperature divergences of the Hartle-Hawking wave function (in case they contribute). However we should notice that these divergences do not look quite the same as what is found in \cite{Anninos:2012ft}.

There is another category of non-fundamental saddle points studied whose complex time contours circle around singular points in the Riemann surface of the scalar field (green contours in Fig.\ref{ctNPafPaf} and Fig.\ref{RiemannSdSd}). This part is more challenging due to the difficulty in adjusting the contours. Still some results are obtained just enough to claim that for boundary topology $T^2$, different ways of circling around singularities lead to different saddle points as in Fig.\ref{M0p95loops}; and for boundary topology $S^2$, while the perturbative analysis shows that letting time contours circling the singularities of $\Phi$ does not result in new saddle points, numerical results as presented in Fig.\ref{LOOP_S2M0p94} show that it is not the case.

Intriguing connection has been empirically noticed between the intrinsic characteristics of the saddle points and the scalar divergence of their contribution to the Hartle-Hawking wave function. It seems that the divergence of $\tilde {\cal I}$ is always accompanied by a singular point in the Riemann surface of the scalar field, which with the increase of $\alpha$, moves the further and further in the positive direction of Lorentzian time, and which tend to collide into the integral contour of the equations of motion (see figures \ref{DPk1},\ref{FIG25},\ref{FIG26},\ref{RSBchA}). A closer look shows that this singularity seems to be connected to another singularity by a branch cut, and that with the increase of $\alpha$ this branch cut is stretched longer and longer. It is not clear especially, if this is just a mathematical feature or it can have some physical meaning.

The perturbative results obtained in Sec.\ref{PERTT2} and Sec.\ref{PertS2} give the correct leading order approximation of the numerical results. This is seen in the many figures (not density plots) where the red dotted lines well overlap solid lines when $\alpha\sim 0$. However the drastic difference between $\cal I$ and $\tilde {\cal I}$ for non-fundamental saddle points cannot be revealed by perturbative results.

The formal holographic renormalizations Eqs (\ref{HolRen2}) and (\ref{Substr2}) have passed the test of ``one-point function'' generation, in that Eq.(\ref{OPFunc}) has been verified numerically for all the saddle points covered by numerical calculation.

\subsubsection*{\underline{\it Issues which need further discussion}}


In the paper, all the claims of scalar/``temperature'' divergence of the Hartle-Hawking wave function are drawn from the investigation of individual saddle-point contributions. The problem is that, whereas plenty of saddles are found, we do not have any clue which ones are actually picked up by the Hartle-Hawking path integral. Considerable amount of studies on this subject have been done in mini-superspace cosmological models of empty spacetime \cite{Halliwell:1989dy,Halliwell:1988ik},\footnote{The same computations can be almost trivially done to the 3d models in this paper setting $\Phi=0$, and similar conclusions can be reached.} showing that which saddle points contribute depends on the initial choice of path integral contour, and that there can be a natural choice of initial contour, which ends up picking up only part of saddles. However it is not clear to which degree of precision the result from mini-superspace formalism can provide guideline in the situation where full gravitational degrees of freedom should be taken into account. Especially the situation becomes quite complicated with the presence of a scalar field. Therefore this paper simply assumes that every saddle point may contribute and studies what the consequence can be if a certain saddle point eventually contributes. Maybe in general, due to the difficulty in directly operating the path integrals of gravity, a plausible alternative is to let the boundary field theory result tell which bulk saddles really contribute to the Hartle-Hawking wave function.


Furthermore, the drastic differences between $\cal I$ and $\tilde {\cal I}$ for certain saddle points seems to imply that in the passage from bulk to boundary, or from $\Psi$ to $\tilde \Psi$, the holographic renormalization changes the normalization property of the wave functions. This sounds contradictory to quantum mechanics, because the holographic renormalization corresponds to a representation change of the quantum system as discussed by the end of Sec.\ref{HR_OSA}, and thus should not change the normalization property of the wave function required by the conservation of probability. However we should remember that in the context of no-boundary quantum cosmology, we are dealing with generally covariant systems where the probabilist interpretation of the wave functions is subtle. In the bulk it has been established either with respect to classical histories \cite{Vilenkin:1988yd} keeping the general covariance manifest, or with respect to the configuration of the universe at a given moment where the general covariance freedom is gauge-fixed to render the wave functions unitary \cite{Barvinsky:1993jf,Barvinsky:2013aya}. However it is not clear how to consistently establish the boundary probabilist interpretation. At least if we take $|\tilde \Psi|^2$ as the boundary probability density, very likely it is inconsistent with that established in the bulk, since the integral transform relating $\Psi$ and $\tilde \Psi$ is manifestly non-unitary with its kernel ${\cal S}_{\rm ct}$ complex.

Finally an important technical limitation of the work is that the numerics work only for scalar mass above about $0.9$, since otherwise the FindFit command fails to extract the value of $\re \beta$. This is an important reason why $m={\sqrt 3\over 2}$ is not studied in the paper for 3d models. In fact it is interesting to extend the study to the saddle points of very low masses (even below ${\sqrt 3\over 2}$) and  to work out the holographic renormalization. It is mostly because the non-fundamental saddle points of small scalar mass show quite different features from those studied so far of scalar mass above $0.93$. I have not shown in the body of the paper, but actually the same calculation as in Sec.\ref{T2NCS} has been done to some low mass models, but only in the bulk. For example when $m=0.25$ for the non-fundamental saddle points of $k=-1,-2,\dots,-5$ (blue contours in figures \ref{ctNPafPaf} and \ref{RiemannSdSd}), no singularity is observed in the density plots of $\Phi$ which move towards the direction of increasing Lorentzian time. Therefore it is very likely that these saddle points behave like the fundamental ($k=0$) ones, and probably they have $\im \beta$ that do not diverge, and have $\cal I$ and $\tilde {\cal I}$ that are both bounded with the increase of $\alpha$. If this is the case, it means that with the increase of mass, there is a transition in the pattern of behavior of such non-fundamental saddle points. It will be interesting to find a way to study the saddle points of low mass to figure out what happens during this transition. 


\subsubsection*{\underline{\it What next}}

Since the formal formalisms obtained only requires the potential to take the form of Eq.(\ref{Vph2O}), we can consider carrying further the study in Sec.\ref{NoQuPsec}, trying different potentials to see whether novel physics can emerge. It will be especially interesting if some exactly solvable potential can be investigated to work out everything analytically. To start with, maybe a good idea is to obtain such models by analytic continuation from black hole solutions with scalar hair, for example \cite{Henneaux:2002wm}. However it is even more desirable to find exactly solvable models with very low scalar masses (at least lower than $\sqrt 3\over 2$). If possible, this will provide direct solution to the problem raised in the last paragraph.

The work in this paper has produced some indication of scalar divergence and ``high temperature'' divergence in the boundary Hartle-Hawking wave function. However the detail of the divergences do not have the same look as those in \cite{Anninos:2012ft, Anninos:2013rza,Castro:2012gc}. For example, the temperature divergence in \cite{Castro:2012gc} arise at loop level, while in this paper, it appears at tree level with the participation of a scalar field. Also very obviously the Hartle-Hawking wave functions here are even as function of scalar deformation $\alpha$ while in \cite{Anninos:2012ft} it is obviously not the case. However there can still be things to do to render the study more relevant to the work in \cite{Anninos:2012ft}. One immediate possibility that can be envisaged is to study the model of boundary topology $S^3$ of scalar mass $\sqrt 2$. Moreover, using the formula in appendix \ref{ACEQGEN}, the generalization to $S^1\times S^2$ maybe possible. If so we can work out its Hartle-Hawking wave function to compare with the result of $S^3$ boundary and examine the effects of different topologies. At the mean time, it will be extremely interesting if some exactly solvable potential can be found.

\section*{Acknowledgements}

The initiation of the topic is attributed to Frederik Denef and Thomas Hertog. The many discussions with them have been very beneficial. Ruben Monten and Yannick Vreys provided invaluable help on numerics without which a large part of the work in this paper is not possible. In the starting phase of the work, Kristof Moors also helped L.Liu to understand the numerical schemes and some important conceptual issues. This work has also benefited from the discussions or communications with Alice Bernamonti, Adam Bzowski, Gabriele Conti, Federico Galli, Herve Partouche and Hongbao Zhang. L.Liu especially thanks F.Denef for being constantly available and supportive during the roughest time of the work, as well as for his comments on the manuscript.
This work is supported by grants from the John Templeton Foundation and from the Odysseus Programme of the Flemish Research Foundation.

\section*{Notes added}

In the past years in studying the scalar-deformed dS$_4$ no-boundary cosmology, R.Monten and Y.Vreys have tremendously improved the numerical tools initially developed in \cite{Hartle:2007gi,Hartle:2008ng} for computing the Hartle-Hawking wave functions. 
When still working on the present paper, L.Liu had many discussions with them, and thereby could apply the newly developed numerical tools to the models in the present paper. 
The work by Monten and Vreys is not yet published, while their paper is in preparation \cite{RMYV} which will include the important findings in 4d among other results. L.Liu hereby emphasizes the importance of \cite{RMYV} to the present paper and the authorship of the numerical algorithms of Monten and Vreys, although the present paper is finalized earlier.

\begin{appendix}

\vspace{2cm}

\section{Action principles for mini-superspace no-boundary saddle points of boundary topology $S^{d_1}\times S^{d_2}$} \label{ACEQGEN}

This appendix presents the action principles for the mini-superspace no-boundary saddle points minimally coupled to a scalar field. The boundary topology is set to be the product of two spheres: $S^{d_1}\times S^{d_2}$ where $d_1+d_2\geq 1$. Therefore the action principles of all the cases studied in this paper can be derived from the result of this appendix by setting $d_1$ and $d_2$ to appropriate values.

\vspace{1cm}

\subsection*{Action}

Let the mini-superspace time coordinate be $\chi$, and let radii of $S^{d_1}$ and $S^{d_2}$ be respectively $a$ and $b$, which are functions of $\chi$ only. The spacetime metric is
\begin{align}
	\ell^{-2}ds^2=-N^2(\chi)d \chi^2+a^2(\chi)d\Omega_{d_1}^2+b^2(\chi)d\Omega_{d_2}^2,
\end{align} 
where $\ell$ is the dS radius related to the cosmological constant as $\Lambda={(d_1+d_2)(d_1+d_2-1) \over 2\ell^2}$, and $d\Omega_{d_{1,2}}^2$ are the line elements of the spheres $S^{d_{1,2}}$ respectively. Let the minimally coupled scalar field be $\Phi=\Phi(\chi)$ with potential $V(\Phi)$. It a function of $\chi$ only due to the mini-superspace formalism. Then we assume, for simplicity, that the kinetic terms in the Lagrangian are normalized like ${\cal L}\propto R+\dot \Phi^2-2V(\Phi)$ where $R$ is the Ricci scalar. Thus the total action for the no-boundary saddle points, which sums up the Einstein-Hilbert term, the Gibbons-Hawking term and the scalar field term, is
\begin{align}
	2\kappa \, S=&\, \int_{\chi_o}^{\chi_*} d\chi \, a^{d_1}b^{d_2}\Bigg[ d_1(d_1-1) \left(-{\dot a^2\over a^2}+{1\over a^2}\right)+ d_2(d_2-1) \left(-{\dot b^2\over b^2}+{1\over b^2}\right) \nonumber \\ & - 2d_1 d_2 {\dot a  \dot b \over ab}-(d_1+d_2 ) ( d_1+d_2-1)+\dot \Phi^2-2V(\Phi) \Bigg]-2 a_o^{d_1} b_o^{d_2}\Bigg(d_1 {\dot a_o\over a_o}+d_2 {\dot b_o \over b_o}\Bigg). \label{abAction}
\end{align}
Here $\kappa= {16\pi G \over \ell^{D-2} \Omega_1 \Omega_2 }$ with $\Omega_{1,2}$ the surface area of the unit spheres $S^{d_{1,2}}$. The lapse is set to unit for simplicity. The south pole is at $\chi=\chi_o$ and the spacetime boundary of the saddle point is at $\chi=\chi_*$. $a_o$ and $b_o$ stand for $a(\chi_o)$ and $b(\chi_o)$ respectively.

\subsection*{Hamiltonian constraint and equations of motion}

The equations derived from varying the action (\ref{abAction}) are
\begin{align}
	d_1(d_1-1)\left({\dot a^2 \over a^2}+ {1 \over a^2}\right)+d_2(d_2-1)\left({\dot b^2 \over b^2}+ {1 \over b^2}\right) +2d_1d_2 {\dot a\over a}{\dot b \over b} \nonumber \\ -(d_1+d_2)(d_1+d_2-1) -\dot\Phi^2 -2V(\Phi)=0, \label{EqVN}\\
	2(d_1-1){\ddot a\over a}+2 d_2 {\ddot b\over b}+(d_1-1)(d_1-2)\left({\dot a^2 \over a^2}+ {1 \over a^2}\right)+d_2(d_2-1)\left({\dot b^2 \over b^2}+ {1 \over b^2}\right) \nonumber \\+ 2d_2(d_1-1){\dot a\over a}{\dot b \over b} -(d_1+d_2)(d_1+d_2-1)+\dot\Phi^2 -2V(\Phi)=0, \label{EqVa} \\
	2(d_2-1){\ddot b\over b}+2 d_1 {\ddot a\over a}+(d_2-1)(d_2-2)\left({\dot b^2 \over b^2}+ {1 \over b^2}\right)+d_1(d_1-1)\left({\dot a^2 \over a^2}+ {1 \over a^2}\right) \nonumber \\+ 2d_1(d_2-1){\dot a\over a}{\dot b \over b} -(d_1+d_2)(d_1+d_2-1)+\dot\Phi^2 -2V(\Phi)=0, \label{EqVb}\\
	\ddot \Phi+d_1{\dot a\over a}\dot \Phi +d_2 {\dot b\over b}\dot \Phi+V'(\Phi)=0.\label{EqVphi}
\end{align}
We can simplify the second order equations (\ref{EqVa}) and (\ref{EqVb}) using the Hamiltonian constraint. This leads to
\begin{align}
	 {\ddot a \over a}+(d_1-1)\left({\dot a^2 \over a^2} + {1 \over a^2}\right)+ d_2{\dot a\over a } {\dot b \over b} - (d_1+d_2)- {2V(\Phi)  \over d_1+d_2-1}=0, \\
	 {\ddot b \over b}+(d_2-1)\left({\dot b^2 \over b^2} + {1 \over b^2}\right)+ d_1{\dot a\over a } {\dot b \over b} - (d_1+d_2)- {2V(\Phi)  \over d_1+d_2-1}=0.
\end{align}

\subsection*{South pole conditions}

Without loss of generality, we let the first sphere (the $a$-sphere) shrink smoothly to zero size at the south pole $\chi_o$. Thus we can expand the metric and the scalar field near the south pole as follows:
\begin{align}
	a(\chi_o+\epsilon)&\, =i\, \epsilon+a_2 \epsilon^2+O(\epsilon^3),\\
	b(\chi_o+\epsilon)&\,=b_o+b_1 \epsilon +b_2 \epsilon^2+O(\epsilon^3),\\
	\Phi(\chi_o+\epsilon)&\,=\Phi_o+c_1 \epsilon +c_2 \epsilon^2 +O(\epsilon^3).
\end{align}
In order to determine the south pole conditions, arising from the no-boundary proposal, we substitute these expansions into the equations (\ref{EqVN})--(\ref{EqVphi}) and read off the coefficients of different orders of $\epsilon$. 
Solving the resulting algebraic equations which we get
\begin{align}
	& b_1=a_2=c_1=0,\\
	& c_2=-{V'(\Phi_o)\over 2(d_1+1)},\\
	& b_2={ \big[ (d_1+d_2)(d_1+d_2-1)+2V(\Phi_o) \big] b_o^2-(d_1+d_2-1)(d_2-1) \over 2(d_1+1)(d_1+d_2-1)b_o}.
\end{align}
Thus the initial conditions for numerical calculation are, in terms of the Lorentzian time,
\begin{align}
	& a(\chi_o+\epsilon)=i \epsilon ,\ \ \dot a(\chi_o+\epsilon)=i\, ;\\
	& b(\chi_o+\epsilon)=b_o,\nonumber \\  &\dot b(\chi_o+\epsilon)={ \big[ (d_1+d_2)(d_1+d_2-1)+2V(\Phi_0) \big] b_0^2-(d_1+d_2-1)(d_2-1) \over (d_1+1)(d_1+d_2-1)b_o}\, \epsilon\, ;\\
	& \Phi(\chi_o+\epsilon)=\Phi_o,\ \ \dot \Phi(\chi_o+\epsilon)=-{V'(\Phi_o)\over d_1+1} \epsilon\, ,
\end{align}
where $\Phi_o$ and $b_o$ are complex, whose phases are to be adjusted such that the solution fit the boundary conditions assigned at $\chi=\chi_*$.

\vspace{1cm}

\subsection*{Saddle-point action for direct numerical computation}

To compute the saddle-point action, we can simplify the expression (\ref{abAction}) using the hamiltonian constraint (\ref{EqVN}), which gives
\begin{align}
	\kappa \, {\cal S}=&\, \int_{\chi_o}^{\chi_*} dt \, a^{d_1}b^{d_2}\Bigg[ {d_1(d_1-1)\over a^2}+  {d_2(d_2-1)\over b^2} \nonumber \\ & -(d_1+d_2 ) ( d_1+d_2-1)-2V(\Phi) \Bigg]- a_o^{d_1} b_o^{d_2}\Bigg(d_1 {\dot a_o\over a_o}+d_2 {\dot b_o \over b_o}\Bigg). \label{OSA183}
\end{align}
The convention adopted in this paper is to use plain letters for any action in general and calligraphic counterparts for their on-shell values. This expression is useful for models of dimension higher than $3$. In 3d models where $d_1=d_2=1$ or $d_1=2$ and $d_2=0$ or $d_1=0$ and $d_2=2$, more efficient ways of evaluation are available which are worked out in Sec.\ref{HR_OSA} and Sec.\ref{NPS2Act}.

\vspace{1cm}

%
\section{Asymptotic expansions near future boundary of asymptotic dS} \label{ABAsydS}

When performing the holographic renormalization, it is important to know the leading behaviors of the metric and scalar field near the dS boundaries. This appendix presents these asymptotic expansions for the models studied in the paper. Let the coordinate time be $\chi$ and the asymptotic boundary be $\re\!\chi\rightarrow \infty$. To study the asymptotic behaviors, we can mimic the procedure in \cite{Hertog:2011ky}, putting $e^{-\chi}=u$ and expanding the metric components and the scalar field in series of $u$. Then inserting the series into the equations of motion we can work out the coefficients  order by order in principle, and we only need to conserve the leading orders. 

Below to be presented are the results for the two types of boundary topologies covered in the paper: $T^2$ and $S^d$. The scalar potential is assumed to have the form $V(\Phi)={1\over 2} \ell^{-2}m^2 \Phi^2+O(\Phi^4)$, and the mass range considered is ${d^2-1 \over 4} <m^2<{d^2 \over 4}$, where $d=2,3,\dots$ is the spatial dimension.

\subsubsection*{Boundary topology $T^2$}

The equation of constraint and the equations of motion are derived from Eqs (\ref{EqVN})--(\ref{EqVphi}) with $d_1=d_2=1$:
\begin{align}
	& 2u^2 a_ub_u-2ab-ab(u^2 \Phi_u^2+m^2 \Phi^2)+O(\Phi^4)=0, \label{eqT2u1} \\
	& 2u^2 a_{uu}+2 u a_u-2a+a(u^2 \Phi_u^2-m^2 \Phi^2)+O(\Phi^4)=0,  \label{eqT2u2}\\
	& 2u^2 b_{uu}+2 u b_u-2b+b(u^2 \Phi_u^2-m^2 \Phi^2)+O(\Phi^4)=0,  \label{eqT2u3} \\
	& a b(u^2 \Phi_{uu}+u\Phi_u)+u^2(a_ub+a b_u)\Phi_u +m^2 ab\, \Phi+O(\Phi^3) =0. \label{eqT2u4}
\end{align}
Then we can expand the fields as power series of $u$, inserting them into the equations to determine the coefficients of the leading orders. First we need to set $\Phi=0$ and obtain the leading order of $a$ and $b$, and then plug these leading orders into Eq.(\ref{eqT2u4}) to find the leading order of $\Phi$, and then plug the leading order of $\Phi$ back into Eqs (\ref{eqT2u1})--(\ref{eqT2u3}) to find the sub-leading orders of $a$ and $b$. By this procedure, we find the following asymptotic behaviors
\begin{align}
	\Phi(u)= &\, u^{\Delta_-}(\hat \alpha+\dots)+ u^{\Delta_+}(\hat\beta+\dots),  \label{Phi152} \\
	a(u)=&\, {C_a\over u}\left(1-{\hat\alpha^2\over 4} u^{2\Delta_-}  +a_2u^2-{\hat \beta^2\over 4} u^{2\Delta_+}+\dots \right), \label{a153} \\
	b(u)=&\, {C_b\over u}\left(1 -{\hat \alpha^2\over 4} u^{2\Delta_-} +b_2u^2 -{\hat \beta^2\over 4} u^{2\Delta_+}+\dots \right),  \label{b154}
\end{align}
where dots stand for higher orders in $u$; $C_a$, $C_b$, $\hat \alpha$, $\hat \beta$ are constants, and $\Delta_\pm=1\pm\sqrt{1-m^2}$, and $a_2$, $b_2$ are complex constants which satisfy
\begin{align}
	a_2+b_2+m^2\hat\alpha \hat \beta=0. \label{163afbtm2}
\end{align}
Note that due to the range of mass ${d^2-1 \over 4} <m^2<{d^2 \over 4}$ ($d=2$), we have ${1\over 2}<\Delta_-<1$ and ${3\over 2}>\Delta_+>1$.

\subsubsection*{Boundary topology $S^d$}

The equations of motion for spatial topology $S^d$ are 
\begin{align}
	& d(d-1) (u^2 a_u^2+1)-d(d-1)a^2-a^2(u^2 \Phi_u^2+m^2 \Phi^2)+O(\Phi^4)=0, \\
	& d(d-1)(u^2 a_{uu}+ u a_u-a)+a \big[(d-1) u^2 \Phi_u^2-m^2 \Phi^2 \big]+O(\Phi^4)=0, \\
	& a(u^2 \Phi_{uu}+u\Phi_u)+d\, u^2a_u \Phi_u +m^2 a\, \Phi +O(\Phi^3)=0.
\end{align}
Following the same scheme in obtaining Eqs (\ref{Phi152})--(\ref{b154}), we find the asymptotic behaviors to be
\begin{align}
	\Phi= &\, u^{\bar \Delta_-}(\hat \alpha+\dots)+ u^{\bar \Delta_+}(\hat \beta+\dots), \label{asy196} \\
	a=&\, {C_a\over u}\left[1+{ u^2\over 4C_a^2}-{\hat \alpha^2\over 4} u^{2\bar \Delta_-}  - {2m^2 \hat \alpha \hat \beta\over d^2(d-1)} u^d -{\hat \beta^2\over 4} u^{2\bar \Delta_+}+\dots \right], \label{asy197}
\end{align}
where $C_a$, $\hat \alpha$, and $\hat \beta$ are complex constants, and $\bar  \Delta_\pm={1\over 2}\! \left(d\pm\sqrt{d^2-4m^2} \right)$.
Due to the mass range chosen, we have ${d-1\over 2}<\bar  \Delta_-< {d\over 2}$ and ${d+1\over 2} > \bar  \Delta_+ > {d\over 2}$. Puting $d=2$ we obtain the results for the case of topology $S^2$ that we study in detail in the paper:
\begin{align}
	\Phi= &\, u^{\Delta_-}(\hat \alpha+\dots)+ u^{\Delta_+}(\hat \beta+\dots), \\
	a=&\, {C_a\over u}\left[1-{\hat \alpha^2\over 4} u^{2\Delta_-} +\left({1\over 4C_a^2}-{1 \over 2}m^2 \hat \alpha \hat \beta\right) u^2 -{\hat \beta^2\over 4} u^{2\Delta_+}+\dots \right],  \label{asy213}
\end{align}
where $C_a$, $\hat \alpha$, and $\hat \beta$ are complex constants, and $\Delta_\pm=1\pm\sqrt{1 -m^2}$.

\end{appendix}
%
%
%

\end{document}